\documentclass[12pt,preprint]{aastex}

\usepackage{color}
\usepackage{epsfig}
\usepackage[fleqn]{amsmath}
\usepackage{multirow}
\usepackage{multicol}
\usepackage{mathptmx}
\usepackage{graphicx}
\usepackage{bm}
\usepackage{mathrsfs}
\shorttitle{Charged  particle motions} \shortauthors{Zhang et al.}
\begin{document}

\title{\large Charged particle motions near non-Schwarzschild black holes with external magnetic fields in modified theories
of gravity}

\author{Hongxing Zhang$^{1,2}$, Naying Zhou$^{1,2}$, Wenfang Liu$^{1}$, Xin Wu$^{1,2,3, \dag}$}
\affil{ 1. School of Mathematics, Physics and Statistics, Shanghai
University of Engineering Science, Shanghai 201620, China \\
2. Center of Application and Research of Computational Physics,
Shanghai University of Engineering Science, Shanghai 201620, China \\
3. Guangxi Key Laboratory for Relativistic Astrophysics, Guangxi
University, Nanning 530004, China} \email{M130120111@sues.edu.cn
(H. Z.); M130120101@sues.edu.cn (N. Z.);  21200007@sues.edu.cn (W.
L.); $\dag$ Corresponding Author Email: wuxin$\_$1134@sina.com,
xinwu@gxu.edu.cn (X. W.)}
\begin{abstract}

A small deformation controlled by four free parameters to the
Schwarzschild metric could be referred to a nonspinning black hole
solution in alternative theories of gravity. Because such a
non-Schwarzschild metric can be changed into a Kerr-like black
hole metric via a complex coordinate transformation, the recently
proposed time-transformed explicit symplectic integrators for the
Kerr type spacetimes are suitable for a Hamiltonian system
describing the motion of charged particles around the
non-Schwarzschild black hole surrounded with an external magnetic
field. The obtained explicit symplectic methods are based on a
time-transformed Hamiltonian split into seven parts, whose
analytical solutions are explicit functions of new coordinate
time. Numerical tests show that such explicit symplectic
integrators for intermediate time-steps perform good long-term
performance in stabilizing Hamiltonian errors regardless of
regular or chaotic orbits. One of the explicit symplectic
integrators  with the techniques of Poincar\'{e} sections and fast
Lyapunov indicators is applied to investigate the effects of the
parameters including the four free deformation parameters on the
orbital dynamical behavior. From the global phase-space structure,
chaotic properties are typically strengthened under some
circumstances as any one of the energy and the magnitudes of
magnetic parameter and negative deformation parameters increases.
However, they are weakened when each of the angular momentum and
positive deformation parameters increases.

\end{abstract}
\section{Introduction}

A Schwarzschild  solution  describing a  nonrotating black hole
and a Kerr  solution  describing a rotating black hole are two
exact solutions of  the Einstein's field equations of general
relativity in vacuum. According to the no-hair theorem,
astrophysical (Kerr) black holes have their masses and spins as
their unique characteristics. The theoretical prediction of the
existence of black holes has been confirmed frequently by  a
wealth of observational evidence, such as X-ray binaries [1,2],
detections of gravitational waves [3,4] and event-horizon-scale
images of M87 [5,6].

Observational tests of strong-field gravity features cannot be
based on a priori hypothesis about the correctness of  general
relativity. Instead, such tests must allow ansatz metric solutions
to deviate from the general relativistic black hole scenarios
predicted by the no-hair theorem. These metric solutions often
come from perturbations of the usual Schwarzschild (or Kerr) black
hole or exact solutions in alternative (or modified) theories of
gravity. A small deformation to the Schwarzschild metric
describing a nonspinning black hole (i.e., a modified
Schwarzschild metric) [7] could be required to satisfy the
modified field equations in dynamical Chern-Simons modified
gravity [8,9]. Applying the Newman-Janis algorithm and a complex
coordinate transformation, Johannsen and Psaltis [10] transformed
such a Schwarzschild-like metric with several free deformation
parameters into a Kerr-like metric, including a set of free
deformation parameters as well as mass and spin. This Kerr-like
metric, which is a parametric deformation of the Kerr solution and
is not a vacuum solution, is regular everywhere outside of the
event horizon. These metric deformations away from the
Schwarzschild or Kerr metric by one or more parameters contain
modified multipole structures. Although the $\gamma$  metric (or
Zipoy-Voorhees metric) describing a static and axially symmetric
field [11,12] is also a parameterizing deviation  from the
Schwarzschild solution for $\gamma\neq 1$, it is an exact solution
of Einstein's equations in vacuum.

In addition to the above-mentioned simply modified theories of
gravity like scalar-tensor gravity, many other forms of modified
theories of gravity can be found in the literature. Some examples
are scalar-tensor theories including Brans-Dicke theory [13,14]
and general scalar-tensor theories [15-19], Einstein-$\ae$ther
theories [20], Bimetric theories [21,22], tensor-vector-scalar
theories [23,24], Einstein-Cartan-Sciama-Kibble theory [25,26],
scalar-tensor-vector theory [27], $f(R)$ theories [28-30], $f(G)$
theory [31,32], Ho\v{r}ava-Lifschitz gravity [33-35] and higher
dimensional theories of gravity [36-38]. Researchers and students
in cosmology and gravitational physics see also review articles
[29,30,39-41] for more information on these modified gravity
theories. Black-hole solutions in modified theories of gravity are
generally unlike those in general relativity, and include many
additional free parameters as well as the parameters predicted by
the no-hair theorem in general relativity. Although a solution in
a modified gravity model can be mathematically equivalent to a
scalar field model, this mathematical correspondence does not
always mean physical equivalence. The two corresponding solutions
may have different physical behaviors. Corrections to the
classical Einsteinian black hole entropy are necessary so as to
constrain the viability of modified gravity theories in the study
of Schwarzschild-de Sitter black holes by the use of the Noether
charge method [42]. However, Not all black-hole solutions in
modified theories of gravity must necessarily dissatisfy the
Einstein field equations. For example, a stationary black-hole
solution of the Brans-Dicke field equations must be that of the
Einstein field equations [43]; this result is still present if no
symmetries apart from stationarity are assumed [44]. The Kerr
metric also remains a solution of certain $f(R)$ theories [45].

A deep understanding of the relevant properties of the standard
general relativistic black hole solutions  and  particle motions
in the vicinity of the black holes is important to study accretion
disk structure, gravitational lensing, cosmology and gravitational
wave theory. Observational data from the vicinity of the circular
photon orbits or the innermost stable circular orbits could be
used as tests of the no-hair theorem. The properties of innermost
stable circular orbits are useful to understand the energetic
processes of a black hole. For this reason, radial effective
potentials and (innermost) stable circular orbits of charged
particles in electromagnetic fields surrounding a black hole have
been extensively investigated in a large variety of papers (see,
e.g., [46-51]). The  motions of charged particles in the
equatorial plane sound simple, but off-equatorial motions of
charged particles in the magnetic fields become very complicated.
In a stationary and axisymmetric black hole solution, there are
three conserved quantities including the energy, angular momentum
and rest mass of a charged particle.  The fourth invariable
quantity related to the azimuthal motion of the particle is
destroyed in general when an electromagnetic field is included
around the black hole. Thus, the particle motion in the spacetime
background is not an integrable system. Chaos describing a
dynamical system with sensitive dependence on initial conditions
can occur in some circumstances. Various aspects of chaotic
motions of charged particles around the standard general
relativistic black holes perturbed by weak external sources like
magnetic fields were discussed in many references (see, e.g.,
[52-59]).

Thanks to the importance of  the deformed (or modified) black hole
solutions in tests of strong-field gravity features of general
relativity, the motions of charged particles in the modified
solutions without or with perturbations of weak external sources
are naturally taken into account by some authors. The authors of
[10] focused on a question how the radii of the innermost stable
circular orbits and circular photon orbits vary with increasing
values of the spin and deviation parameters in a Kerr-like metric
of a rapidly rotating black hole. They demonstrated that their
Kerr-like metric is suitable for strong-field tests of the no-hair
theorem in the electromagnetic spectrum. Charged particle motions
around non-Schwarzschild or non-Kerr black hole immersed in an
external uniform magnetic field were considered in [60-62]. The
influence of a magnetic field on the radial motion of a charged
test particle around a black hole surrounded with an external
magnetic field in Ho\v{r}ava-Lifshitz gravity was investigated in
[63-65]. The radial motions of charged particles in the $\gamma$
spacetime in the presence of an external magnetic field was
studied in [66]. In fact, the $\gamma$ spacetime is nonintegrable
and can allow for the onset of chaos if the external magnetic
field is not included in [67]. The authors of [68] gave some
insight into the effect of one deformation parameter on chaos of
charged particles in the vicinity of non-Schwarzschild black hole
with external magnetic field.

Numerical integration methods are vital to detect the chaotical
behavior of charged particles in the vicinity of  the standard
general-relativistic or modified black hole solutions without or
with perturbations from weak external sources. They should have
good stability and high precision so as to provide reliable
results to detecting the chaotical behavior. The most appropriate
long-term integration solvers for Hamiltonian systems are a class
of symplectic integrators which respect the symplectic structures
of Hamiltonian dynamics [69,70]. The motions of charged particles
near the black holes without or with weak external sources can be
described by Hamiltonian systems, and thus allow for the
applicability of symplectic methods. If the Hamiltonian systems
are split into two parts, explicit symplectic integration
algorithms are not available in general. However, implicit
symplectic integrators, such as the  implicit midpoint rule
[71,72] and implicit Gauss-Legendre Runge-Kutta symplectic schemes
[54,73,74], are always suitable for their applications to these
Hamiltonian systems that do not need any separable forms. When the
Hamiltonians are separated into one part with explicit analytical
solutions and another part with implicit solutions, explicit and
implicit combined symplectic methods could be constructed in
[75-79]. The implicit algorithms are more computationally
demanding than the explicit ones in general, therefore, the
explicit symplectic integrations should be developed as much as
possible. Recently, the authors of [80-82] successfully
constructed the explicit symplectic integrators for the
Schwarzschild type black holes without or with external magnetic
fields by splitting the corresponding Hamiltonians into several
parts having analytical solutions as explicit functions of proper
time. More recently, the time-transformed explicit symplectic
integrators were designed for the Kerr family spacetimes [83-85].

The idea for constructing the time-transformed explicit symplectic
integrators as well as the explicit symplectic integrators
introduced in [80-83] allows for the applicability of many
standard general-relativistic or modified black hole solutions
with or without  perturbations of weak external sources. In spite
of this, there is no universal rule on  how to construct explicit
symplectic integrators for Hamiltonians corresponding to the
spacetimes. Specific Hamiltonian problems have different
separations, or different choices of time-transformed Hamiltonians
and their splitting forms. As is claimed above, the
non-Schwarzschild metric with four free deformation parameters
could produce a Kerr-like metric through a complex coordinate
transformation [10]. Now, there is a question whether the
time-transformed explicit symplectic integrators for the Kerr type
spacetimes [83] are applicable to such a deformed
non-Schwarzschild black hole immersed in an external magnetic
field. We plan to address this question in this paper. In
addition, we mainly pay attention to the effects of the four free
deformation parameters on the chaotical behavior. The present work
is unlike Ref. [68] in which one deformation parameter is added to
the non-Schwarzschild metric and no explicit symplectic
integrators are considered.

The remainder of this paper is organized as follows. A  metric
deformation to the Schwarzschild spacetime is introduced in
Section 2. Time-transformed explicit symplectic integrators are
described in Section 3. Orbital dynamical properties are discussed
in Section 4. Finally, the main results are concluded in Section
5.

\section{Deformed Schwarzschild metric}

In Schwarzschild coordinates $(t,r,\theta ,\varphi)$, a
Schwarzschild-like metric $ds^2 =g_{\alpha \beta }dx^\alpha
dx^\beta$ is written in [7,10] as
\begin{eqnarray}
   ds^2 &=& -f(1+h)dt^2+f^{-1}(1+h)dr^2 +r^2d\theta ^2+r^2\sin^2\theta d\varphi ^2, \\
f &=& 1-\frac{2M}{r}, \nonumber \\
h &=& k_0+\frac{k_1M}{r}+\frac{k_2M^2}{r^2}+\frac{k_3M^3}{r^3}.
\end{eqnarray}
Here, $M$ denotes a mass of the black hole. The speed of light $c$
and the gravitational constant $G$ are taken as geometric units,
$c=G=1$. Deformation function $h$ is a  perturbation to the
Schwarzschild metric, where $k_0$, $k_1$, $k_2$ and $k_3$ are
deformation parameters. It comes from modified multipole
structures related to spherical deformations of the star. When the
action through algebraic, quadratic curvature invariants coupled
to scalar fields is  modified, such small deformations to the
Schwarzschild metric are obtained from the modified field
equations and the scalar field's equation in the dynamical theory.
Clearly, Eq. (1) with $h=0$ corresponds to the Schwarzschild
metric. When $h\neq 0$, Eq. (1) looks like the Schwarzschild
metric but can be transformed into a Kerr-like black-hole metric
by the Newman-Janis algorithm [86] and a complex coordinate
transformation [10].

Suppose the black hole is immersed in an external electromagnetic
field with a four-vector potential
\begin{eqnarray}
    A_\mu =\frac{1}{2}\delta _\mu ^\varphi  Br^2\sin^2\theta,
\end{eqnarray}
where $B$ is a constant strength of the uniform magnetic field.
The motion of a test particle with mass $m$ and charge $q$ is
described in the following Hamiltonian
\begin{eqnarray}
H=\frac{1}{2m}g^{\mu \nu}(p_\mu -qA_\mu )(p_v-qA_\nu).
\end{eqnarray}
Here, $p_\mu$ is a generalized momentum, which is determined by
\begin{eqnarray}
\dot{x}^\mu=\frac{\partial H}{\partial p_\mu}=\frac{1}{m}g^{\mu
\nu}(p_\nu-qA_\nu),
\end{eqnarray}
equivalently,
\begin{eqnarray}
p_\mu=m\dot{x}^\nu g_{\mu \nu}+qA_\mu.
\end{eqnarray}
The 4-velocity $\dot{x}^\mu$ is a derivative of the coordinate
$x^\mu$ with respect to proper time $\tau$. Because the
Hamiltonian equations satisfy Eq. (5) and
\begin{eqnarray}
\dot{p}_\mu=-\frac{\partial H}{\partial x^\mu},
\end{eqnarray}
$p_t$ and $p_\varphi$ are two constants of motion:
\begin{eqnarray}
p_t &=& m\dot{t} g_{tt}=-m\dot{t}f(1+h)=-E, \\
p_\varphi &=& m\dot{\varphi} g_{\varphi\varphi}+qA_\varphi =
mr^2\dot{\varphi}\sin^2 \theta+\frac{1}{2}qBr^2\sin^2\theta
    =L.
\end{eqnarray}
$E$ is an energy of the particle, and $L$ is an angular momentum
of the particle.

For simplicity, dimensionless operations are given to the related
quantities as follows: $t \rightarrow tM$, $ \tau \rightarrow \tau
M$, $ r\rightarrow rM$, $ B\rightarrow B/M$, $E\rightarrow mE$,
$p_r\rightarrow mp_r$, $L\rightarrow mML$, $p_\theta\rightarrow
mMp_\theta$, $q\rightarrow mq$ and $H\rightarrow mH$. In this way,
$M$ and  $m$ in Eqs. (1)-(9) are taken as geometric units,
$m=M=1$. The Hamiltonian (4) has two degrees of freedom
$(r,\theta)$ in a four-dimensional phase space $(r,\theta, p_r,
p_\theta)$, and can be rewritten as a dimensionless form
\begin{eqnarray}
    H =-\frac{E^2}{2f(1+h)}+\frac{1}{2r^2\sin^2\theta }(L-\frac{1}{2}Qr^2\sin^2\theta )^2 +\frac{fp_r^2}{2(1+h)}+\frac{p_\theta ^2}{2r^2},
\end{eqnarray}
where $Q=Bq$.

Besides the two constants (8) and (9), the conserved Hamiltonian
quantity
\begin{eqnarray}
    H=-\frac{1}{2}
\end{eqnarray}
is a third constant of the system (10). The third constant of
motion exists due to the invariance of the 4-velocity or the rest
mass of the particle in the time-like spacetime (1). Given $Q=0$,
the system (10) holds a fourth constant of motion and therefore is
integrable and nonchaotic. When $Q\neq 0$, the system (10) has no
fourth constant and then becomes nonintegrable. In this case,
analytical solutions cannot be given to the system (10), but
numerical solutions can.

\section{Explicit symplectic integrations}

First, time-transformed explicit symplectic methods for the system
(10) is introduced. Then, their performance is numerically
evaluated.

\subsection{Design of algorithms}

As is claimed above, the metric (1) seems to be the Schwarzschild
metric but the system (10) is not suitable for the application of
the explicit  symplectic methods suggested in [80-82] because the
Hamiltonian (10) is unlike the Hamiltonians of the Schwarzschild
type spacetimes (including the Reissner-Nordstr\"{o}m metric, the
Reissner-Nordstr\"{o}m-(anti)-de Sitter solution and these
spacetimes perturbed by external magnetic fields), which can be
separated into several parts having analytical solutions as
explicit functions of proper time $\tau$. Since the
Schwarzschild-like metric (1) can correspond to a Kerr-like metric
via some coordinate transformation [12], the time-transformed
explicit symplectic methods for the Kerr type spacetimes proposed
in [73] are guessed to be applicable to the system (10). The
implementation of the algorithms is detailed below.

Extending the phase-space variables $(p_r,p_\theta;  r ,\theta )$
of the Hamiltonian (10) to $(p_r,p_\theta,p_0; r,\theta,q_0)$,
where $\tau $ is viewed as a new coordinate $q_0=\tau$ and its
corresponding momentum is $p_0$ with $p_0=-H=1/2\neq p_t$, we have
an extended phase-space Hamiltonian
\begin{eqnarray}
    J =H+p_0.
\end{eqnarray}
It is clear that $J$ is always identical to zero, $J=0$. Taking a
time transformation
\begin{eqnarray}
    d\tau &=& g(r)dw, \\
    g(r) &=& 1+h,
\end{eqnarray}
we get a new time transformation Hamiltonian
\begin{eqnarray}
    \mathcal{H} = g(r)J &=& -\frac{E^2}{2f}+\frac{(1+h)(L-\frac{1}{2}Qr^2\sin^2\theta )^2}{2r^2\sin^2\theta}+\frac{p_r^2}{2}-\frac{p_r^2}{r} \nonumber \\
                 && +\frac{(1+k_0)p_\theta ^2}{2r^2}+\frac{k_1p_\theta ^2}{2r^3} +\frac{k_2p_\theta ^2}{2r^4}+\frac{k_3p_\theta
                 ^2}{2r^5}+p_0g(r).
\end{eqnarray}
The Hamiltonian $\mathcal{H}$ has new coordinate time variable $w$
and the phase-space variables $(p_r,p_\theta,p_0; r,\theta,q_0)$.
Because $J=0$, $ \mathcal{H}=0$.

Similar to the Hamiltonians of the Schwarzschild type spacetimes
in Refs. [80-82], the time-transformed Hamiltonian $\mathcal{H}$
can be split in the following form
\begin{eqnarray}
\mathcal{H} =\mathcal{H}_1 +\mathcal{H}_2 +\mathcal{H}_3
+\mathcal{H}_4 +\mathcal{H}_5 +\mathcal{H}_6 +\mathcal{H}_7,
\end{eqnarray}
where sub-Hamiltonians read
\begin{eqnarray}
    \mathcal{H}_1 &=&-\frac{E^2}{2f}+\frac{(1+h)(L-\frac{1}{2}Qr^2\sin^2\theta )^2}{2r^2\sin^2\theta } +p_0(1+h), \\
    \mathcal{H}_2&=&\frac{p_r^2}{2},\\
    \mathcal{H}_3&=&-\frac{p_r^2}{r},\\
    \mathcal{H}_4&=&\frac{(1+k_0)p_\theta ^2}{2r^2},\\
    \mathcal{H}_5&=&\frac{k_1p_\theta ^2}{2r^3},\\
    \mathcal{H}_6&=&\frac{k_2p_\theta ^2}{2r^4},\\
    \mathcal{H}_7&=&\frac{k_3p_\theta ^2}{2r^5}.
\end{eqnarray}
Each of the seven sub-Hamiltonians is analytically solvable and
its solutions are explicit functions of the new coordinate time
$w$. $\mathcal{A}$, $\mathcal{B}$, $\mathcal{C}$, $\mathcal{D}$,
$\mathcal{E}$, $\mathcal{F}$ and $\mathcal{G}$ are differential
operators, which correspond to $\mathcal{H}_1$,  $\mathcal{H}_2$,
$\mathcal{H}_3$, $\mathcal{H}_4$,  $\mathcal{H}_5$,
$\mathcal{H}_6$ and $\mathcal{H}_7$, respectively. These operators
are written as
\begin{eqnarray}\nonumber
\mathcal{A} &=&-\frac{\partial \mathcal{H}_1 }{\partial r}
\frac{\partial}{\partial p_r}-\frac{\partial \mathcal{H}_1 }
{\partial \theta } \frac{\partial}{\partial p_\theta
}+\frac{\partial \mathcal{H}_1 }
{\partial p_0 } \frac{\partial}{\partial q_0 } \\
&=&f_1 \frac{\partial}{\partial p_r}+f_2 \frac{\partial}{\partial
p_\theta }+(1+h) \frac{\partial}{\partial q_0 },\\
\nonumber f_1 &=&\frac{k_1}{2r^2}+\frac{k_2}{r^3} +
\frac{3k_3}{2r^4} -\frac{E^2}{r^2(\frac{2}{r}- 1) ^2}   \\
\nonumber &&+\frac{(L -
\frac{Qr^2\sin^2\theta}{2})^2(\frac{k_1}{r^2} + \frac{2k_2}{r^3}
+\frac{3k_3}{r^4})}{2r^2\sin^2\theta }\\ \nonumber
&&+(L-\frac{Qr^2\sin^2\theta }{2})[Q+\frac{(L -
\frac{Qr^2\sin^2\theta}{2})}{r^2\sin^2\theta }]  \\
\nonumber && \cdot(\frac{k_0+1}{r} + \frac{k_1}{r^2}+
\frac{k_2}{r^3} +\frac{k_3}{r^4})-p_0\frac{\partial h}{\partial r},\\
\nonumber f_2&=&(L-\frac{Qr^2\sin^2\theta }{2})[Q+\frac{(L -
\frac{Qr^2\sin^2\theta}{2})}{r^2\sin^2\theta }]\\ \nonumber &&
\cdot(k_0 +1+ \frac{k_1}{r}+ \frac{k_2}{r^2}
+\frac{k_3}{r^3})\cot\theta,
\end{eqnarray}
\begin{eqnarray}
    \mathcal{B} &=&p_r\frac{\partial}{\partial p_r},
\end{eqnarray}
\begin{eqnarray}
\mathcal{C} &=&-\frac{2}{r}p_r\frac{\partial}{\partial
r}-\frac{p_r^2}{r^2}\frac{\partial}{\partial p_r},
\end{eqnarray}
\begin{eqnarray}
\mathcal{D} &=&\frac{(1+k_0)p_\theta
}{r^2}\frac{\partial}{\partial \theta }-\frac{(1+k_0)p_\theta
^2}{r^3}\frac{\partial}{\partial p_r},
\end{eqnarray}
\begin{eqnarray}
\mathcal{E} &=& \frac{k_1p_\theta}{r^3}\frac{\partial}{\partial
\theta }-\frac{3}{2}\frac{k_1p_\theta
^2}{r^4}\frac{\partial}{\partial p_r},
\end{eqnarray}
\begin{eqnarray}
\mathcal{F} &=&\frac{k_2p_\theta}{r^4}\frac{\partial}{\partial
\theta }-2\frac{k_1p_\theta ^2}{r^5}\frac{\partial}{\partial p_r},
\end{eqnarray}
\begin{eqnarray}
\mathcal{G} &=&\frac{k_3p_\theta}{r^5}\frac{\partial}{\partial
\theta }-\frac{5}{2}\frac{k_3p_\theta
^2}{r^6}\frac{\partial}{\partial p_r}.
\end{eqnarray}

The solutions $\textbf{z}=(r,\theta, q_0,p_r,p_\theta)^{T}$ for
the time-transformed Hamiltonian $\mathcal{H}$ advancing a new
coordinate time step $\Delta w=\sigma$ from the initial solutions
$\textbf{z}(0)=(r_0,\theta _0, q_{00}, p_{r0}, p_{\theta 0})^{T}$
can be given by
\begin{eqnarray}
\textbf{z}=S_2^\mathcal{H}(\sigma)\textbf{z}(0),
\end{eqnarray}
where $S_2^\mathcal{H}$ is symmetric products of exponents of the
seven operators and has the expressional form
\begin{eqnarray}
S_2^\mathcal{H}(\sigma)&=& e^{\frac{\sigma}{2}\mathcal{G} }\times
e^{\frac{\sigma}{2}\mathcal{F} }\times
e^{\frac{\sigma}{2}\mathcal{E} } \times
e^{\frac{\sigma}{2}\mathcal{D}}\times
e^{\frac{\sigma}{2}\mathcal{C}}\times
e^{\frac{\sigma}{2}\mathcal{B}}\times e^{\mathcal{\sigma
A}}\times e^{\frac{\sigma}{2}\mathcal{B}} \nonumber \\
&& \times e^{\frac{\sigma}{2}\mathcal{C} }\times
e^{\frac{\sigma}{2}\mathcal{D} } \times
e^{\frac{\sigma}{2}\mathcal{E} }\times
e^{\frac{\sigma}{2}\mathcal{F} }\times
e^{\frac{\sigma}{2}\mathcal{G} }.
\end{eqnarray}
Such  symmetric products are a component of symplectic operators
at second order. The symplectic method $S_2$ is an extension to
the works of [83-85] regarding the time-transformed explicit
symplectic methods for the Kerr spacetimes. Of course, such
symmetric products of order 2 easily yield a fourth-order
construction of Yoshida [86]
\begin{eqnarray}
S_4^{\mathcal{H}}=S_2^{\mathcal{H}}(\gamma \sigma)\times
S_2^{\mathcal{H}}(\delta \sigma)\times S_2^{\mathcal{H}}(\gamma
\sigma),
\end{eqnarray}
where $\gamma =1/(1-\sqrt[3]{2}) $ and $ \delta =1-2\gamma$.

\subsection{Numerical Evaluations}

Let us choose parameters $E=0.9965$, $L=4$, $Q=6\times 10^{-4}$,
$k_0=10^{-3}$, $k_1=10^{-2}$, $k_2=10^{-1}$ and $k_3=1$. The
initial conditions are $p_r=0$ and $\theta=\pi/2$. The initial
value $r=15$ for Orbit 1, and $r=50$ for Orbit 2. The initial
values $p_{\theta}>0$ for the two orbits are determined by Eq.
(11).

Given the time step $\sigma=1$, the errors of the Hamiltonian $J$
for the second-order method $S_2$ and the fourth-order method
$S_4$ solving Orbit 1 have no secular drifts. The errors are three
orders of magnitude smaller for $S_4$ than for $S_2$ before the
integration time $w=10^7$, as shown in Figure 1(a). With the
integration spanning this time and tending to $w=10^8$, the errors
still remain bounded for $S_2$, but exhibit long-term growths for
$S_4$. The secular drifts of the Hamiltonian errors for $S_4$ are
due to roundoff errors. When the number of integration steps is
small, the truncation errors are more important than the roundoff
errors. As the integration is long enough, the roundoff errors are
dominant errors and cause the Hamiltonian errors to grow with
time. However, such error drifts for $S_4$ lose when a larger time
step $\sigma=4$ is adopted. If Orbit 1 is replaced with Orbit 2,
the Hamiltonian errors for each of the two methods are not
explicitly altered.

In what follows, $S_4$ with the time step $\sigma=4$ is used.
Figure 1(b) describes the relationship between the proper time
$\tau$ and the new coordinate time $w$ when Orbit 1 is tested.
Clearly, $w$ is almost equal to $\tau$. This result coincides with
the theoretical result $g\approx 1+k_0\approx 1$ when $r\gg 2$ and
$k_0\approx 0$. Therefore, the time transformation function $g$ in
Eq. (14) mainly plays an important role in implementing the
desired separable form of the time-transformed Hamiltonian
$\mathcal{H}$ rather than  adaptive control to time steps.

\section{Regular and chaotic dynamics of orbits}

The regularity of Orbit 1 and the chaoticity of Orbit 2 are
clearly shown through the Poincar\'{e} section at the plane
$\theta=\pi/2$ with $p_{\theta}>0$ in Figure 1(c). The phase-space
of Orbit 1 is a Kolmogorov-Arnold-Moser (KAM) torus, which belongs
to the characteristic of a regular quasi-periodic orbit. For Orbit
2, many discrete points are densely, randomly filled with an area
and are regarded as the characteristic of a chaotic orbit. The
Hamiltonian errors for $S_4$ acting on Orbit 1 are approximately
same as those for $S_4$ acting on Orbit 2. This fact indicates
that the algorithmic performance in the Hamiltonian error behavior
is regardless of the regularity or chaoticity of orbits.

Now, we continue to use the technique of Poincar\'{e} section to
trace the orbital dynamical evolution. The parameters are the same
as those in Figure 1 but $Q=8\times 10^{-4}$, $k_0=10^{-4}$, and
different values $E$ are given. When $E=0.991$ in Figure 2(a), the
plotted seven orbits are ordered. As the energy increases, e.g.,
$E=0.9925$, three of the orbits are chaotic in Figure 2(b). For
$E=0.9975$ in Figure 2(c), chaos is present almost elsewhere in
the whole phase space. These results indicate an increase of the
energy leading to enhancing the strength of chaos from the global
phase-space structure. However,  the chaotic properties are
weakened as the particle angular momenta $L$ increase, as shown in
Figure 3.

Besides the technique of Poincar\'{e} section, Lyapunov exponents
for measuring an exponential rate of the separation between two
nearby orbits with time are often used to detect chaos from order.
The largest Lyapunov exponent is defined in [88] by
\begin{eqnarray}
\lambda=\lim_{w\rightarrow\infty}\frac{1}{w}\ln\frac{d(w)}{d0},
\end{eqnarray}
where $d0$ is the starting separation between the two nearby
orbits  and $d(w)$ is the distance between the two nearby orbits
at time $w$. However, it takes long enough time to obtain
stabilizing values of the Lyapunov exponents. Instead, a fast
Lyapunov indicator (FLI), as a quicker method to distinguish
between the ordered and chaotic two cases, is often used. It comes
from a slightly modified version of the largest Lyapunov exponent,
and is calculated in [88] by
\begin{eqnarray}
FLI=\log_{10}\frac{d(w)}{d0}.
\end{eqnarray}
An exponential growth of FLI with time $\log_{10}w$ means the
chaoticity of an bounded orbit, whereas a power law growth of FLI
shows the regularity of an bounded orbit. When the integration
time arrives at $10^6$, the FLIs in Figure 4(a) can clearly
identify the regular and chaotic properties of three energies
corresponding to the orbits with the initial separation $r=15$ in
Figure 2. The regular and chaotic properties of three angular
momenta corresponding to the orbits with the initial separation
$r=70$ in Figure 3 are also described the FLIs in Figure 4(b).
Clearly, the angular momentum $L=4.4$ corresponds to the
regularity, whereas the angular momenta $L=3.85$ and $L=4$
correspond to chaos. Chaos is stronger for $L=3.85$ than for
$L=4$. As far as the Poincar\'{e} sections and FLIs are concerned,
they are two popular methods to detect chaos from order. The
technique of Poincar\'{e} sections can clearly, intuitively
describe the global phase-space structure, but is mainly
applicable to conservative systems with two degrees of freedom or
four dimensional phase spaces. The method of FLIs is suitable for
any dimensions.

Taking the parameters $L = 4$, $k_0 = 10^{-4}$, $k_1 = 10^{-2}$,
$k_2 = 10^{-1}$ and $k_3 = 1$, we employ the technique of
Poincar\'{e} sections to plot the global phase-space structures
with $E = 0.9915$ for three positive values of the magnetic
parameter $Q$ in Figures 5 (a)-(c).  When $Q = 5\times10^{-4}$,
all orbits are regular KAM tori in Figure 5(a). Given $Q =
8\times10^{-4}$ in Figure 5(b), many tori are twisted and a few
orbits can be chaotic. When $Q=10^{-3}$ in Figure 5(c), the number
of chaotic orbits increases and the strength of chaos is enhanced.
In other words, an increase of the positive magnetic parameter is
helpful to induce the occurrence of chaos. How does a negative
magnetic parameter affect the chaotic behavior as the magnitude of
the negative magnetic parameter increases? The key to this
question can be found in Figures 5 (d)-(f) with $E = 0.9975$. No
chaos exists for $Q = - 10^{-4}$ in Figure 5(d). Three chaotic
orbits are plotted for $Q=-8\times 10^{-4}$ in Figure 5(e). More
orbits can be chaotic when $Q = - 10^{-3}$ in Figure 5(f). That is
to say, the chaotic properties from the global phase-space
structures are typically strengthened with an increase of the
absolute value of the negative magnetic parameter. In short, chaos
becomes stronger as the magnitude of the positive or negative
magnetic parameter ($|Q|$) varies from small to large. This result
is also supported by the FLIs in Figure 6. Here, the FLI for a
given value of $Q$ is obtained after the integration time
$w=2\times10^6$. All FLIs that are not less than 6 correspond to
the onset of chaos, while those that are less than this value turn
out to indicate the regularity of orbits. When
$Q>8.5\times10^{-4}$ in Figure 6(a) or $Q<-7.5\times10^{-4}$ in
Figure 6(b), a dynamical transition from order to chaos occurs.

Now, let us focus on the dependence of chaos on the deformation
parameters. Chaos becomes weaker when the  deformation parameter
$k_0$ is positive and increases in Figures 7 (a)-(c). However, it
gets stronger when the deformation parameter $k_0$ is negative and
its magnitude increases in Figures 7 (d)-(f). The effects of the
deformation parameter $k_0$ on chaos described by the technique of
Poincar\'{e} sections are consistent with those described by the
method of FLIs in Figure 8. The effects of the other deformation
parameters on chaos are shown through the methods of Poincar\'{e}
sections and FLIs in Figures 9-14. They are similar to the effect
of the deformation parameter $k_0$ on chaos. Precisely speaking,
an increase of any one of the positive deformation parameters
$k_1$, $k_2$ and $k_3$ weakens the chaotic properties, while an
increase of each of the magnitudes of the negative deformation
parameters $k_1$, $k_2$ and $k_3$ strengthens the chaotic
properties. The result regarding the effects of the four
deformation parameters on the chaotic properties is similar to the
result of [69] for describing the effect of deformation parameter
$k_3$ on the chaotic properties.

The above demonstrations clearly show how small changes of these
parameters affect the dynamical transitions from order to chaos.
The main result is that chaos in the global phase space is
strengthened as any one of the energy $E$, magnetic parameter
$|Q|$, absolute values of the negative deformation parameters
($|k_0|$, $|k_1|$, $|k_2|$ and $|k_3|$) increases, but weakened
when any one of the angular momentum $L$ and positive deformation
parameters $k_0$, $k_1$, $k_2$ and $k_3$ increases. Here, an
interpretation is given to the result.  Expanding $1/f$ in the
Taylor series, we rewrite Eq. (17) at the equatorial plane
$\theta=\pi/2$  as
\begin{eqnarray}
\mathcal{H}_1 &\approx&
\frac{1}{2}[(1+k_0)(1-LQ)-E^2+\frac{k_2}{4}Q^2]
-\frac{E^2}{r}+\frac{Q^2}{8}
(1+k_0)r^2 \nonumber \\
&& +\frac{L^2}{2r^2}(1+k_0)+\frac{L^2k_1}{2r^3}
+\frac{1-LQ}{2}(\frac{k_1}{r}+\frac{k_2}{r^2} +\frac{k_3}{r^3})
+\frac{k_3}{2r}Q^2.
\end{eqnarray}
The second term corresponds to the black hole gravity to the
particles. The third term yields an attractive force from a
contribution of the magnetic field regardless of whether $Q>0$ or
$Q<0$. The fourth term provides an inertial centrifugal force due
to the particle angular momentum. The fifth, sixth and seventh
terms come from coupled interactions among the metric deformation
perturbations, angular momentum and magnetic field. For
$1-LQ\approx1$, they have repulsive force effects to the charged
particles when $k_1>0$, $k_2>0$ and $k_3>0$, but attractive force
effects when $k_1<0$, $k_2<0$ and $k_3<0$. A small increase of the
energy $E$ or the magnetic field $|Q|$ means enhancing the
attractive force effects and therefore the motions of particles
can become more chaotic in some circumstances. As the angular
momentum $L$ increases, the repulsive force effects are
strengthened and chaos is weakened. With a minor increase of
relatively small positive deformation parameter $k_0$, the
magnetic field attractive force and the centrifugal force will
increase, but the centrifugal force has a larger increase than the
magnetic field force for the parameters chosen in Figure 7. This
leads to weakening the strength of chaos. However, as the absolute
value $|k_0|$ with $k_0<0$ increases, the centrifugal force has a
larger decrease than the magnetic field force and chaos becomes
stronger. Increases of the other positive deformation parameters
$k_1$, $k_2$ and $k_3$ cause the repulsive forces to increase, and
chaos to get weaker. However, the attractive force effects are
enhanced and chaos gets stronger as the magnitudes of negative
deformation parameters $k_1$, $k_2$ and $k_3$ increase.

\section{Conclusions}

When a nonrotating compact object has spherical deformations, it
is suffered from metric deformation perturbations. Such small
deformation perturbations to the Schwarzschild metric could be
regarded as a nonrotating black hole solution  departure from the
standard Schwarzschild spacetime  in modified theories of gravity.
The non-Schwarzschild spacetime with four free deformation
parameters is integrable. However, the dynamics of charged
particles moving around the Schwarzschild-like black hole is
nonintegrable when the inclusion of an external asymptotically
uniform magnetic field destroys the fourth invariable quantity
related to the azimuthal motion of the particles.

Although the deformation perturbation metric looks like the
Schwarzschild metric, it can be changed into a Kerr-like black
hole metric via some appropriate coordinate transformation.
Therefore, the time-transformed explicit symplectic integrators
for the Kerr type spacetimes introduced in [84] should be
similarly applicable to the deformation perturbation Schwarzschild
black hole surrounded with external magnetic field. In fact, we
can design explicit symplectic methods for a time-transformed
Hamiltonian, which is split into seven parts with analytical
solutions as explicit functions of new coordinate time. A main
role for the time transformation function is the implementation of
such desired separable form of the time-transformed Hamiltonian
rather than that of adaptive time-steps control. It is shown
numerically that the obtained time-transformed explicit symplectic
integrators perform good long-term stable error behavior
regardless of regular or chaotic orbits when intermediate
time-steps are chosen.

One of the obtained time-transformed explicit symplectic
integrators combined with the techniques of Poincar\'{e} sections
and FLIs is used to well show how small changes of the parameters
affect the dynamical transitions from order to chaos. Chaos in the
global phase space can be strengthened under some circumstances as
any one of the energy and the absolute values of the (positive or
negative) magnetic parameter and negative deformation parameters
increases. However, it is weakened as any one of the angular
momentum and positive deformation parameters increases.

\textbf{Author Contributions}: Conceptualization, Methodology,
Supervision, Xin Wu; Software, Writing
--- original draft, Hongxing Zhang; Software, Naying Zhou and Wenfang
Liu.

\textbf{Funding}: This research has been supported by the National
Natural Science Foundation of China (Grant No. 11973020 (C0035736)
and the National Natural Science Foundation of Guangxi (No.
2019JJD110006).

\textbf{Data Availability Statement}: Not applicable.

\textbf{Acknowledgments}: Authors are very grateful to four
referees for valuable comments and useful suggestions.

\textbf{Conflicts of Interest}: The authors declare no conflict of
interest.


\begin{figure*}
    \centering{
        \includegraphics[width=12pc]{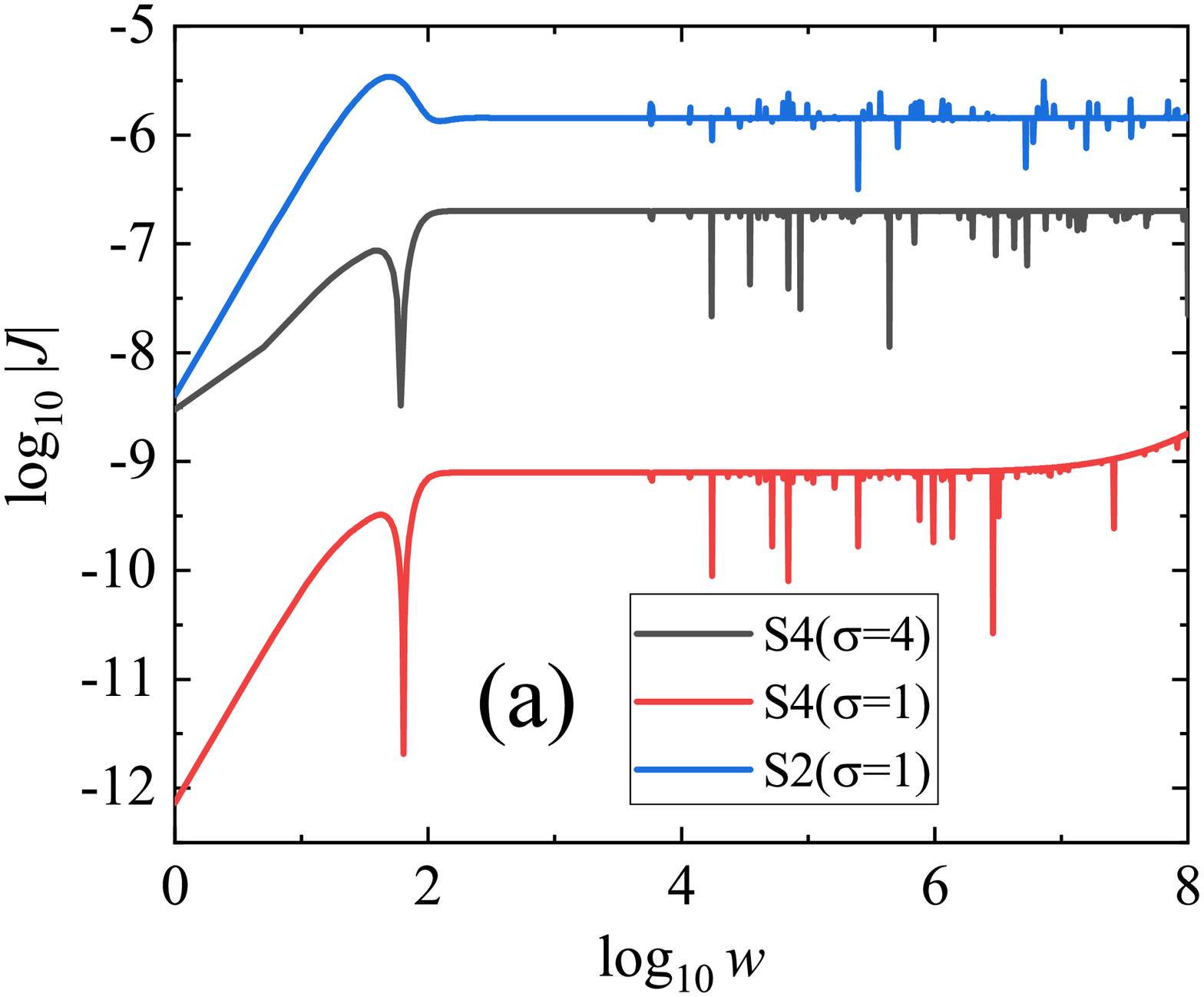}
        \includegraphics[width=12pc]{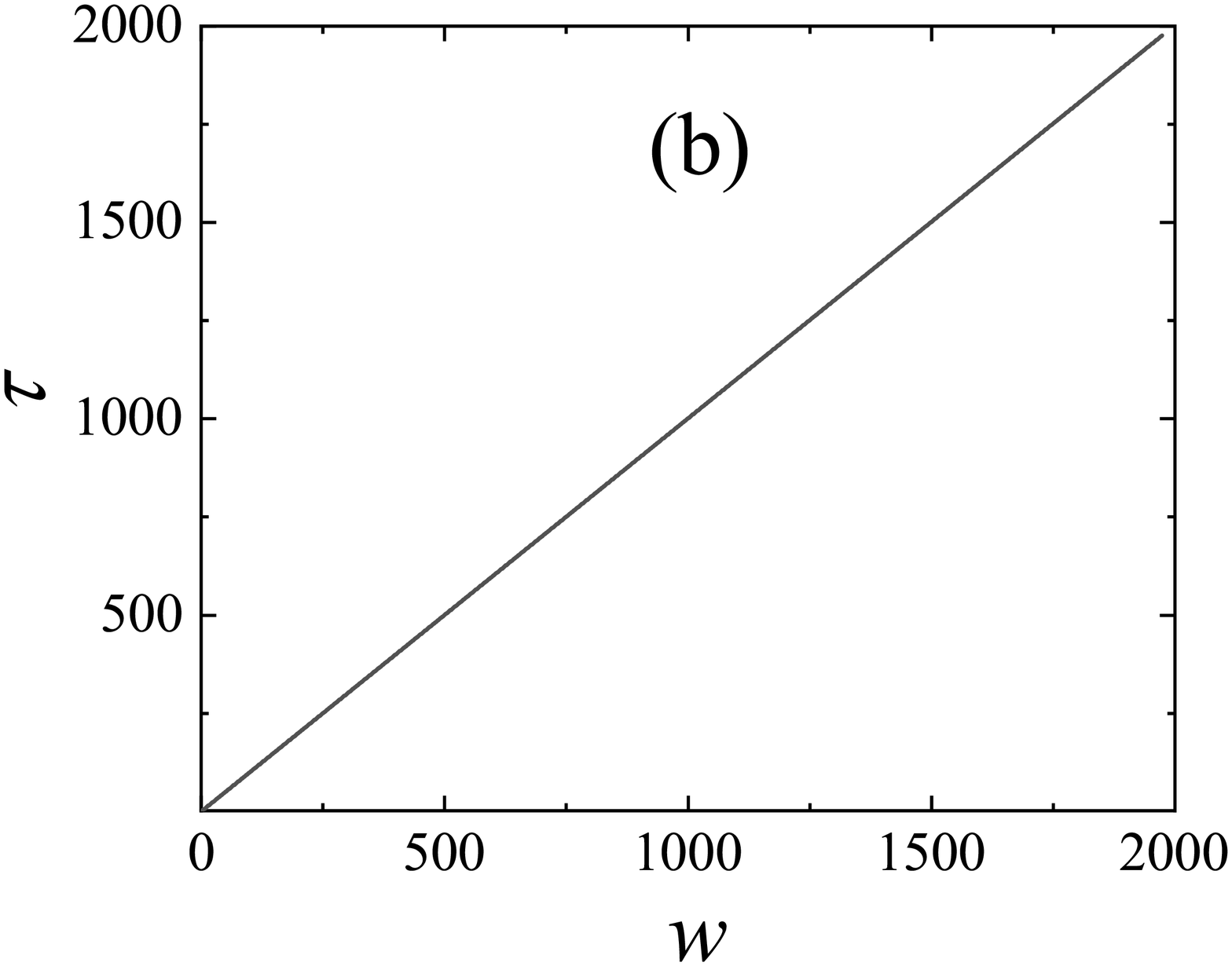}
        \includegraphics[width=12pc]{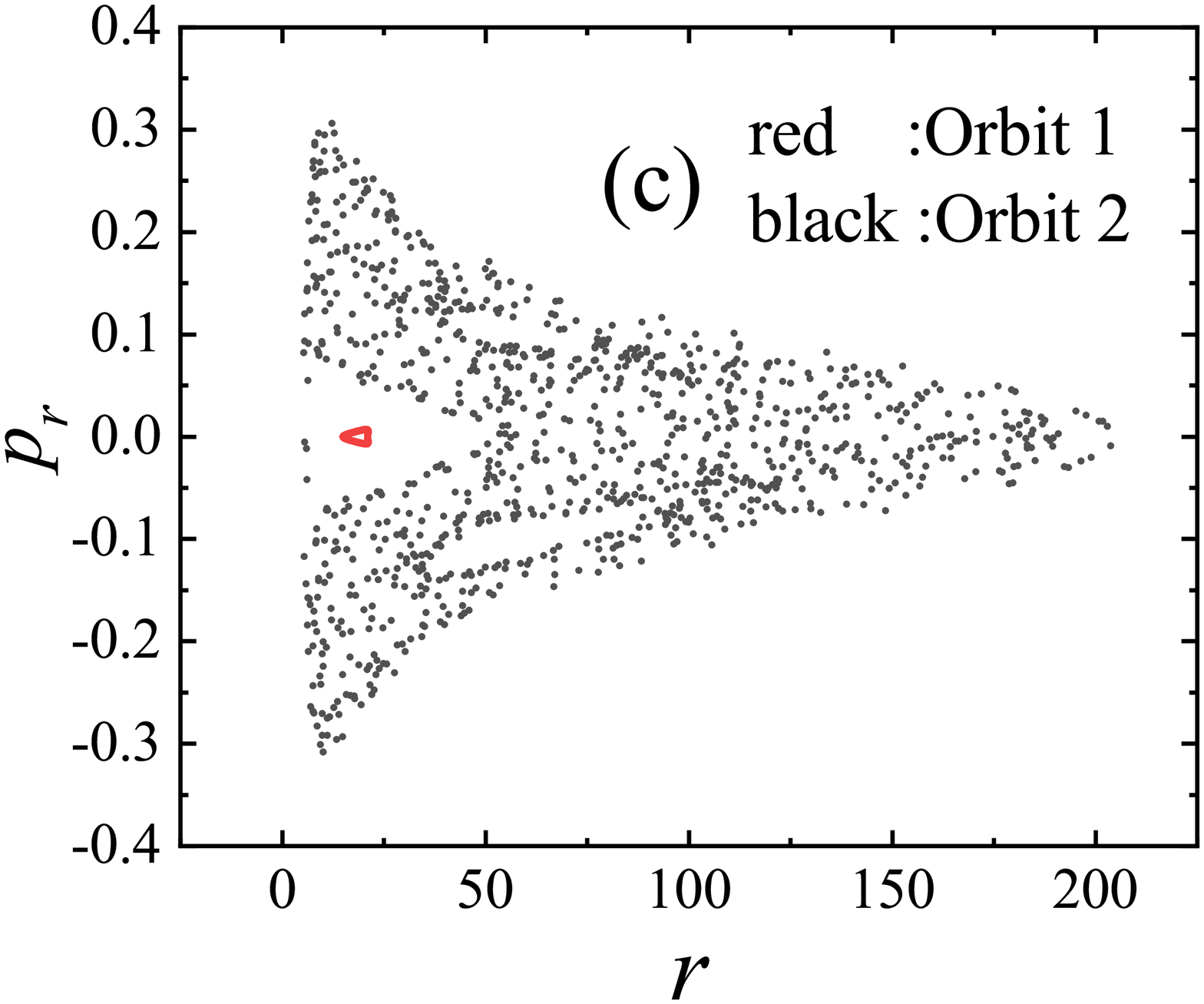}
\caption{(a) Errors of the Hamiltonian $J$ in Eq. (12). S2
($\sigma=1$) represents the second-order method S2 with time step
$\sigma=1$; S4 ($\sigma=1$) means the fourth-order method S4 with
new coordinate time step $\sigma=1$, and S4 ($\sigma=4$) stands
for the fourth-order method S4 with time step $\sigma=4$. Orbit 1
with the initial separation $r=15$ is tested. Orbit 1 has the
other initial conditions  $p_r=0$, $\theta=\pi/2$  and
$p_\theta>0$ determined by $J=0$. The parameters are $E=0.9965$,
$L=4$, $Q=6\times 10^{-4}$, $k_0=10^{-3}$, $k_1=10^{-2}$,
$k_2=10^{-1}$ and $k_3=1$. The error for S4 ($\sigma=1$) is three
orders of magnitude smaller than for S2 ($\sigma=1$). The error
remains bounded for S2 ($\sigma=1$), but it has a secular drift
for S4 ($\sigma=1$) due to roundoff errors. The secular drift in
the error loses for S4 ($\sigma=4$). (b) Relation between proper
time $\tau$ and new coordinate time $w$. This shows that  $\tau$
and $w$ are almost the same. (c) Poincar\'{e} sections at the
plane $\theta=\pi/2$ with $p_\theta>0$. Orbit 1 is ordered,
whereas Orbit 2 with the initial separation $r=50$ is chaotic.
Panels (b) and (c) come from the results provided by the algorithm
S4 ($\sigma=4$).  } }
\end{figure*}

\begin{figure*}
    \centering{
        \includegraphics[width=12pc]{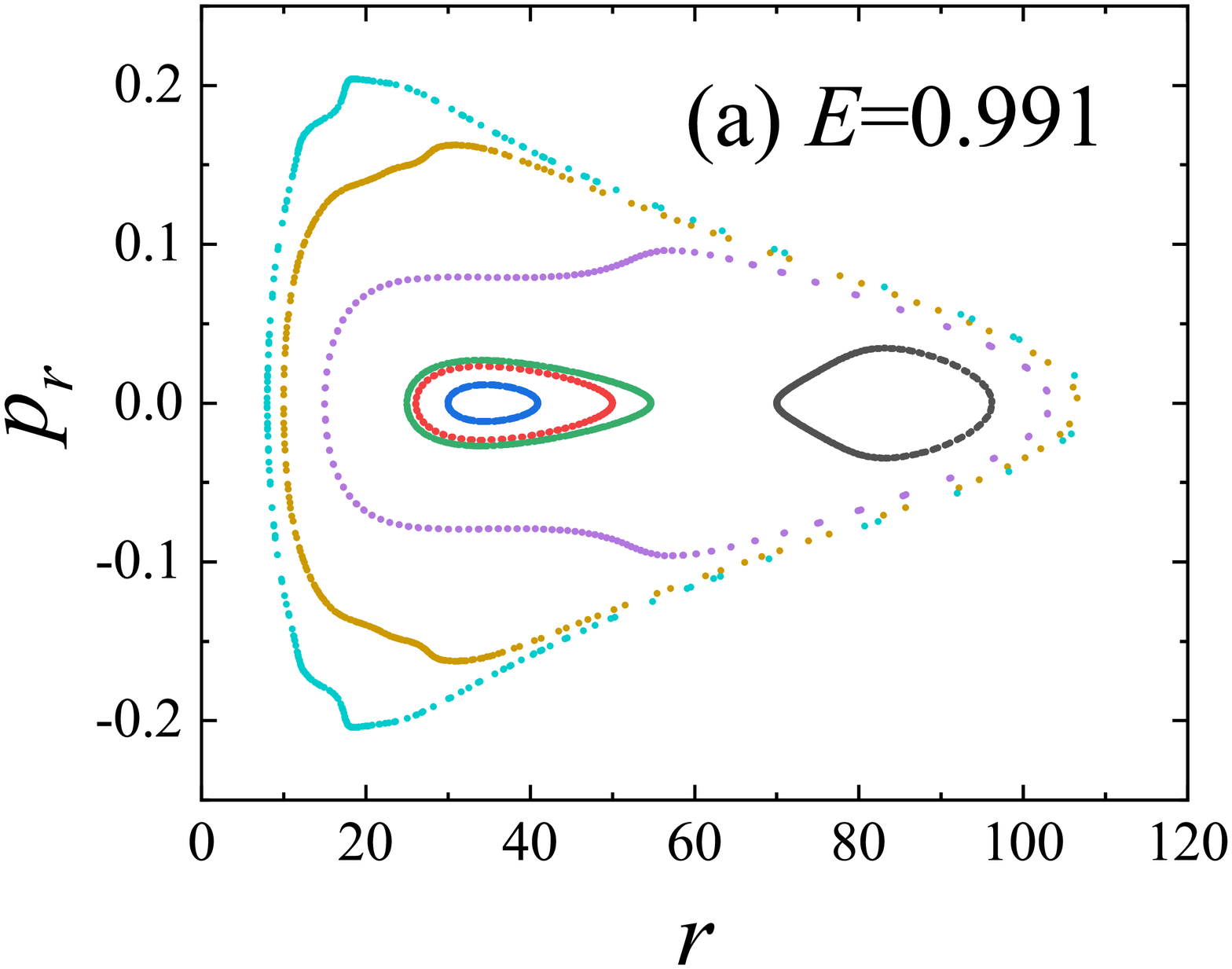}
        \includegraphics[width=12pc]{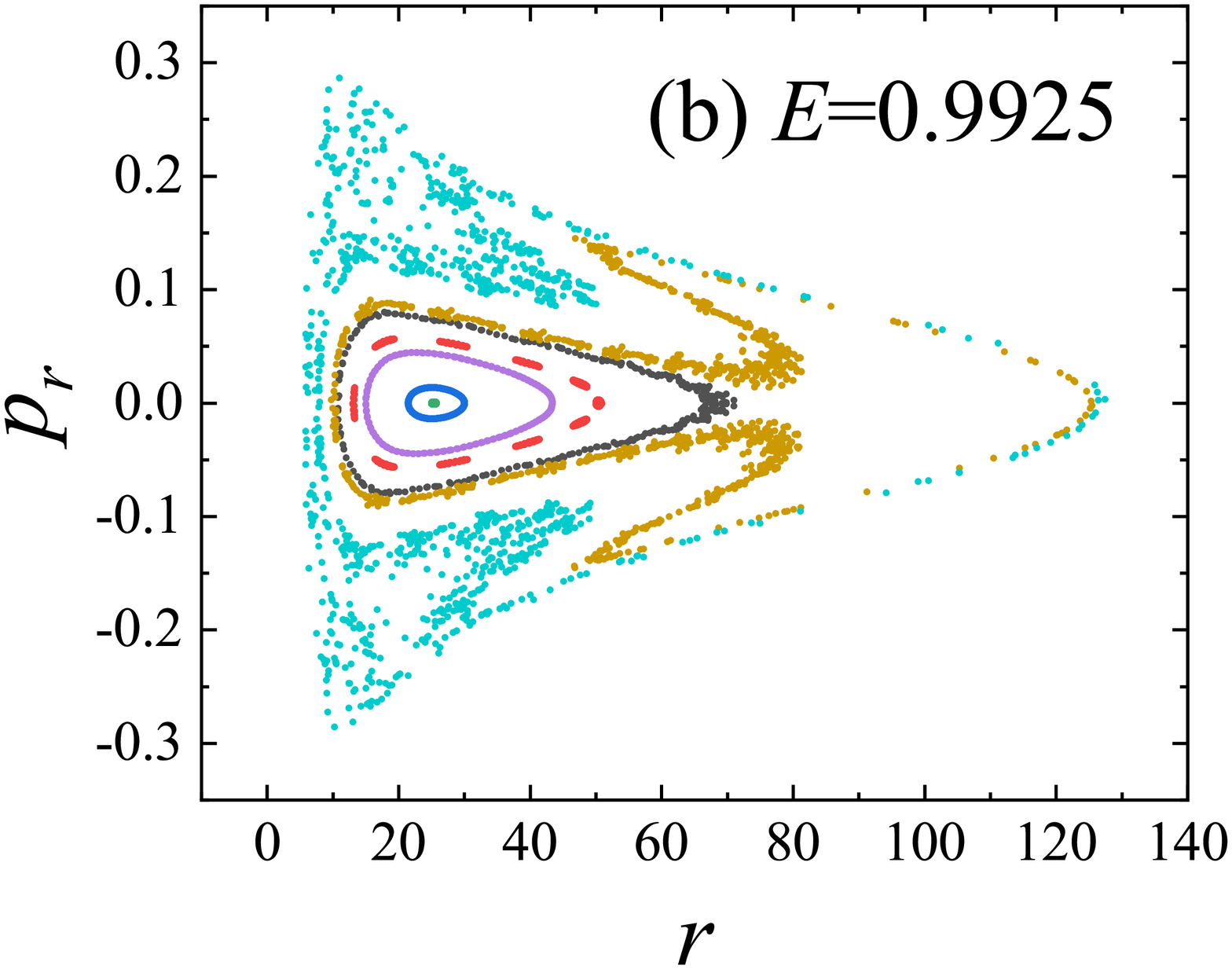}
        \includegraphics[width=12pc]{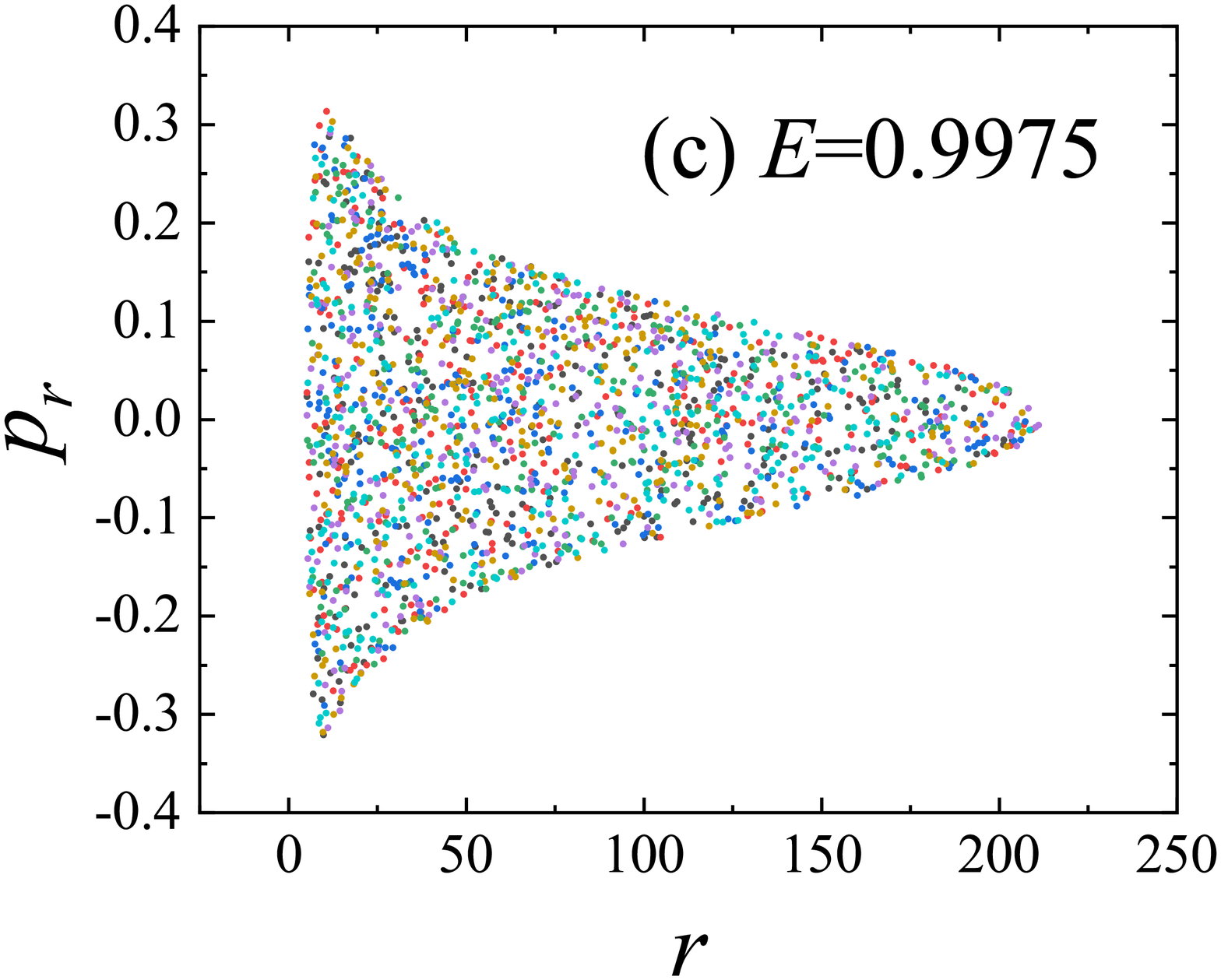}
        \caption{Poincar\'{e} sections. The parameters are the same
        as those in Figure 1(c) but $Q=8\times 10^{-4}$,
        $k_0=10^{-4}$ and energies $E$ are different. The energies are (a)
        $E=0.991$, (b) $E=0.9925$ and (c) $E=0.9975$. The three
        sub-figures show that the chaoticity becomes strong with the energy increasing.}
        }
\end{figure*}

\begin{figure*}
    \centering{
        \includegraphics[width=12pc]{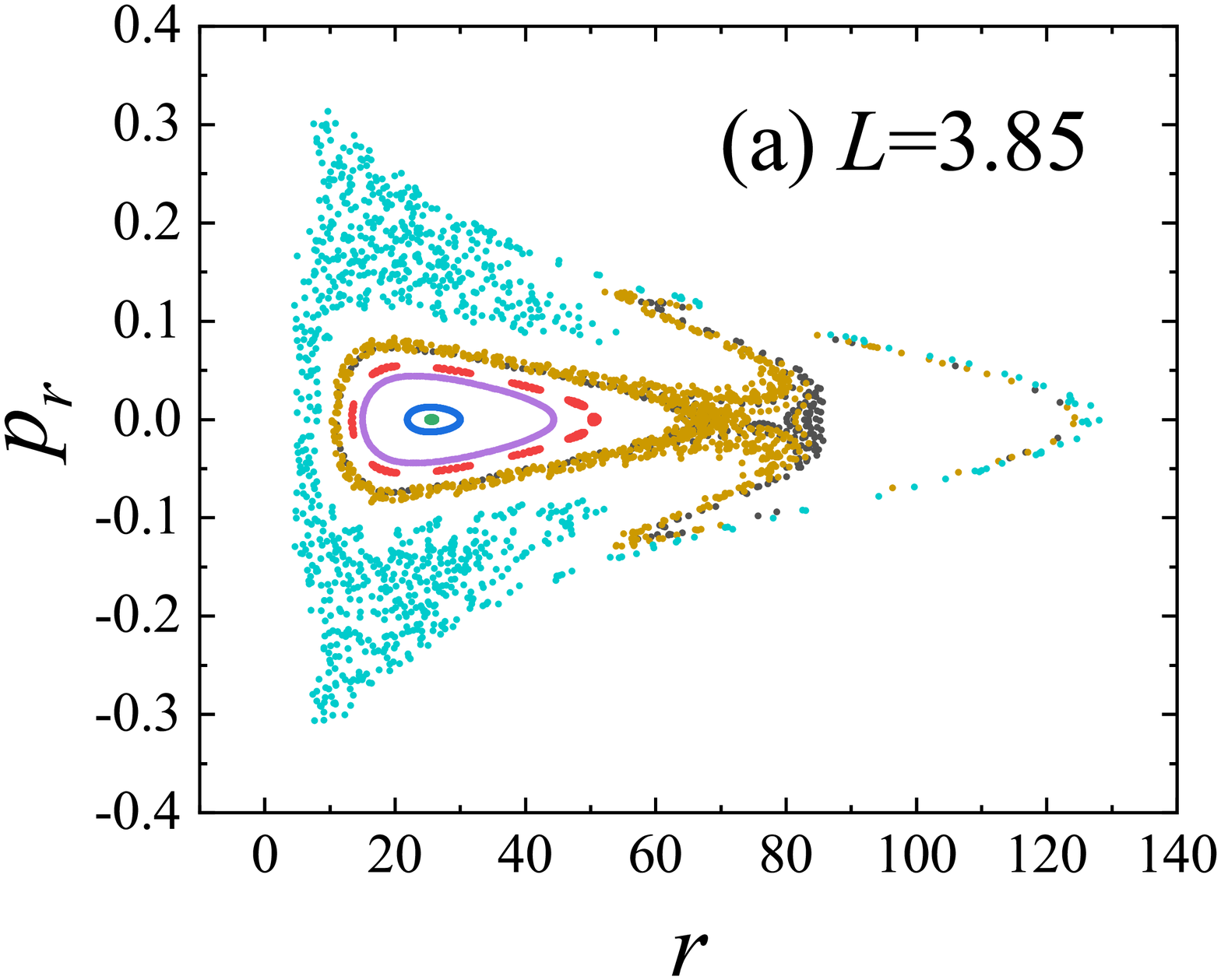}
        \includegraphics[width=12pc]{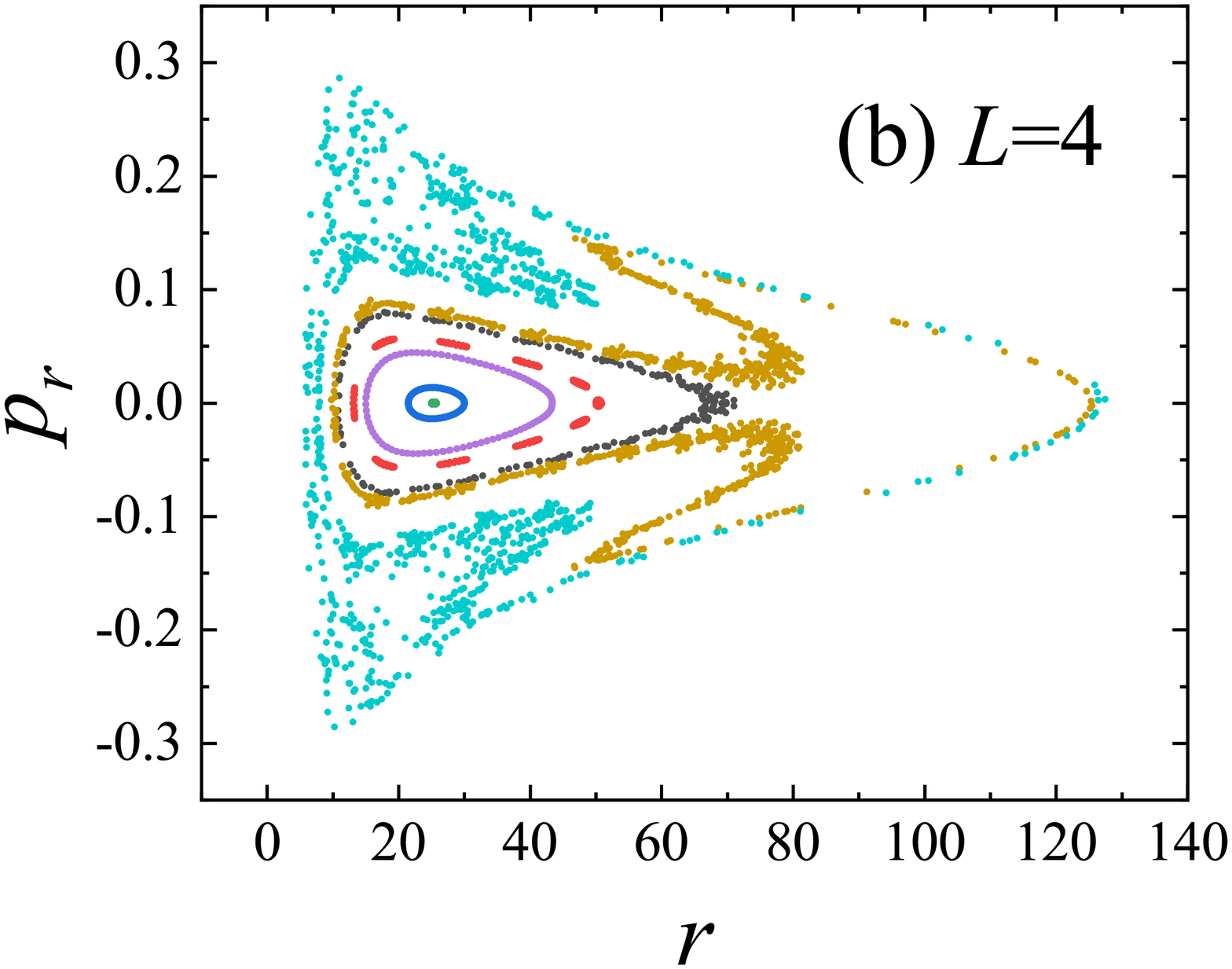}
        \includegraphics[width=12pc]{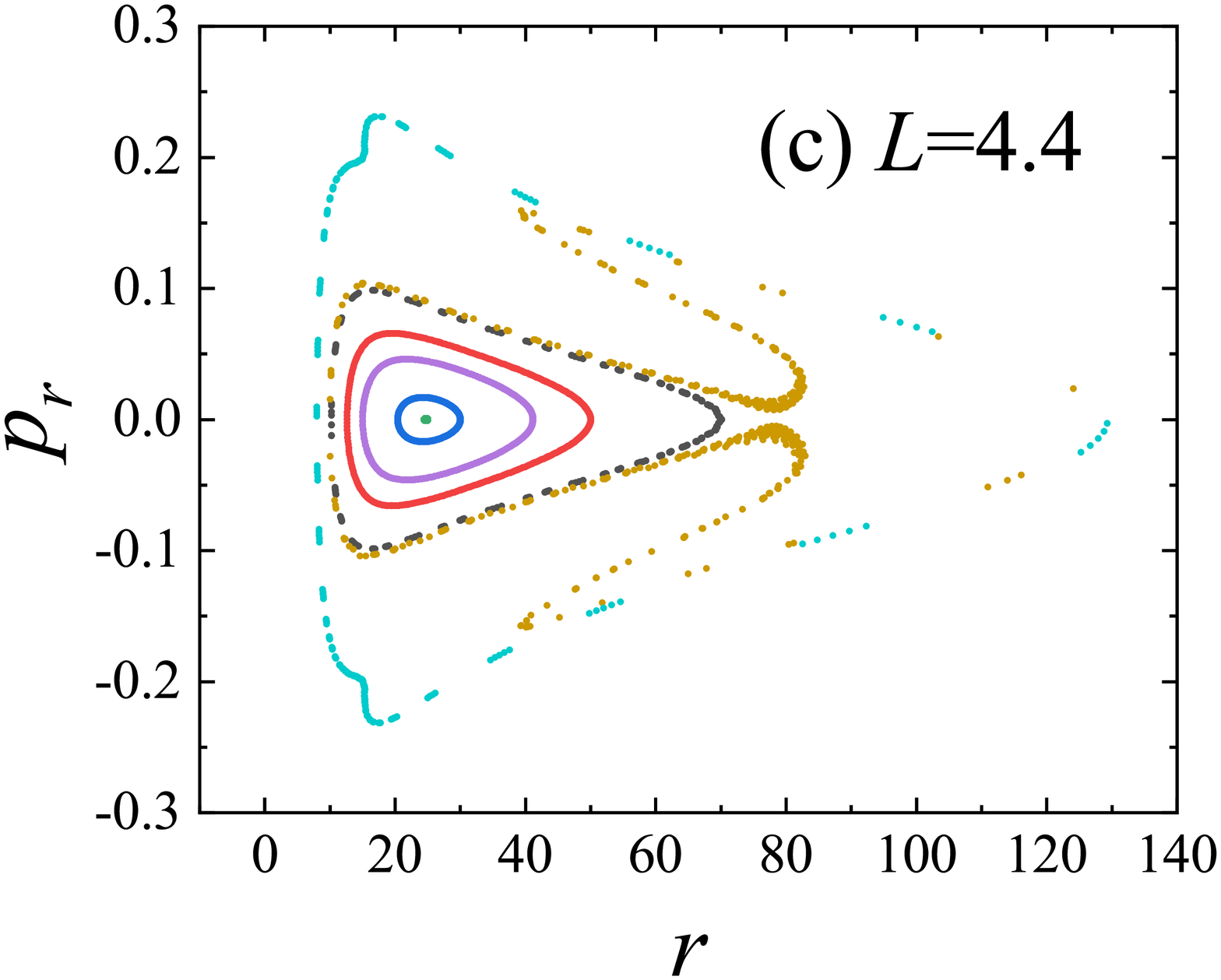}
        \caption{Poincar\'{e} sections. The parameters are $E=0.9925$, $Q=8\times 10^{-4}$, $k_0=10^{-4}$, $k_1=10^{-2}$, $k_2=10^{-1}$ and
        $k_3=1$. The angular momenta are (a) $L=3.85$, (b) $L=4$
        and (c) $L=4.4$. It is clearly shown that chaos is gradually
        weakened as the angular momentum increases. } }
\end{figure*}

\begin{figure*}
    \centering{
        \includegraphics[width=18pc]{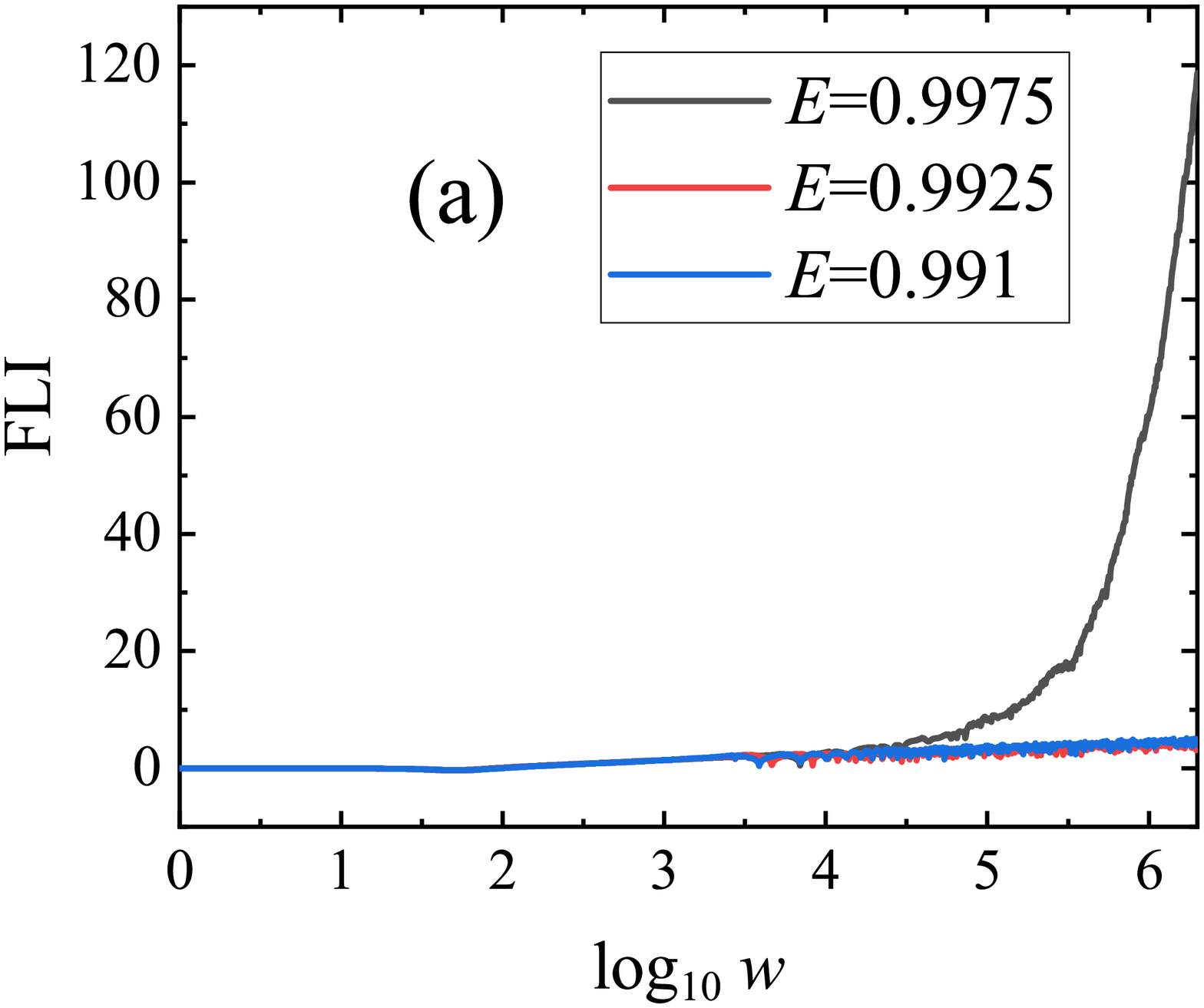}
        \includegraphics[width=18pc]{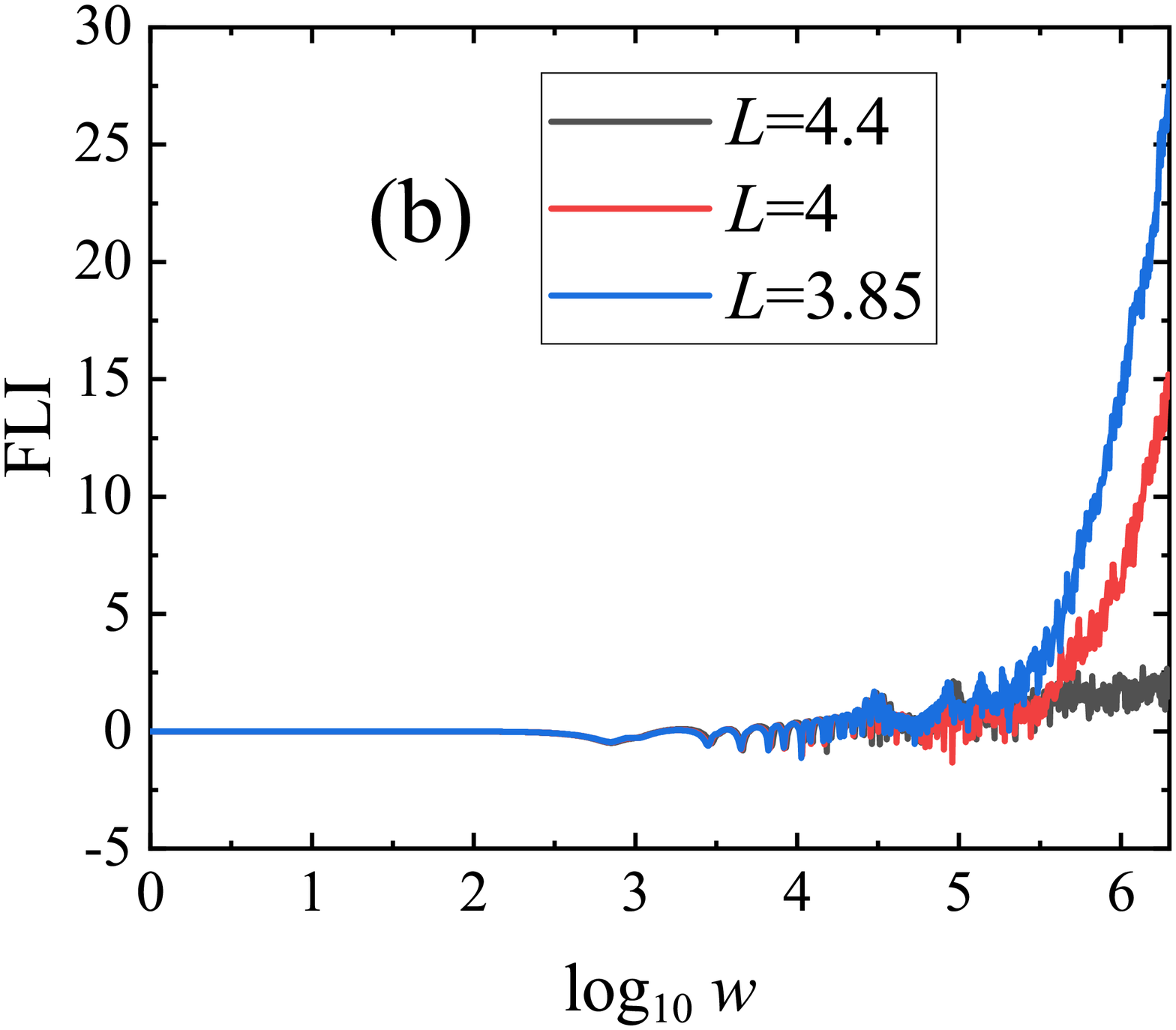}
        \caption{Fast Lyapunov indicators (FLIs). (a) The initial
        separation is $r=15$; the other initial conditions and
        parameters are those of Figure 2. The FLIs for $E=0.991$
        and  $E=0.9925$ correspond to the regular behavior, but
        the FLI for $E=0.9975$ shows the chaotic behavior. (b) The initial
        separation is $r=70$; the other initial conditions and
        parameters are those of Figure 3. The FLI for $L=4.4$ indicates the
        regularity. $L=3.85$ corresponds to stronger chaos than
        $L=4$. } }
\end{figure*}

\begin{figure*}
    \centering{
        \includegraphics[width=12pc]{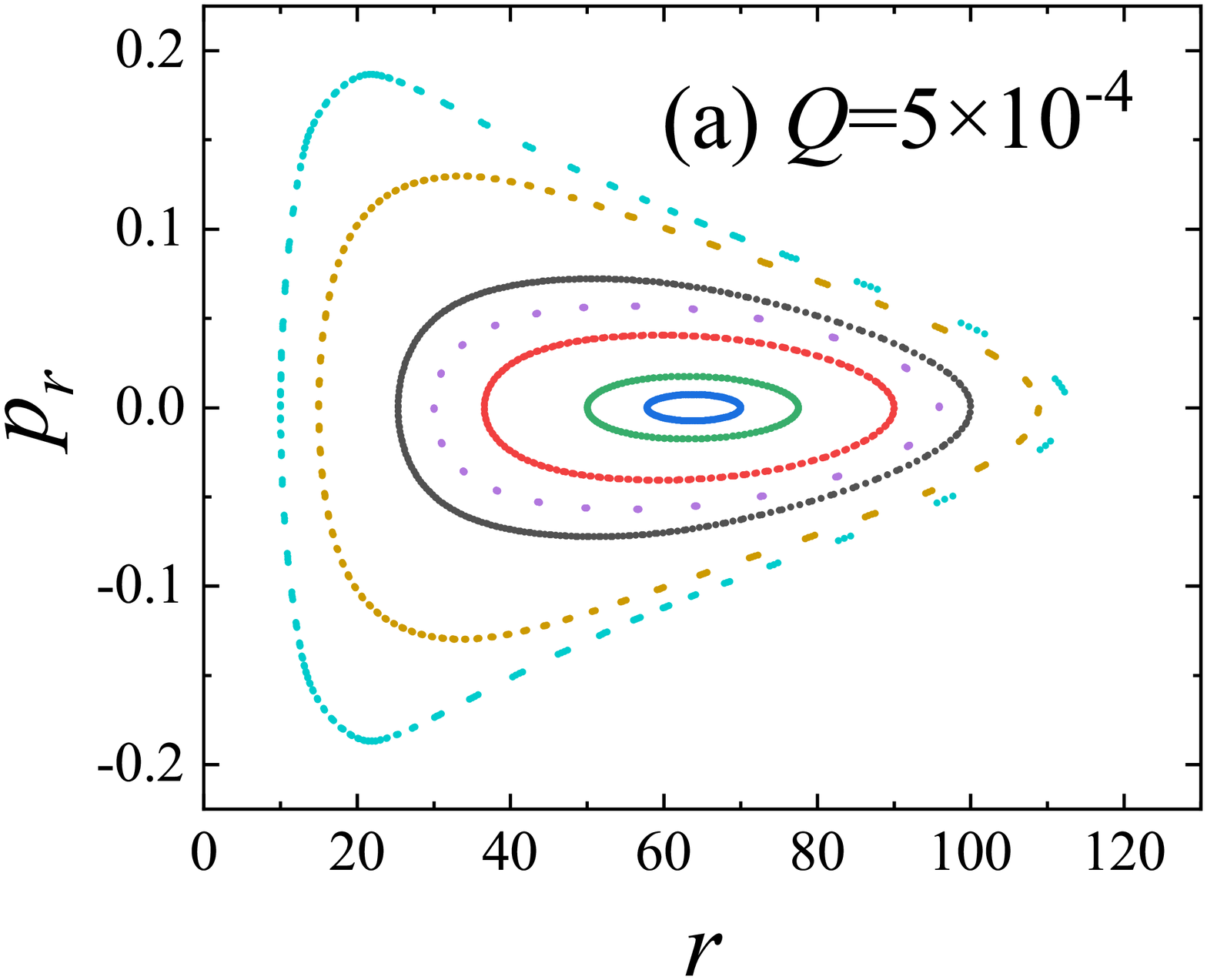}
        \includegraphics[width=12pc]{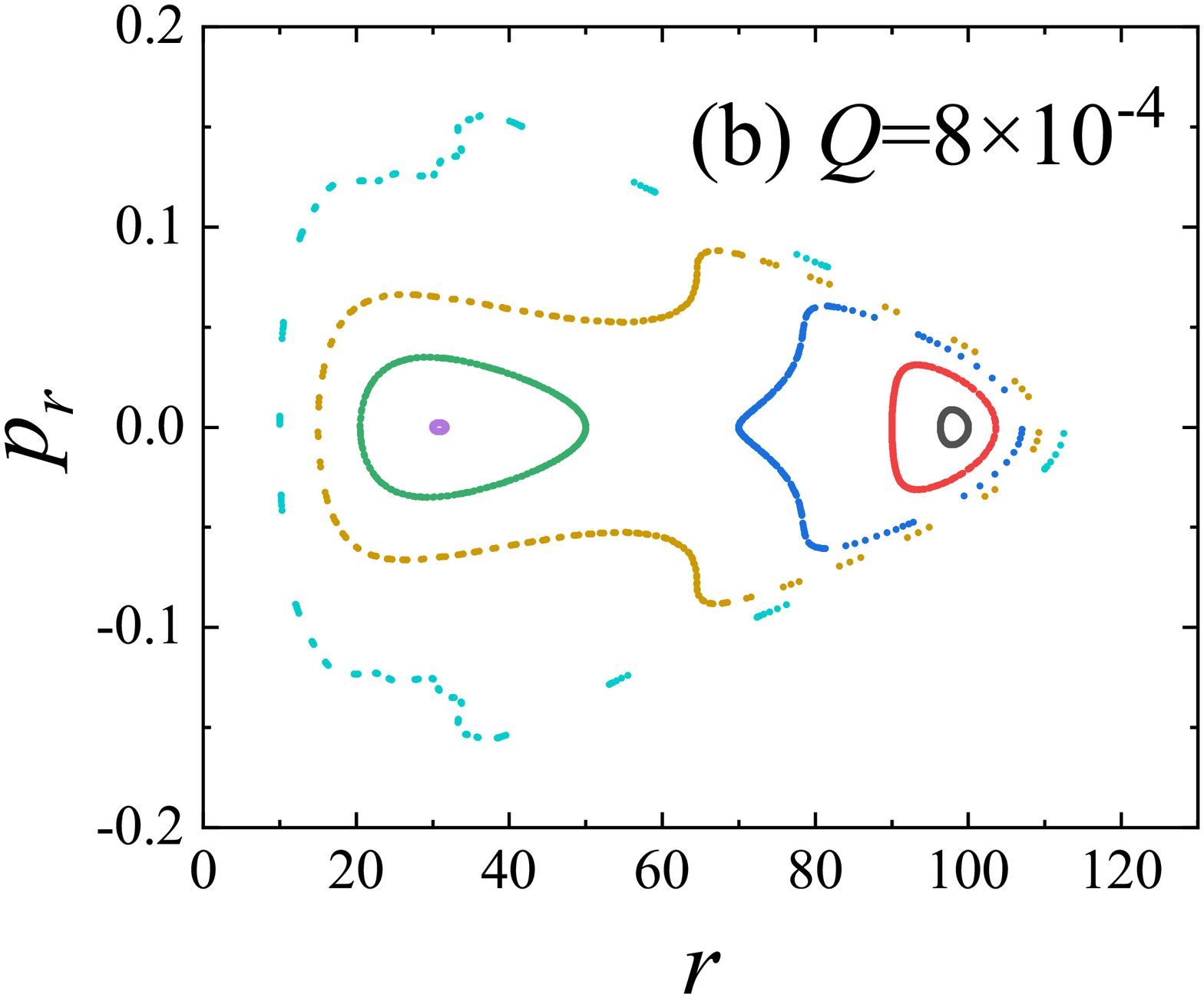}
        \includegraphics[width=12pc]{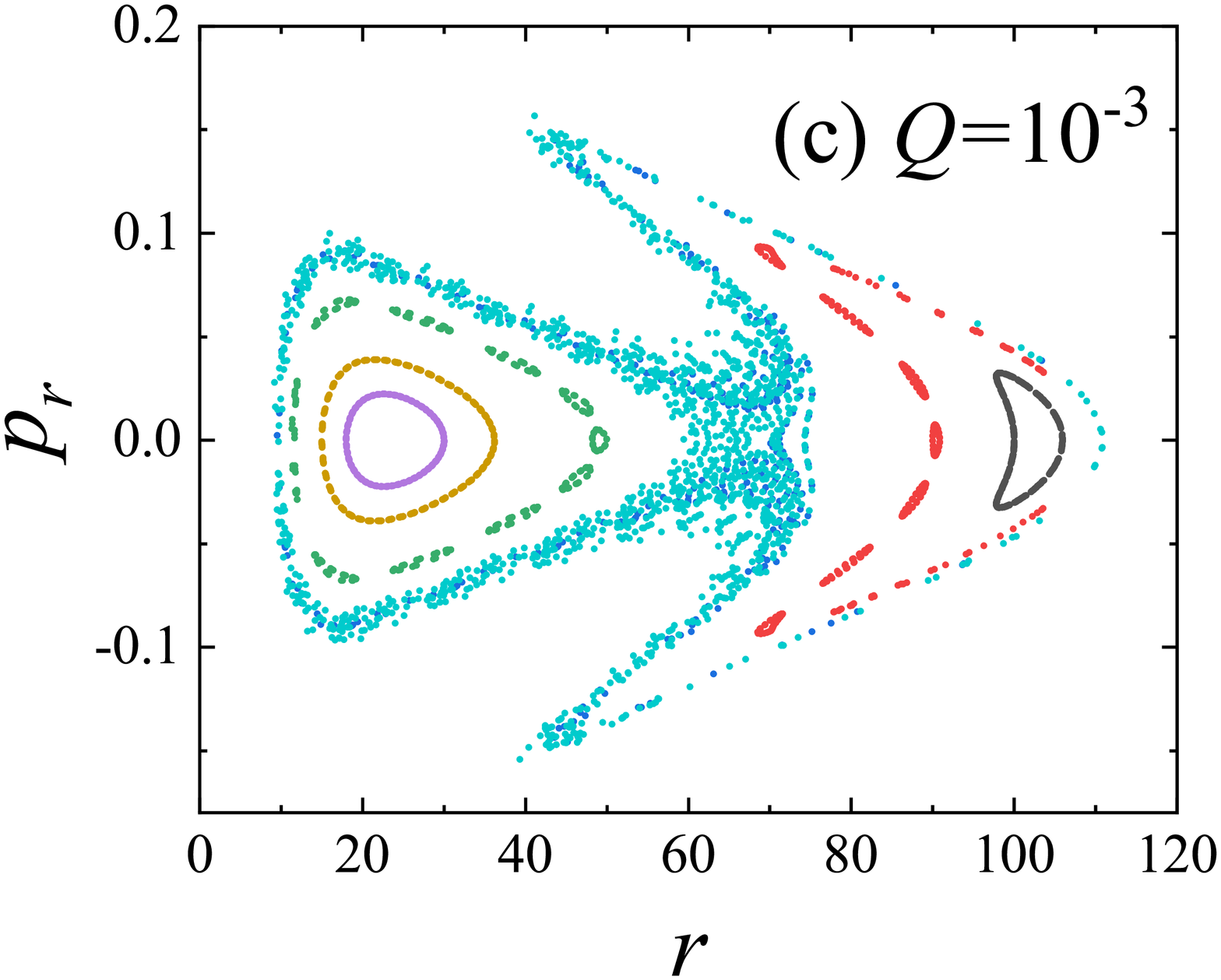}
        \includegraphics[width=12pc]{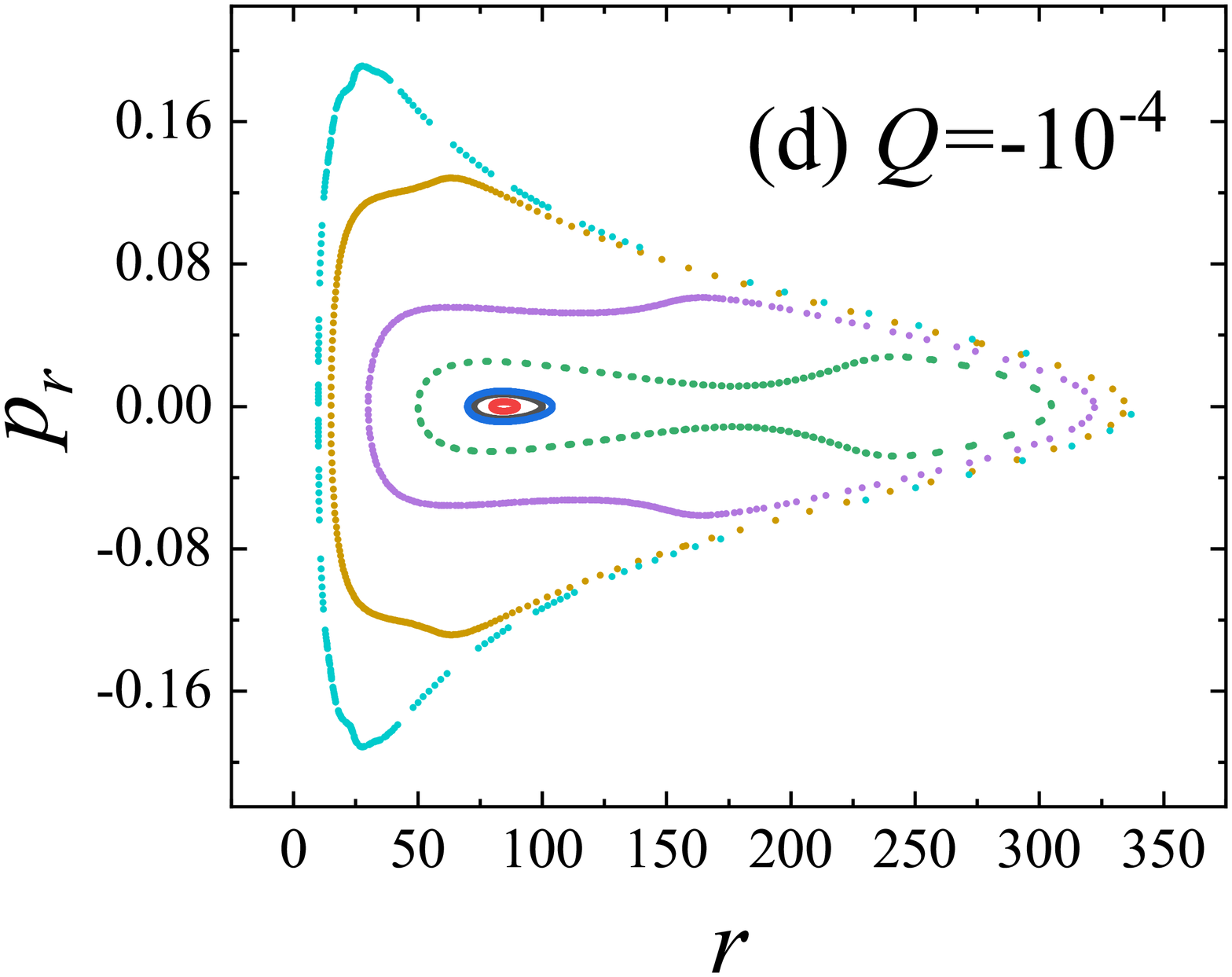}
        \includegraphics[width=12pc]{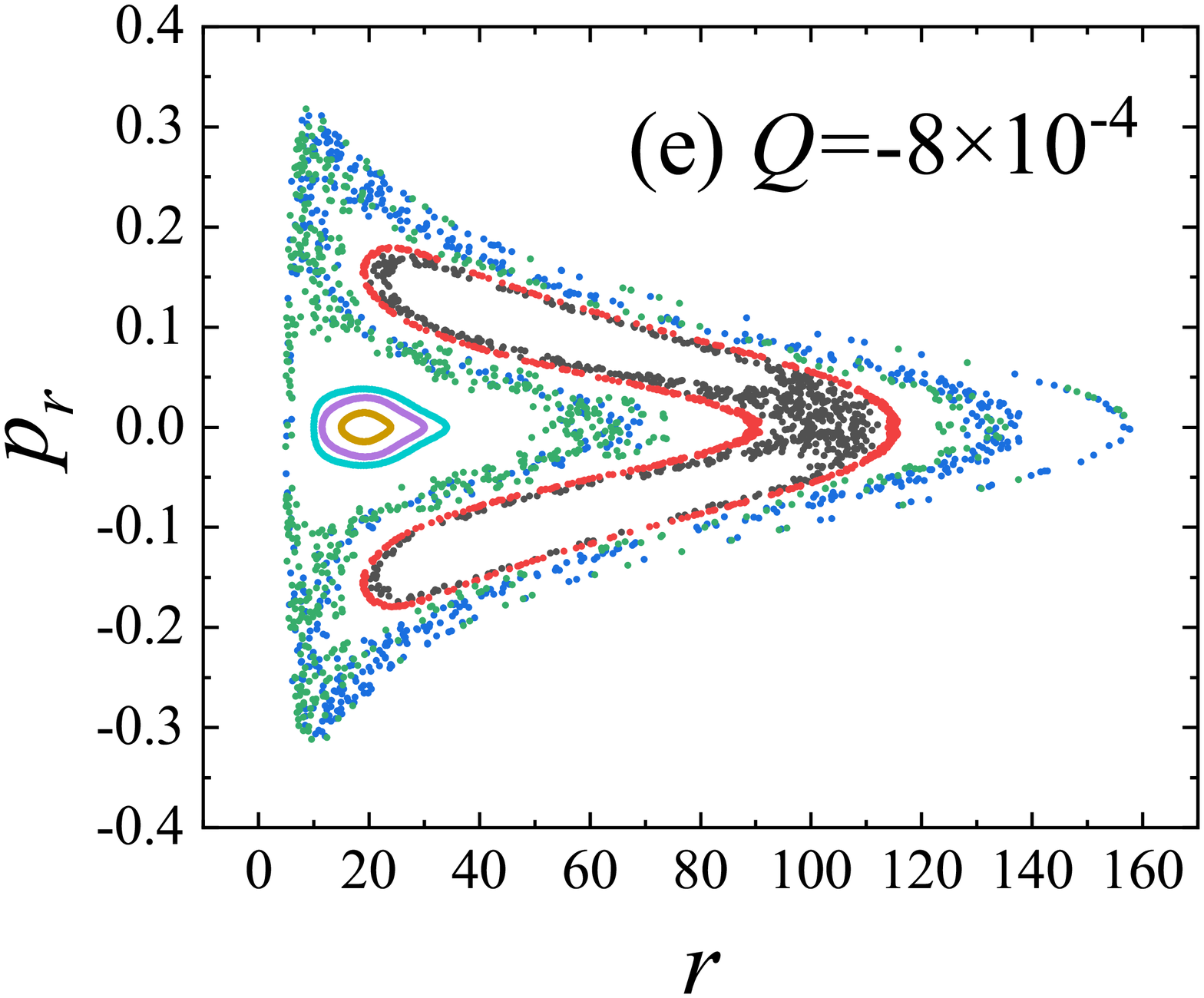}
        \includegraphics[width=12pc]{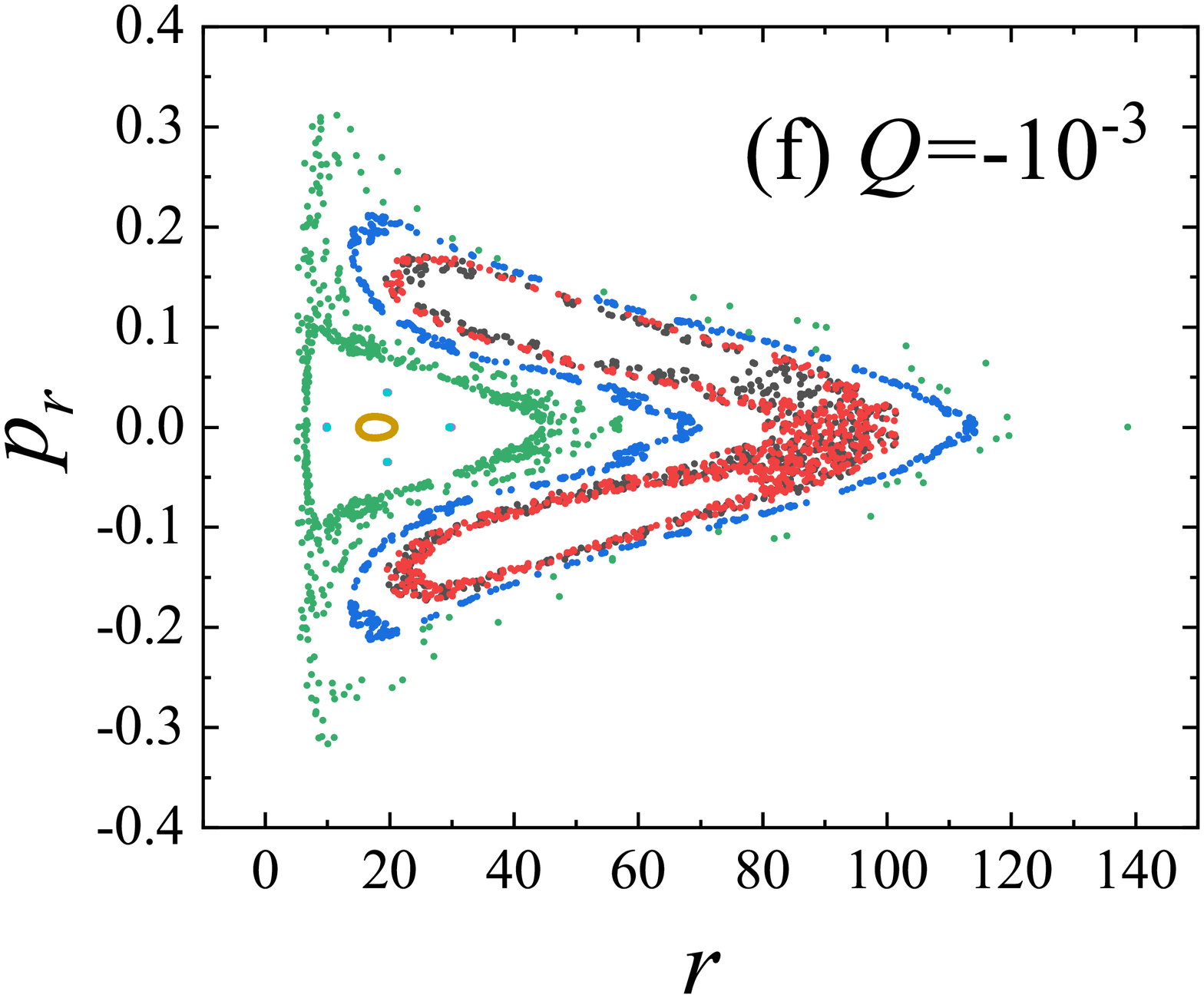}
        \caption{Poincar\'{e} sections for different values of the magnetic parameter
        $Q$. The other parameters are $L=4$, $k_0=10^{-4}$, $k_1=10^{-2}$, $k_2=10^{-1}$ and
        $k_3=1$. (a)-(c):  $E=0.9915$ and $Q>0$; the strength of
        chaos is enhanced with an increase of $Q$. (d)-(f): $E=0.9975$ and
        $Q<0$; chaos is strong as $|Q|$ increases. } }
\end{figure*}

\begin{figure*}
    \centering{
        \includegraphics[width=18pc]{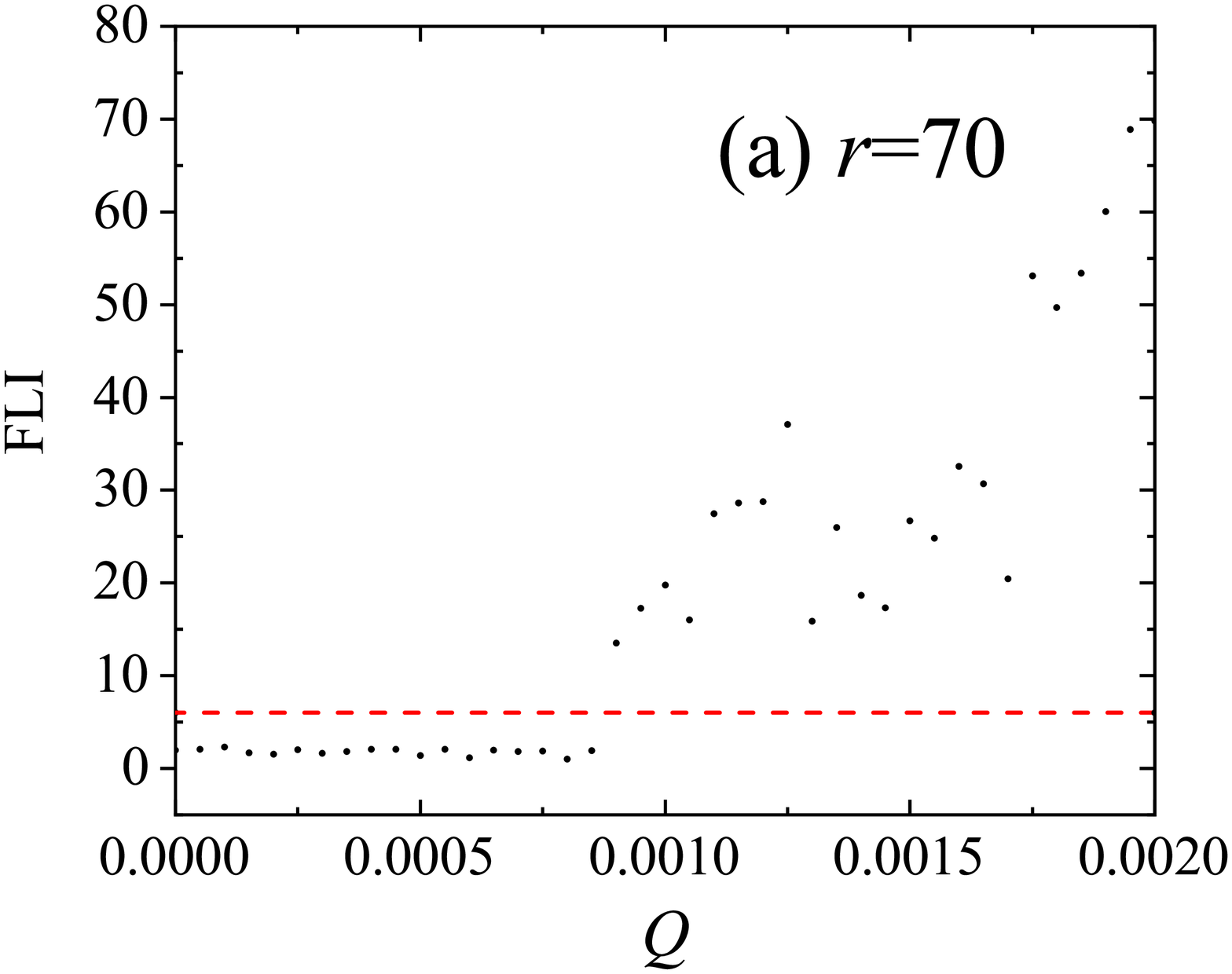}
        \includegraphics[width=18pc]{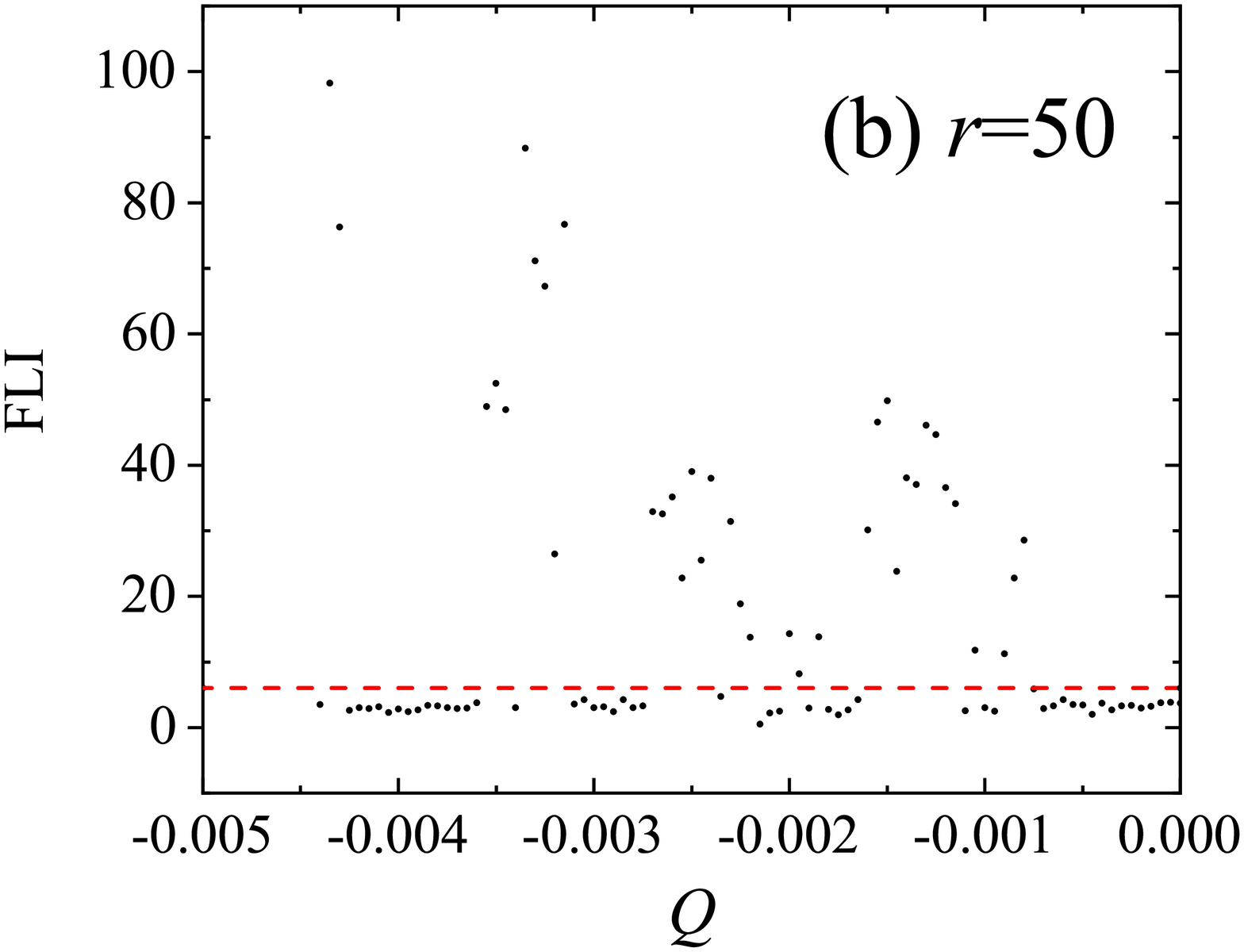}
\caption{(a): Dependence of FLI on the positive magnetic parameter
$Q$ in Figures 5 (a)-(c). The initial separation is $r=70$. The
FLI for each value of $Q$ is obtained after the integration time
$w=2\times10^6$. The FLIs $\geq 6$  mean chaos, and the FLIs $<6$
show the regularity. When $Q>8.5\times 10^{-4}$, chaos begins to
occur. (b): Dependence of FLI on the negative magnetic parameter
$Q$ in Figures 5 (d)-(f). The initial radius is $r=50$. When
$Q<-7.5\times 10^{-4}$, there is  a dynamical transition from
order to chaos. } }
\end{figure*}

\begin{figure*}
    \centering{
        \includegraphics[width=12pc]{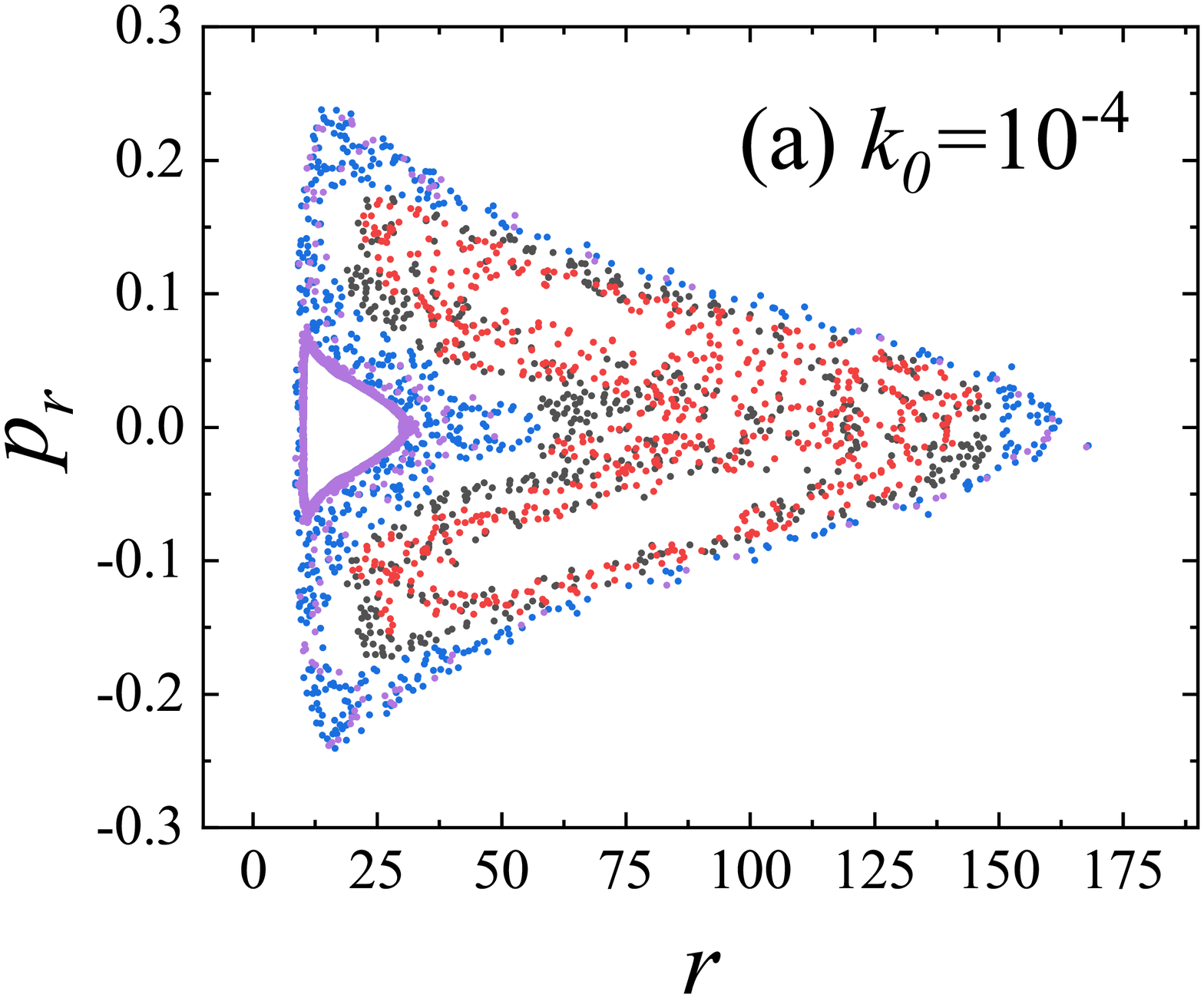}
        \includegraphics[width=12pc]{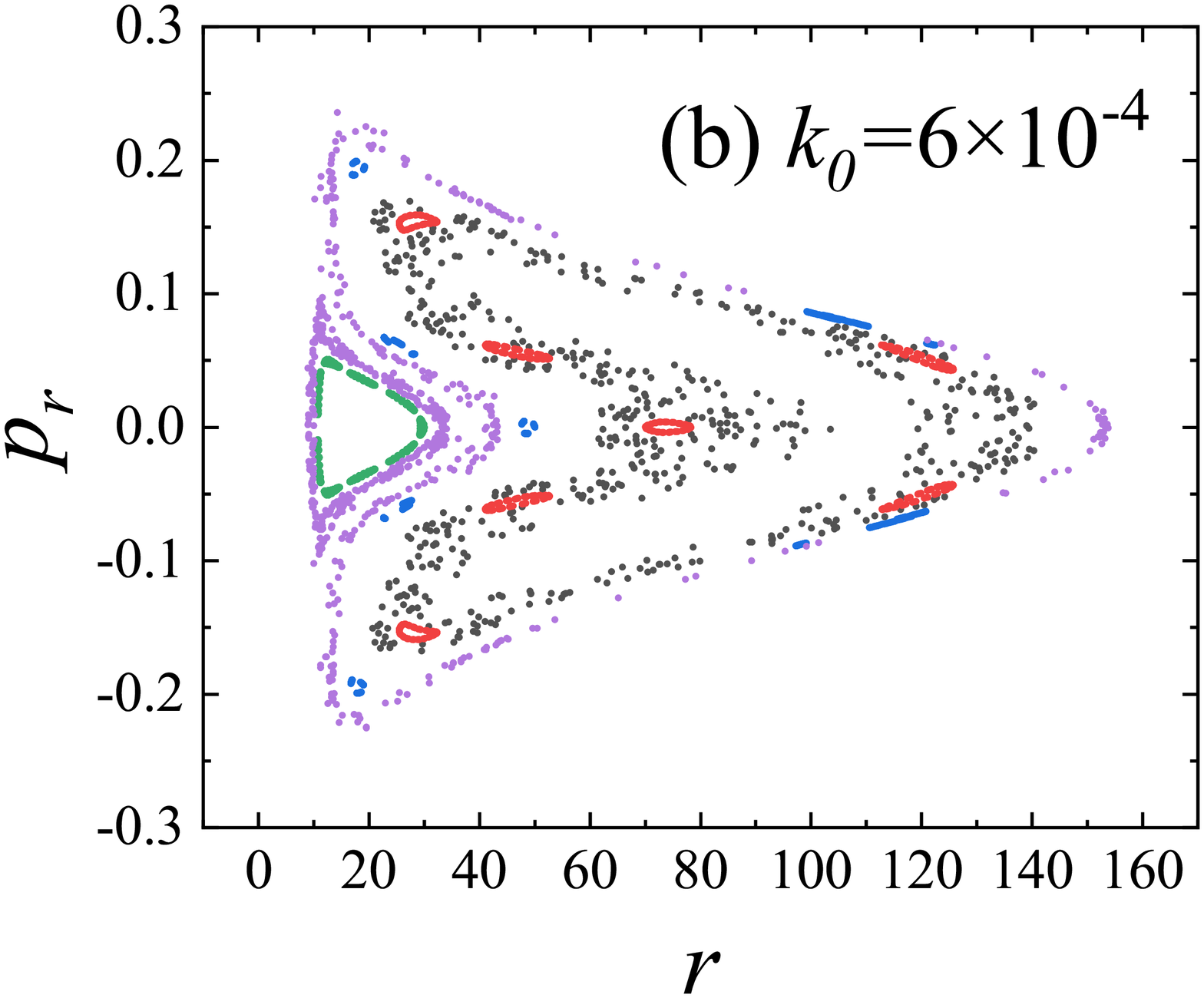}
        \includegraphics[width=12pc]{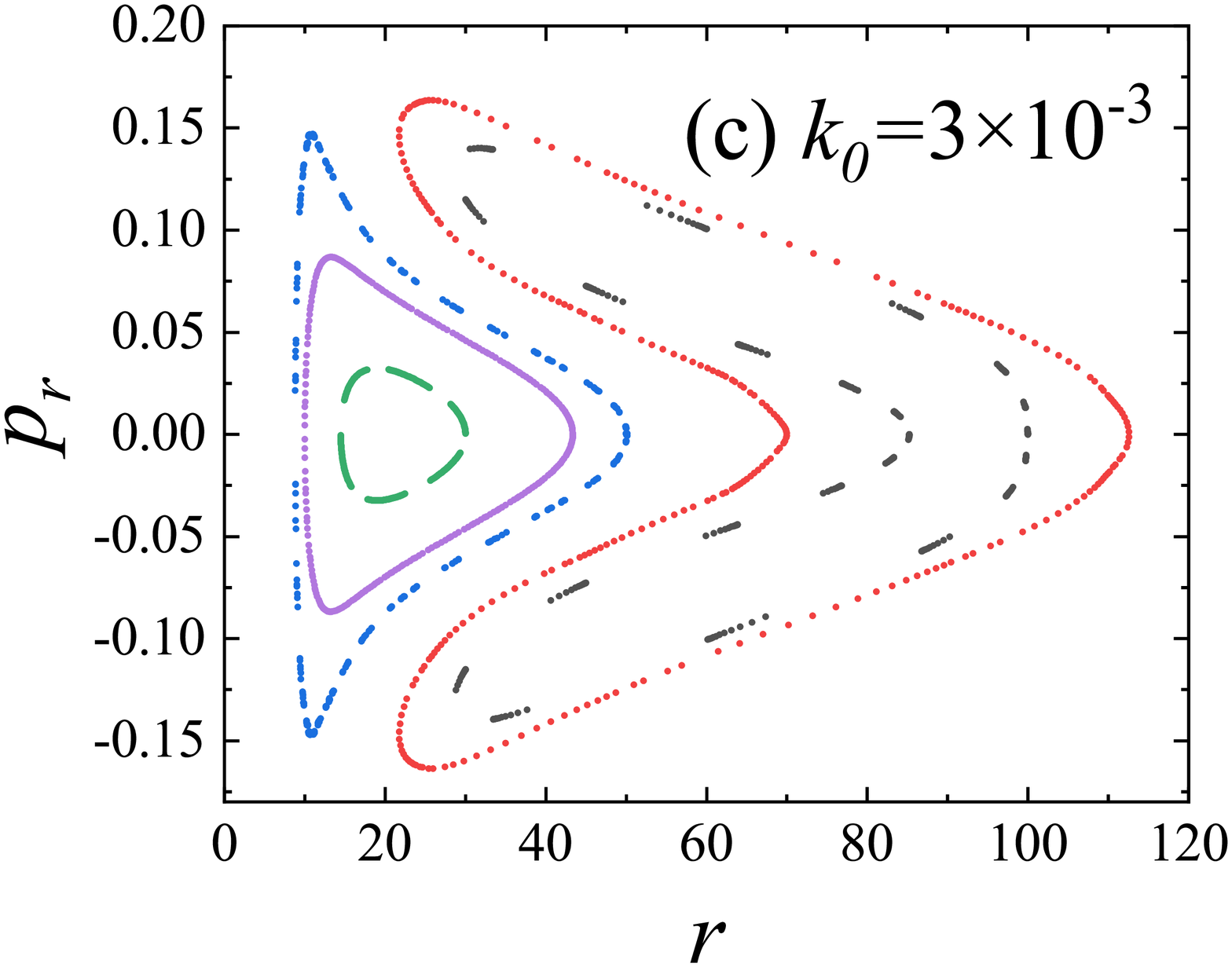}
        \includegraphics[width=12pc]{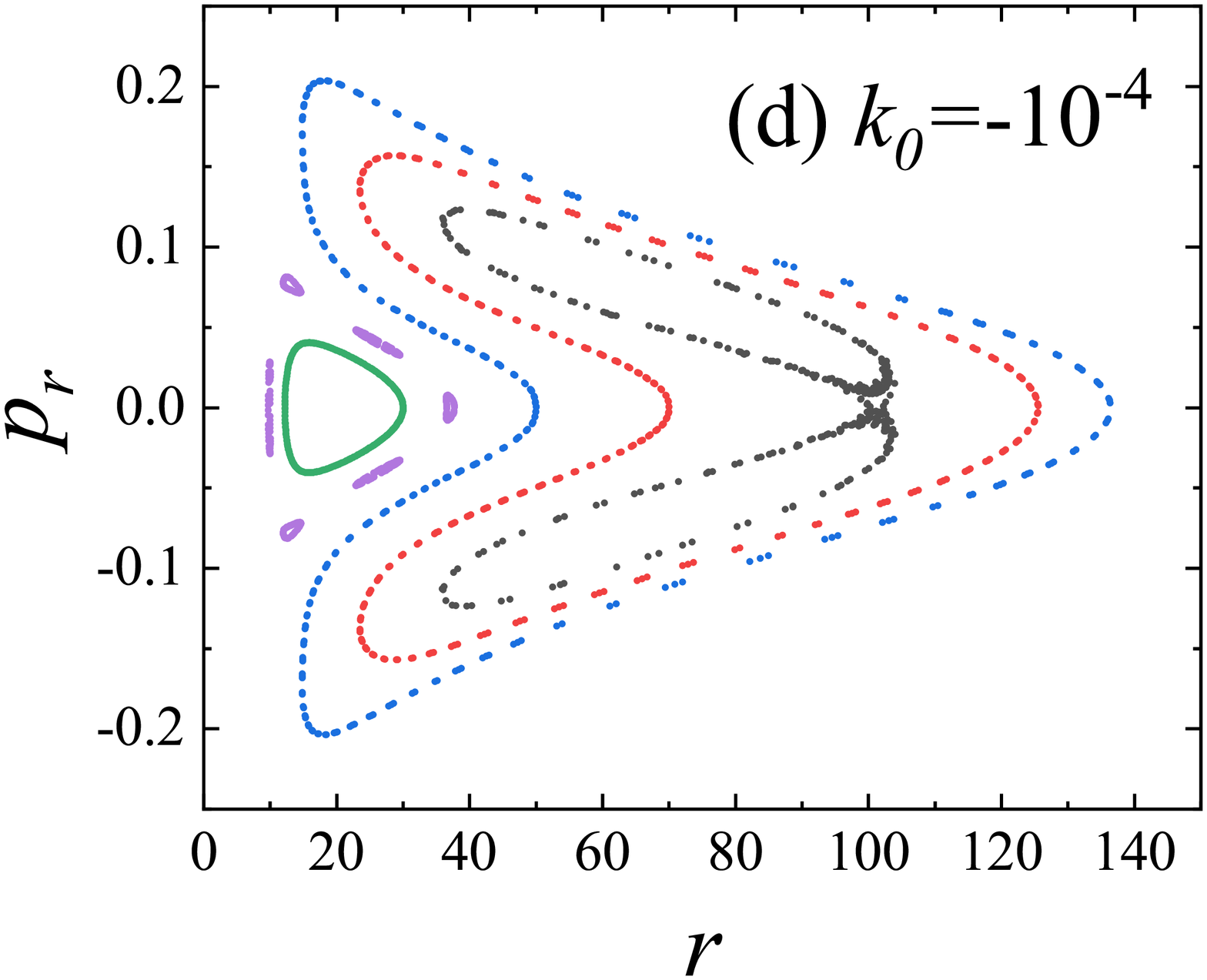}
        \includegraphics[width=12pc]{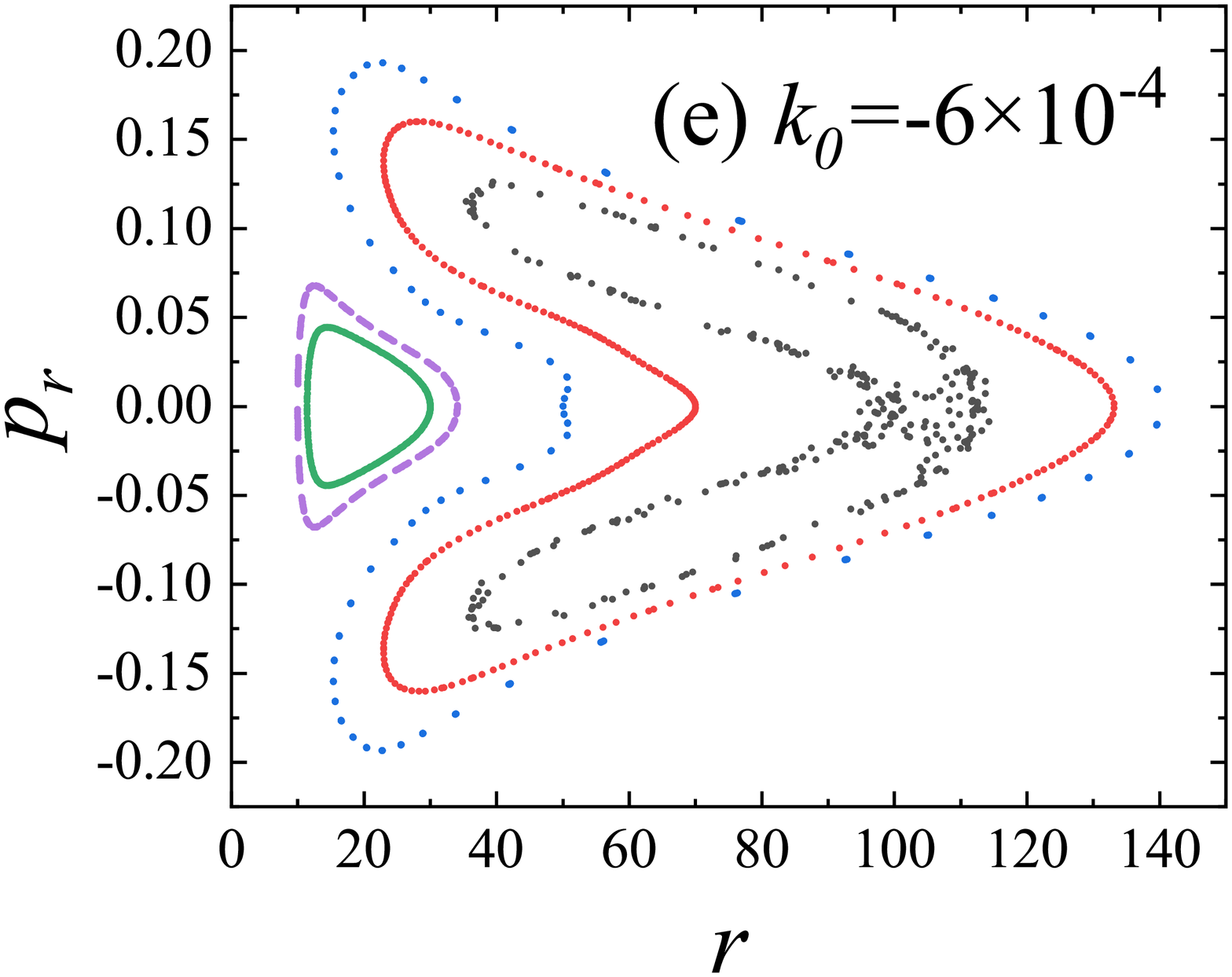}
        \includegraphics[width=12pc]{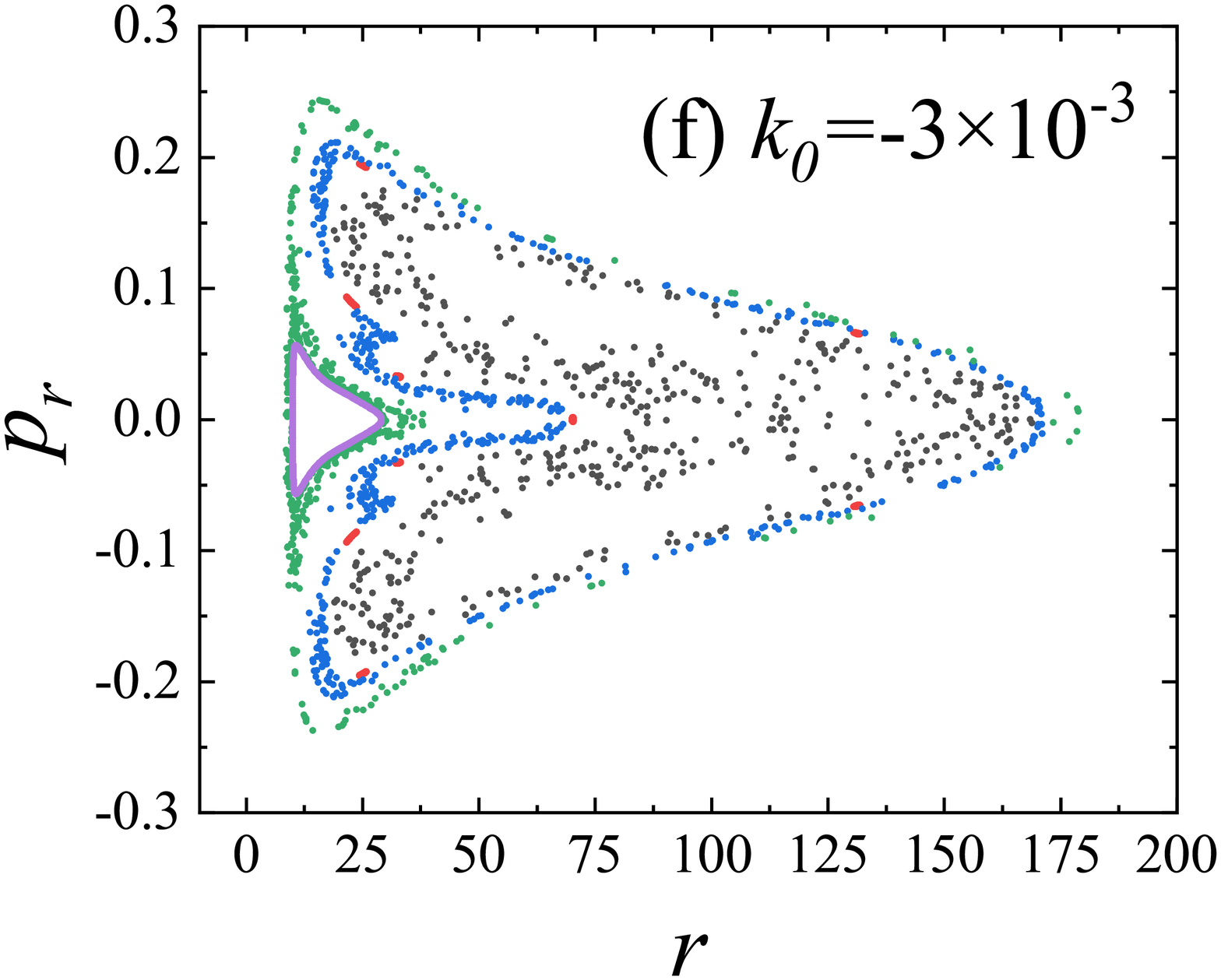}
        \caption{Poincar\'{e} sections for different values of the deformation parameter
        $k_0$. The parameters are $L=4.6$, $Q=8\times10^{-4}$, $k_1=10^{-2}$, $k_2=10^{-1}$ and
        $k_3=1$. (a)-(c):  $E=0.995$ and $k_0>0$. The strength of
        chaos is weakened with an increase of $k_0$. (d)-(f): $E=0.994$ and $k_0<0$.
        Chaos is enhanced as $|k_0|$ increases.  } }
\end{figure*}

\begin{figure*}
    \centering{
\includegraphics[width=18pc]{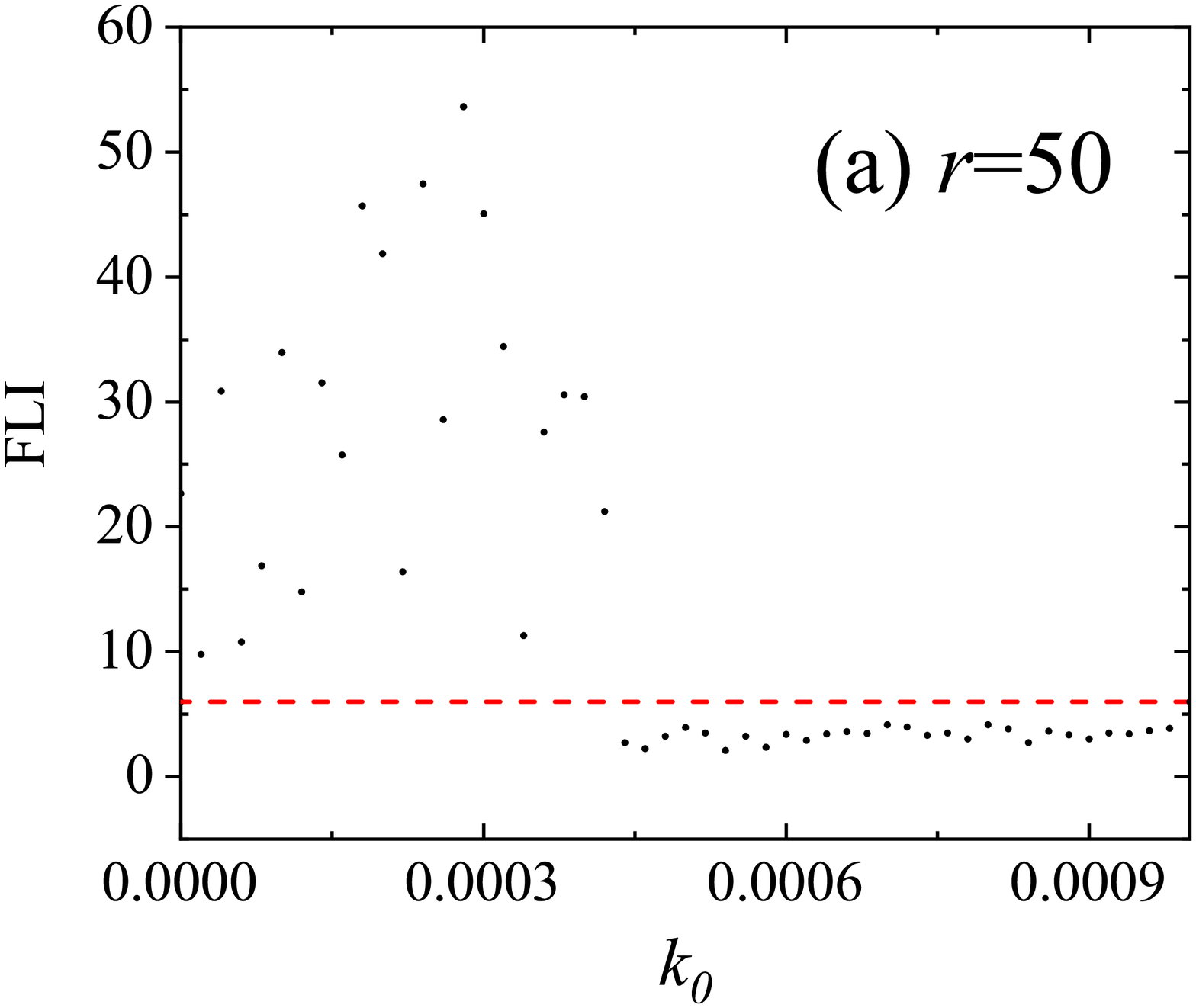}
\includegraphics[width=18pc]{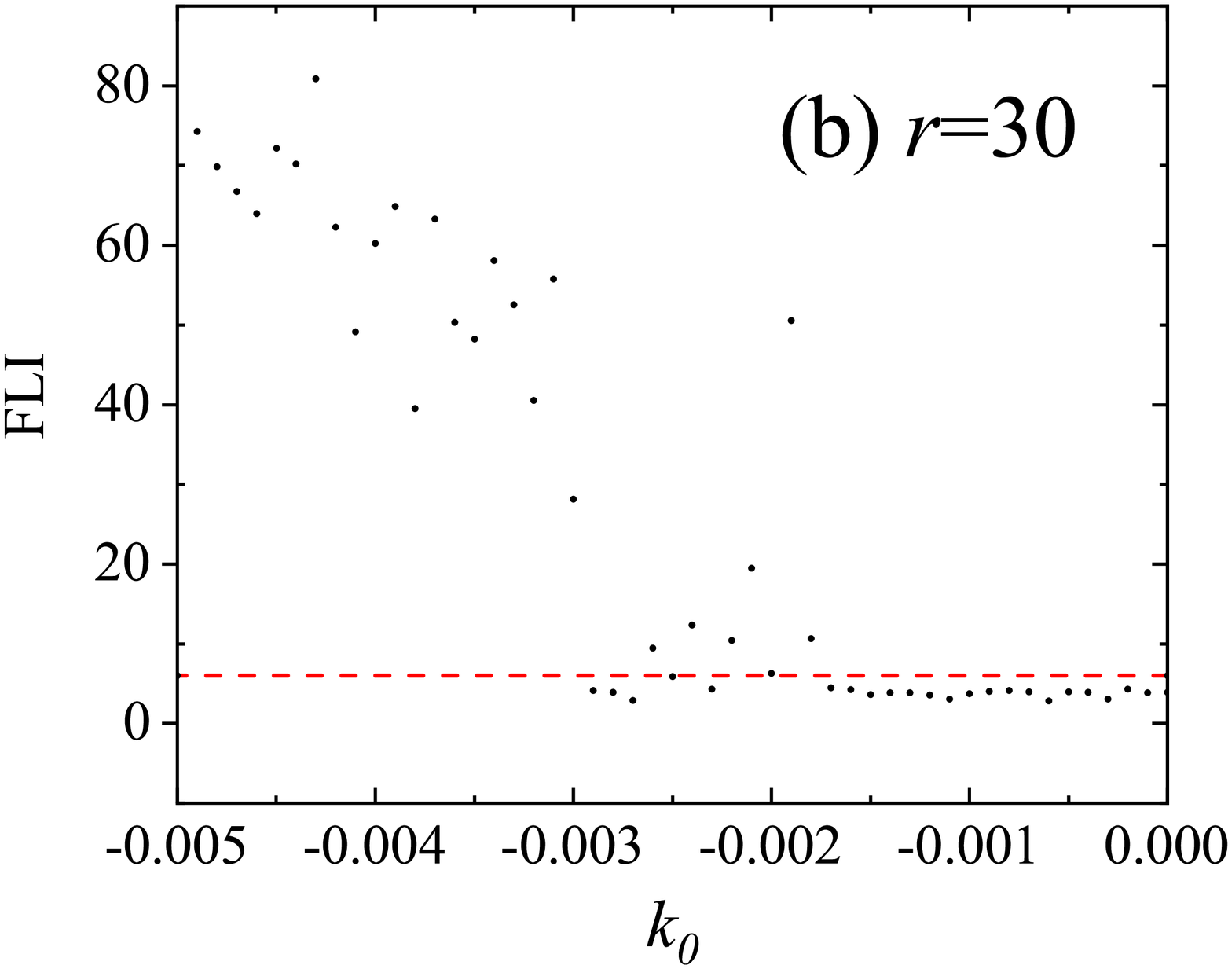}
\caption{(a): Dependence of FLI on the positive deformation
parameter $k_0$ in Figures 7 (a)-(c). The initial separation is
$r=50$. When $k_0>4.2\times 10^{-4}$, chaos begins to lose. (b):
Dependence of FLI on the negative deformation parameter $k_0$ in
Figures 7 (d)-(f). The initial radius is $r=30$. When
$k_0<1.7\times 10^{-3}$, the chaotic properties are strengthened.
} }
\end{figure*}

\begin{figure*}
    \centering{
        \includegraphics[width=12pc]{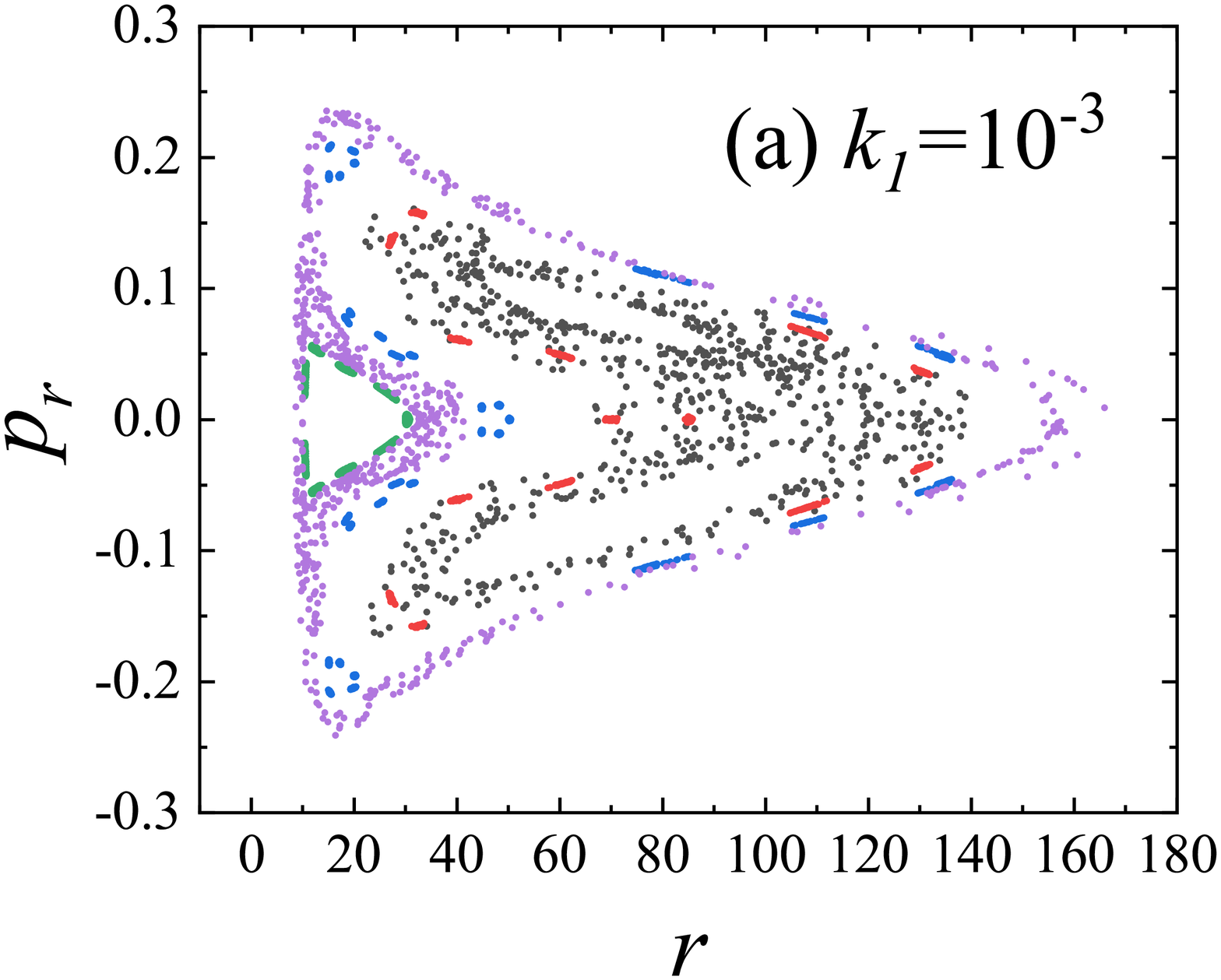}
        \includegraphics[width=12pc]{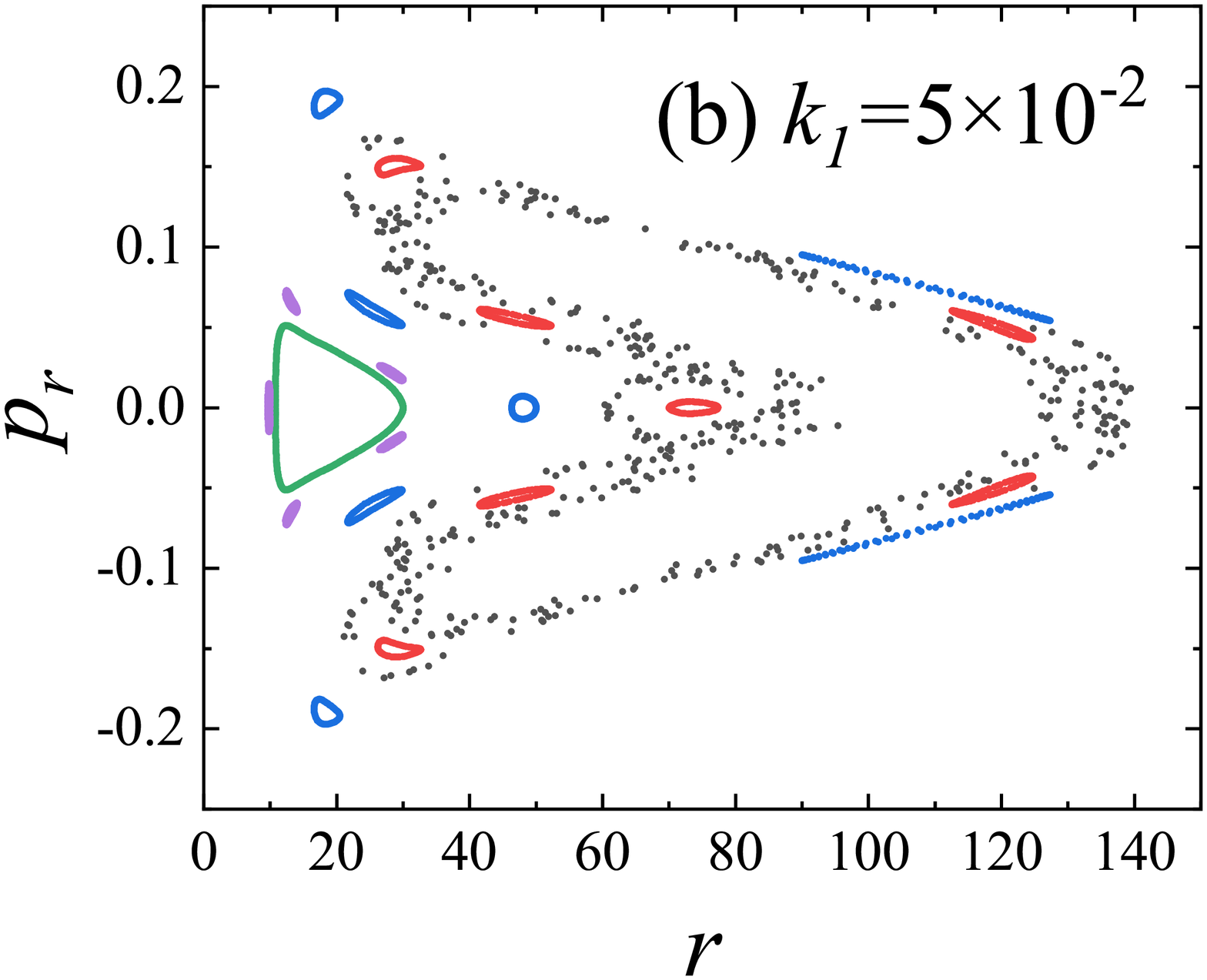}
        \includegraphics[width=12pc]{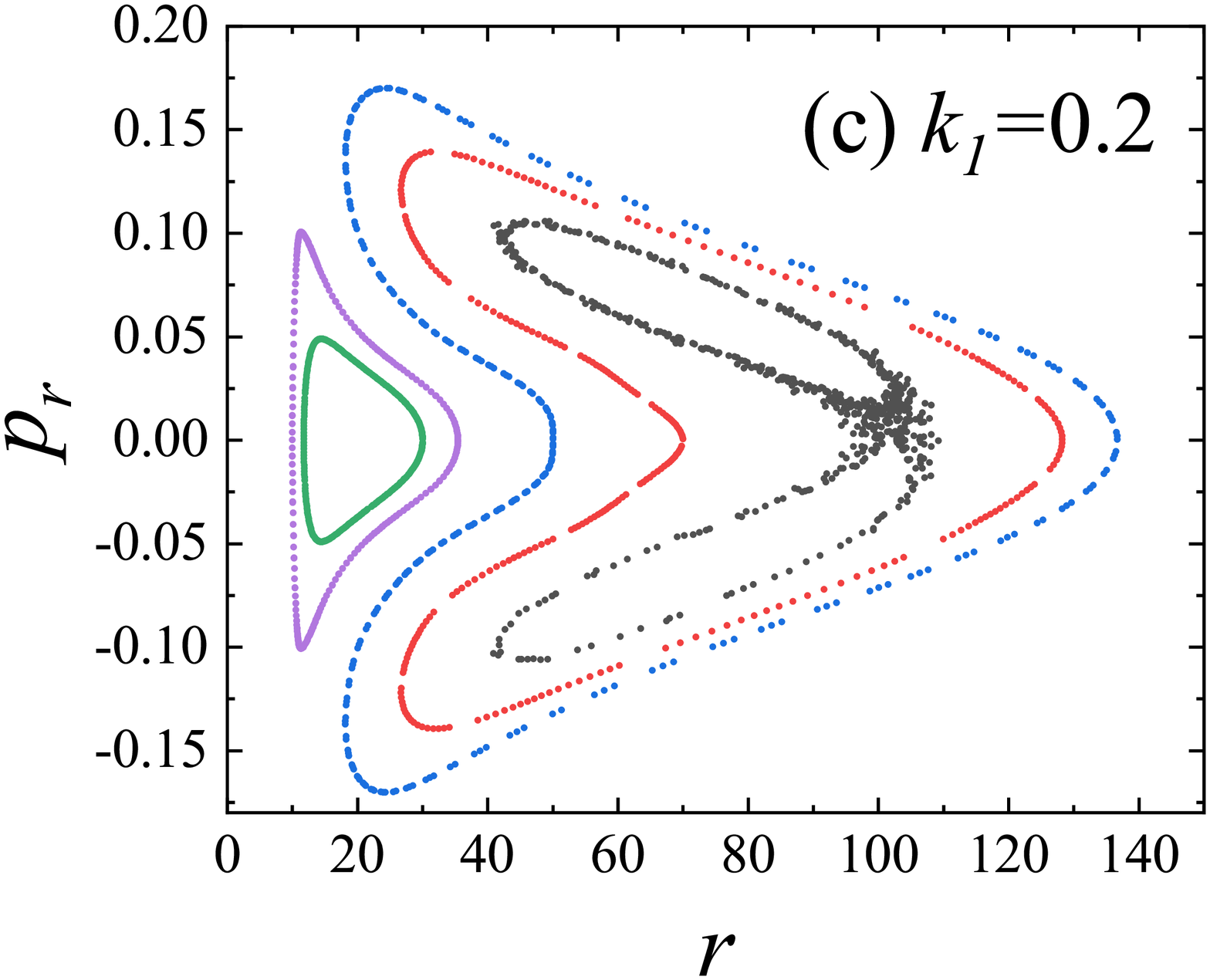}
        \includegraphics[width=12pc]{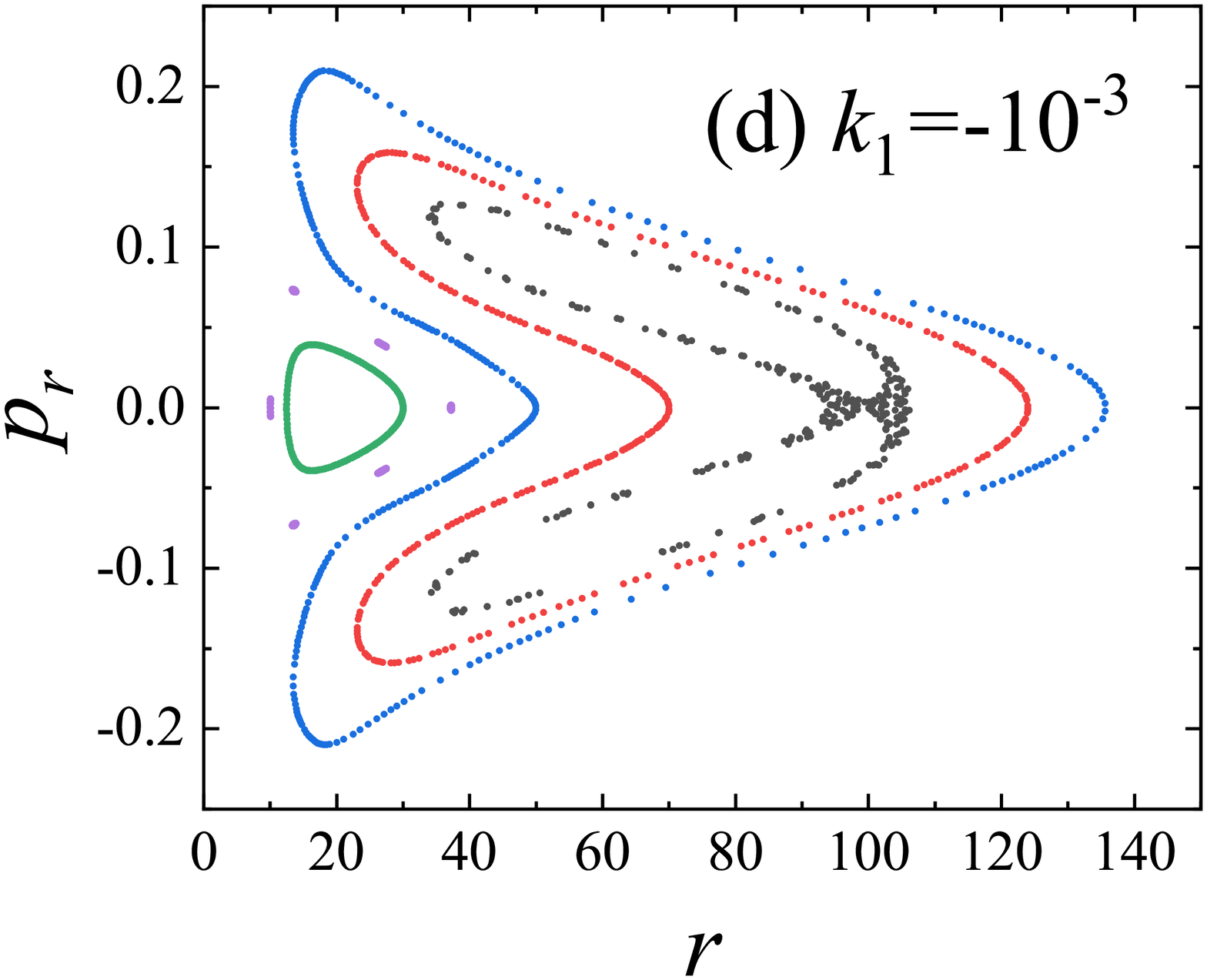}
        \includegraphics[width=12pc]{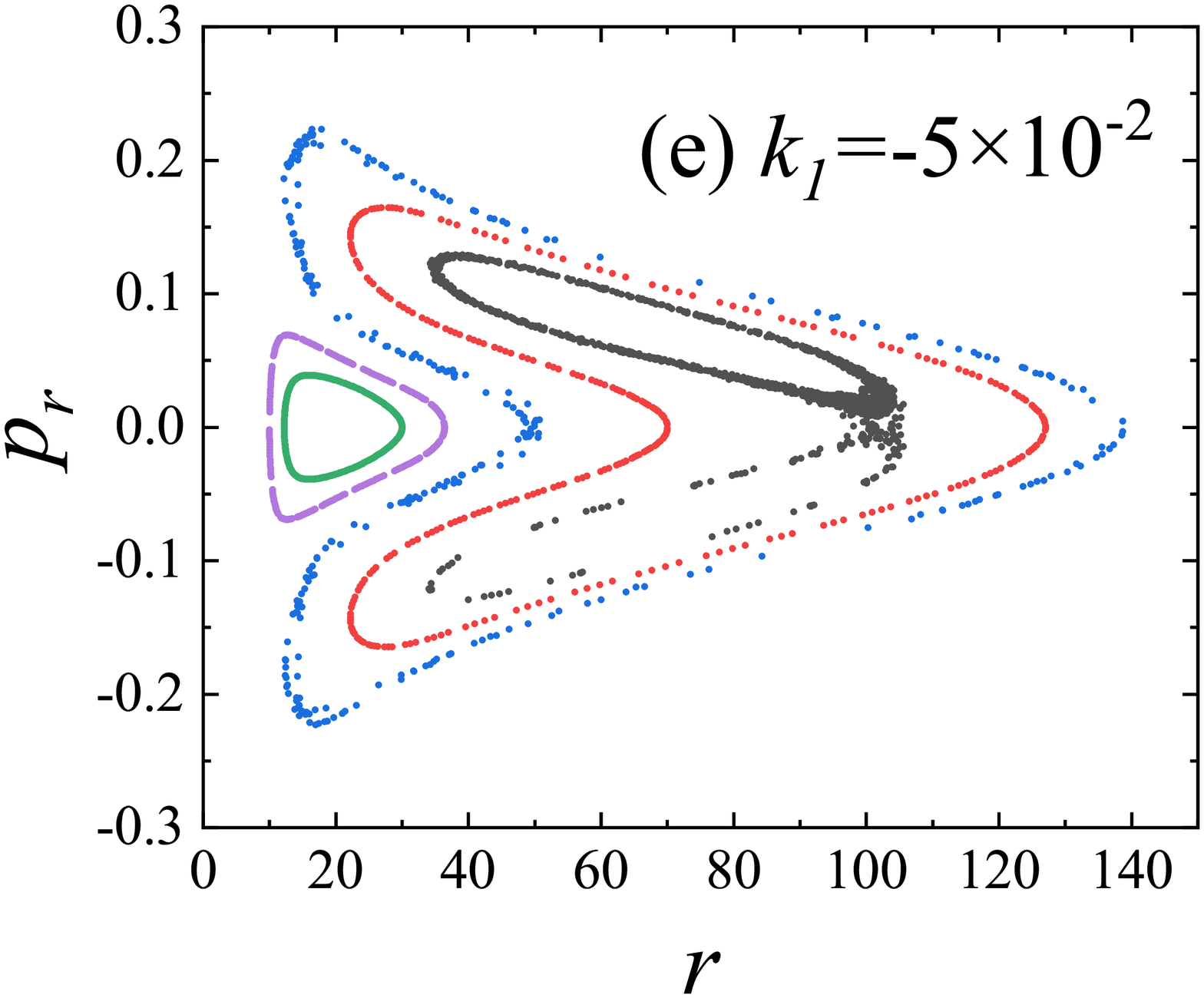}
        \includegraphics[width=12pc]{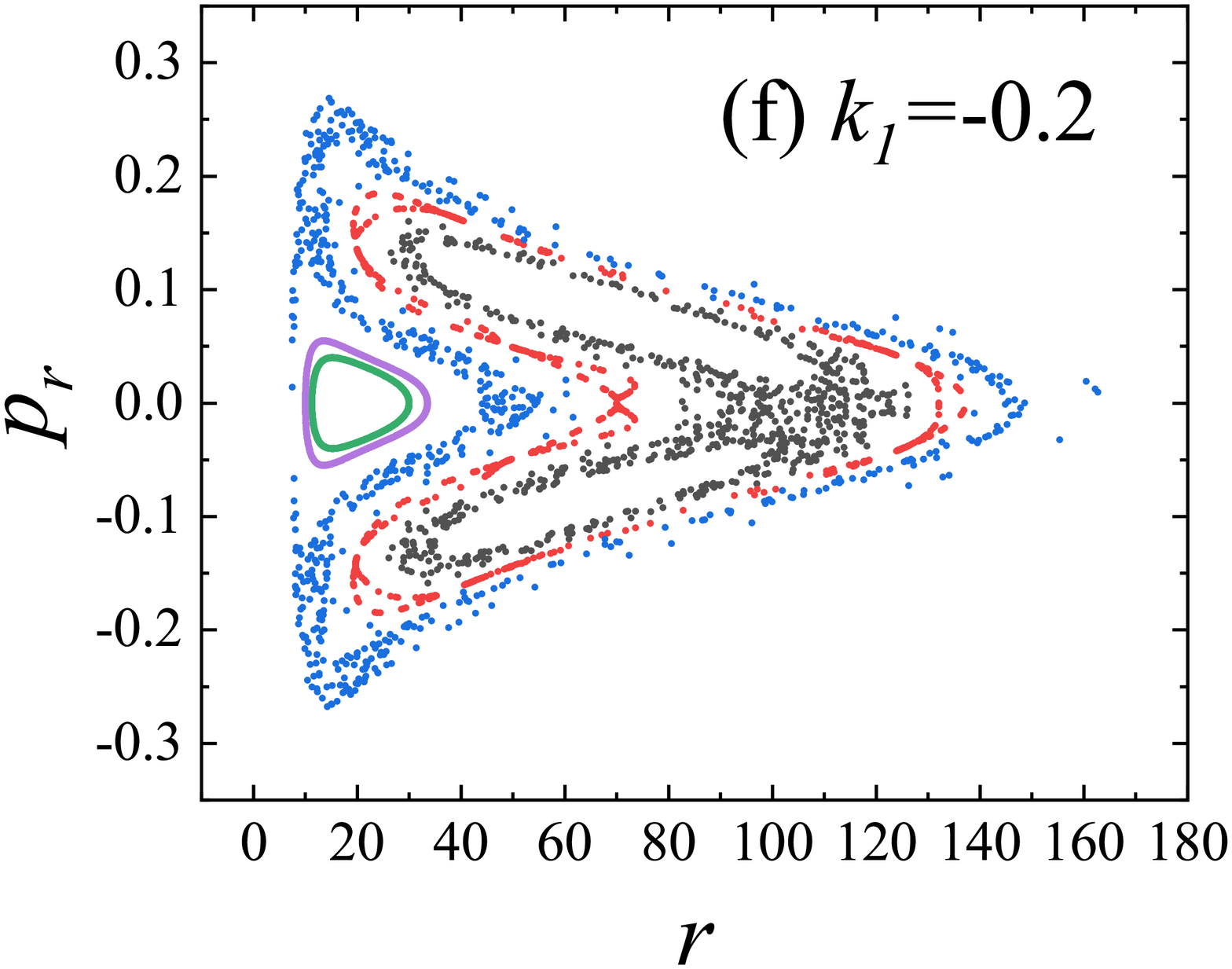}
        \caption{Same as Figure 7, but $k_0$ in Figure 7 is replaced with $k_1$.
        (a)-(c) : $k_0=5\times10^{-4}$. (d)-(f) : $k_0=10^{-4}$.  } }
\end{figure*}

\begin{figure*}
    \centering{
        \includegraphics[width=18pc]{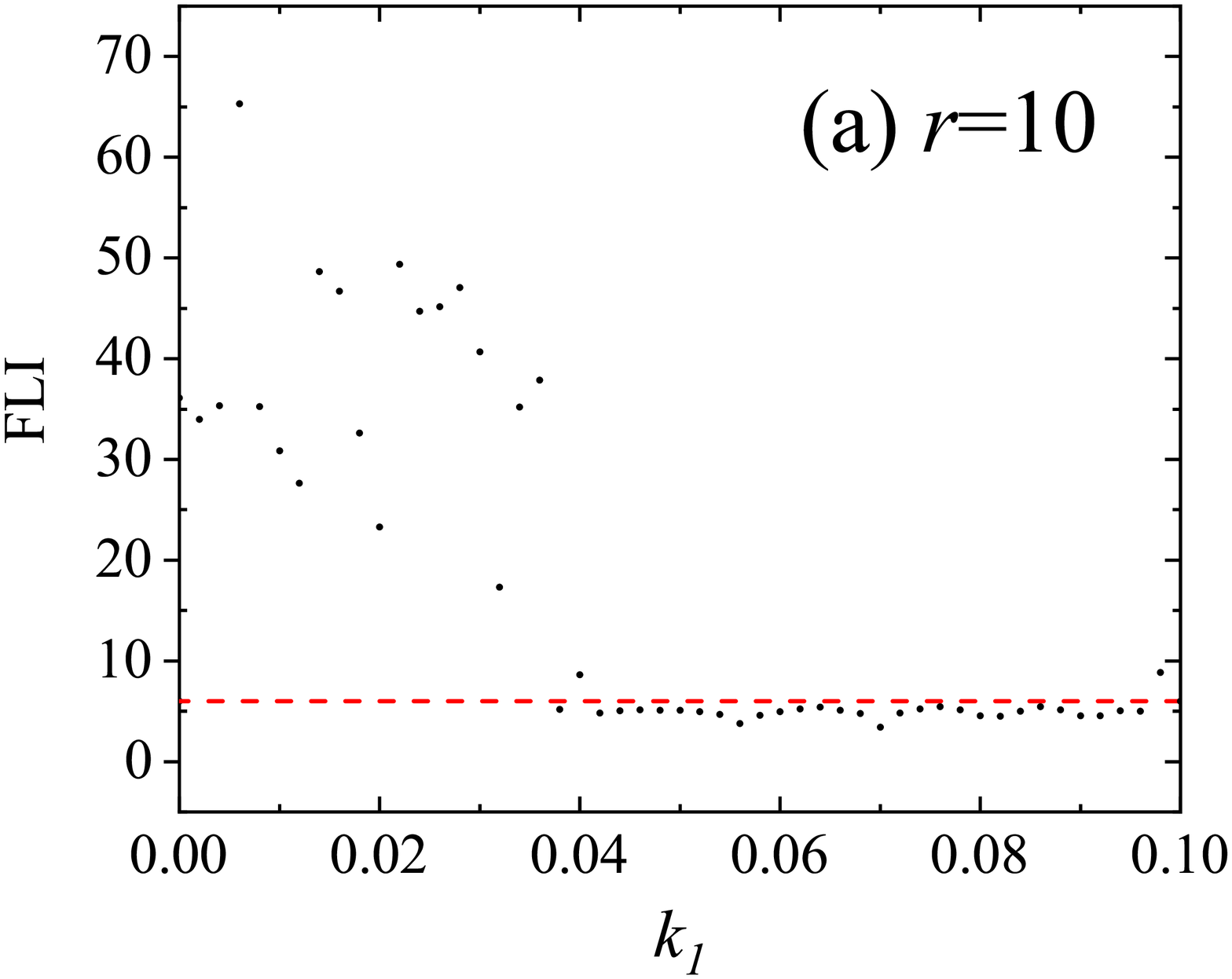}
        \includegraphics[width=18pc]{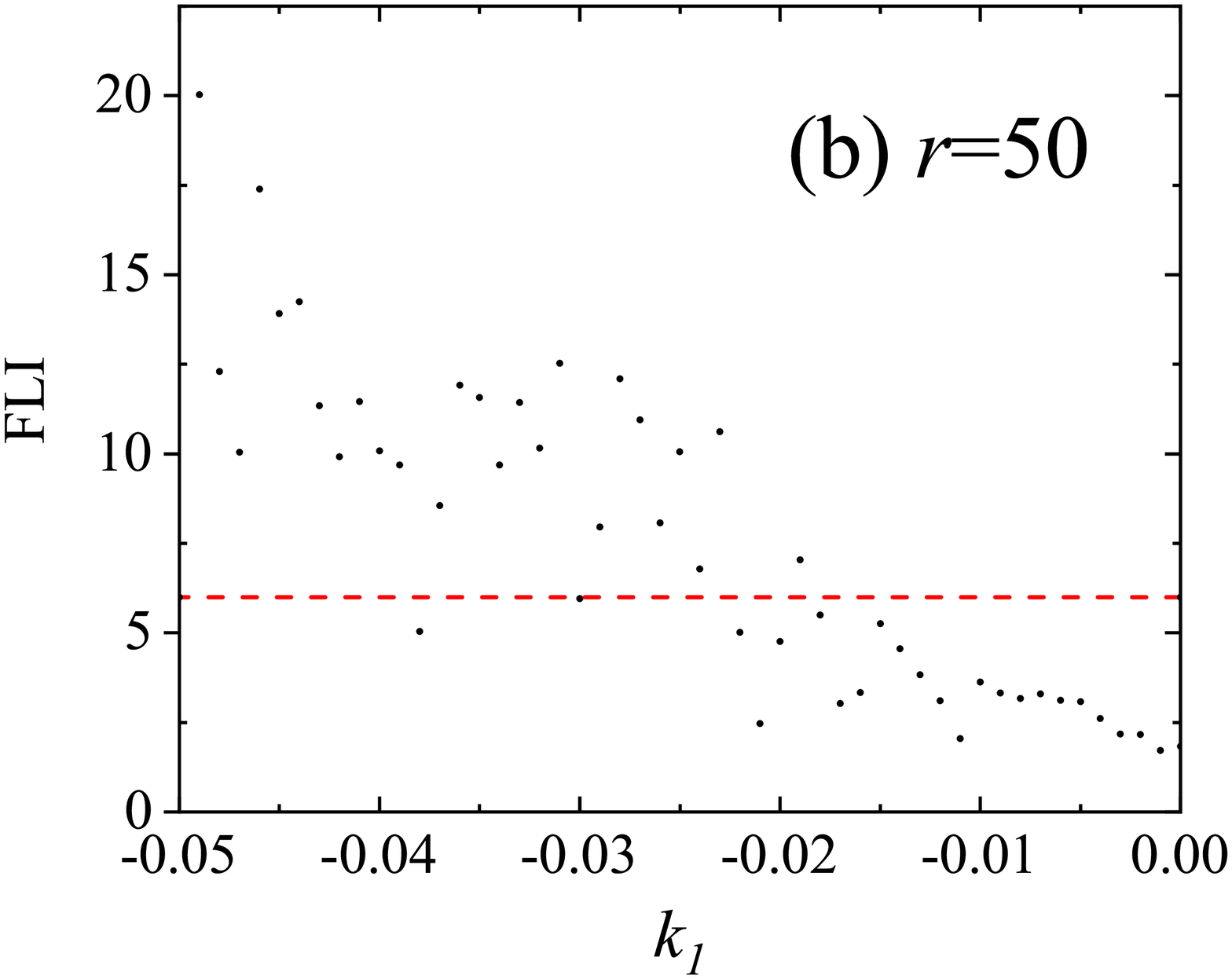}
        \caption{Same as Figure 8, but $k_0$ in Figure 8 is replaced with $k_1$.
        (a): $k_0=5\times10^{-4}$ and the initial radius is $r=10$;
        chaos is ruled out as $k_1>0.042$. (b): $k_0=10^{-4}$ and the initial radius is $r=50$;
        chaos is enhanced as $k_1<-0.018$. } }
\end{figure*}

\begin{figure*}
    \centering{
        \includegraphics[width=12pc]{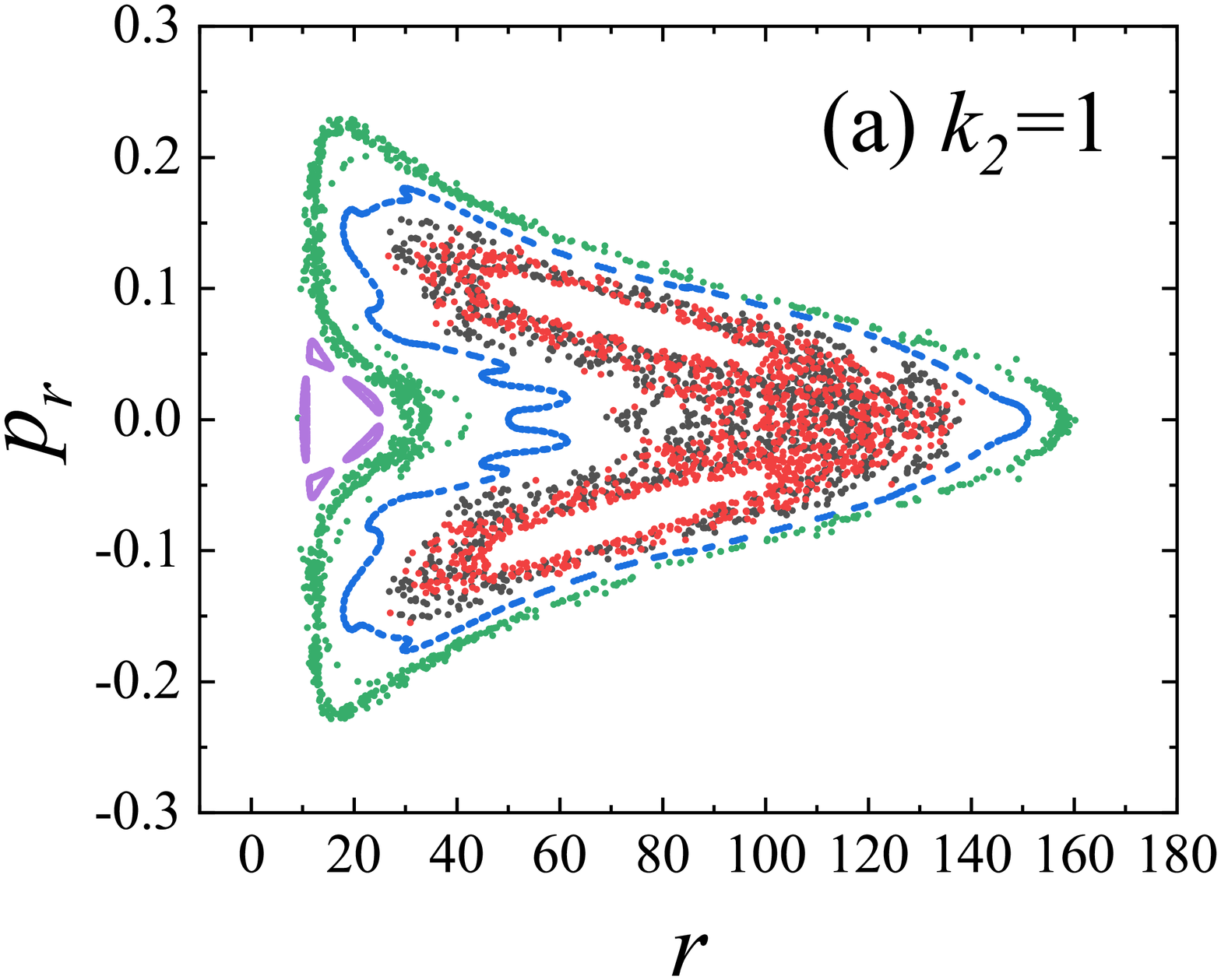}
        \includegraphics[width=12pc]{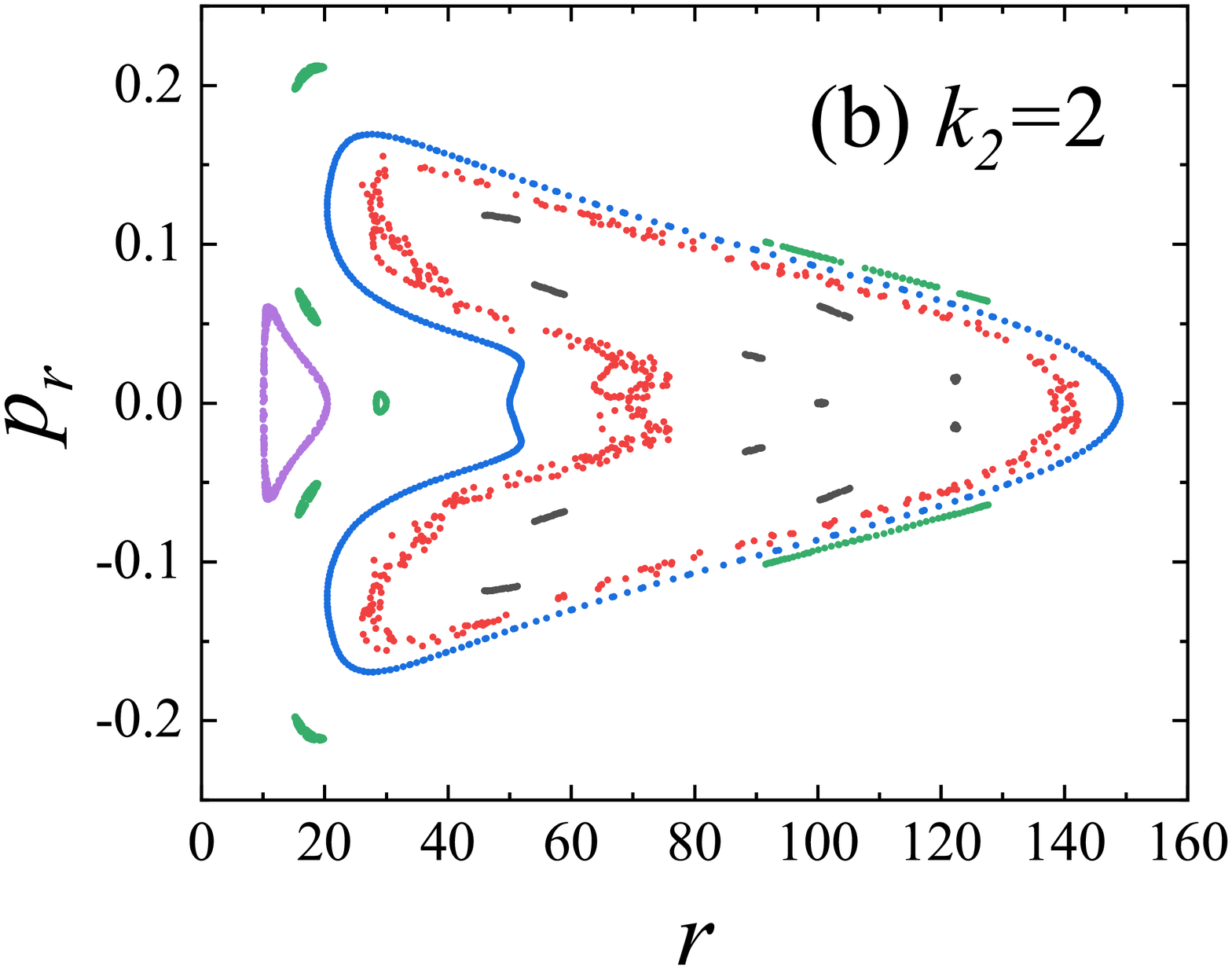}
        \includegraphics[width=12pc]{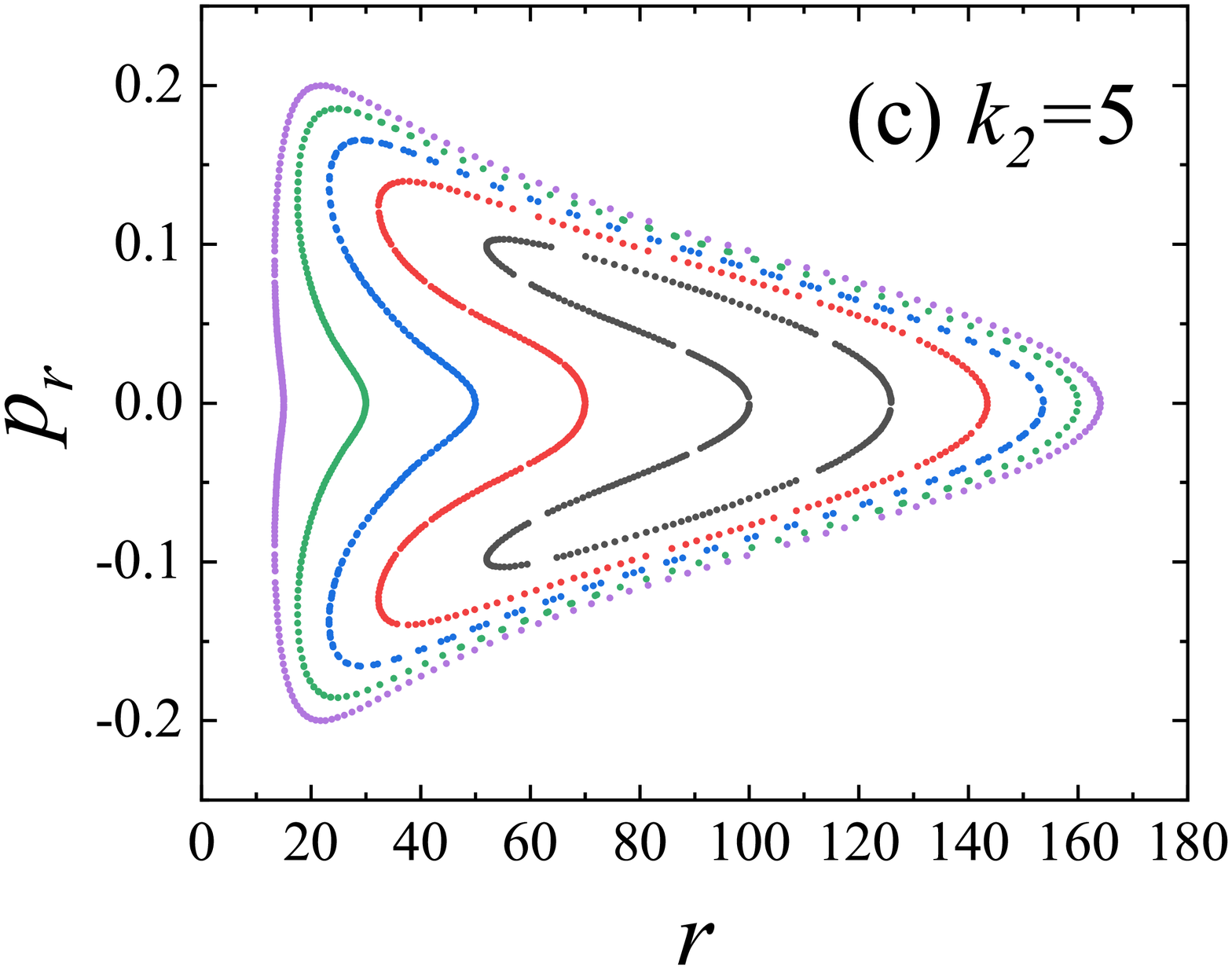}
        \includegraphics[width=12pc]{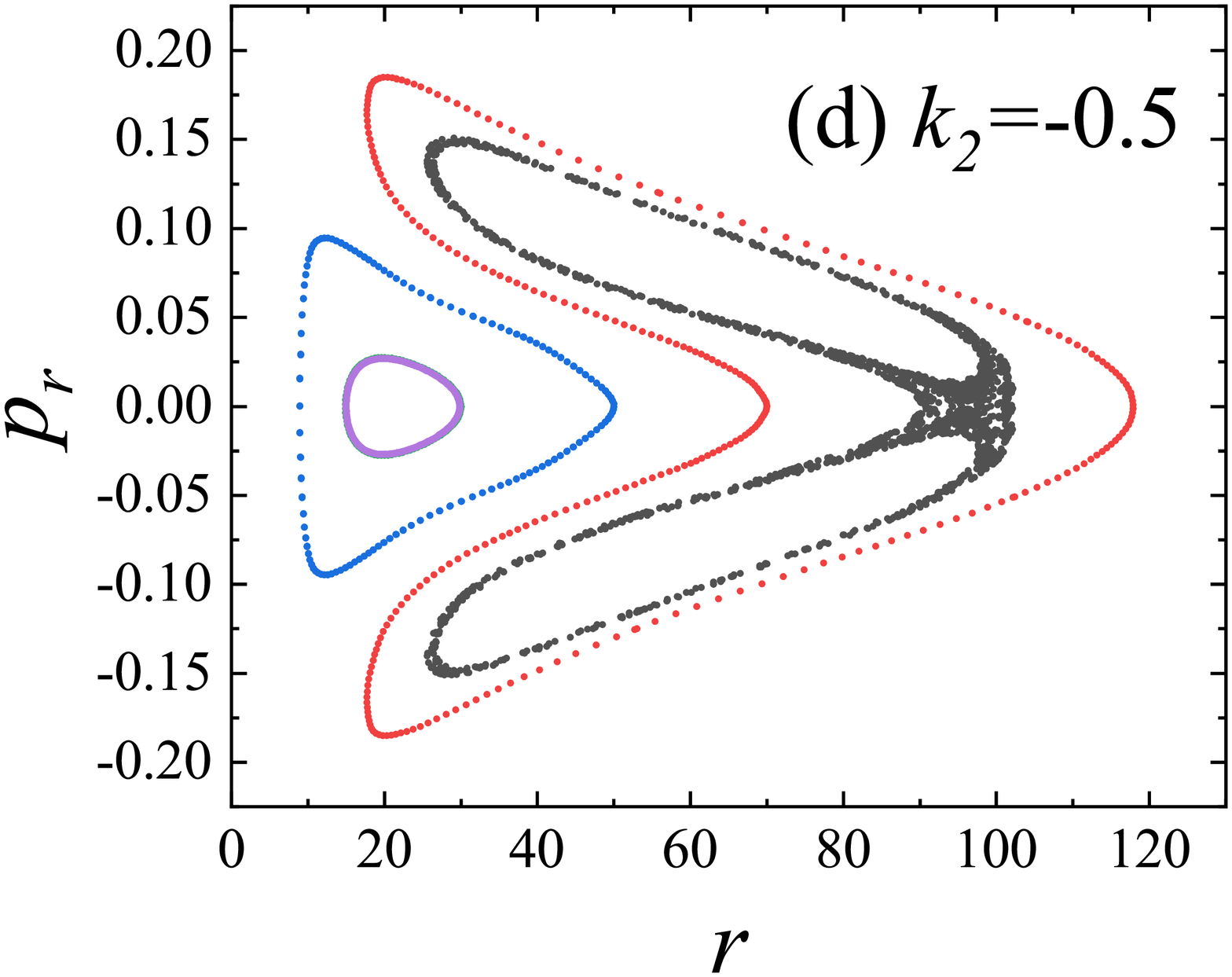}
        \includegraphics[width=12pc]{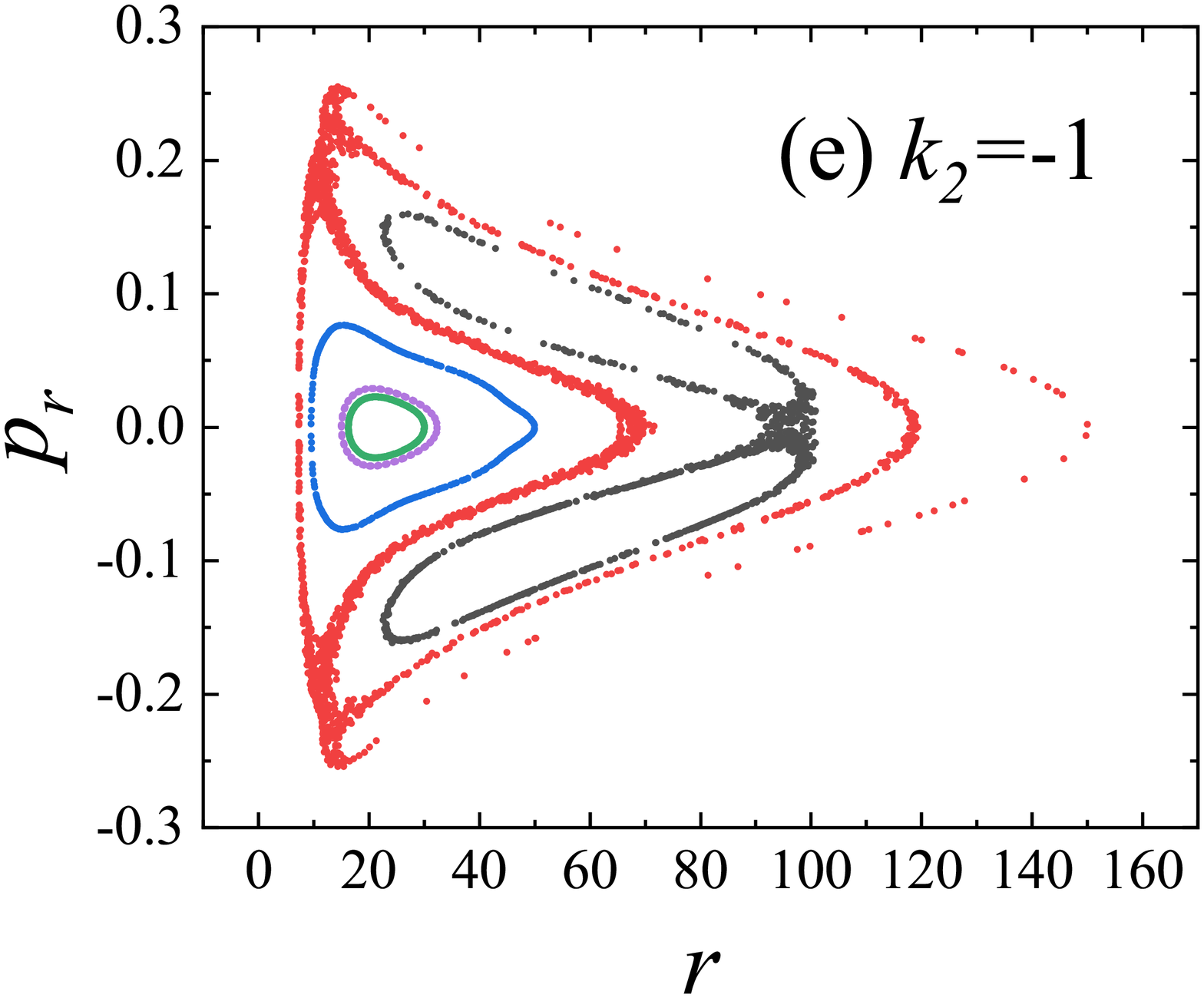}
        \includegraphics[width=12pc]{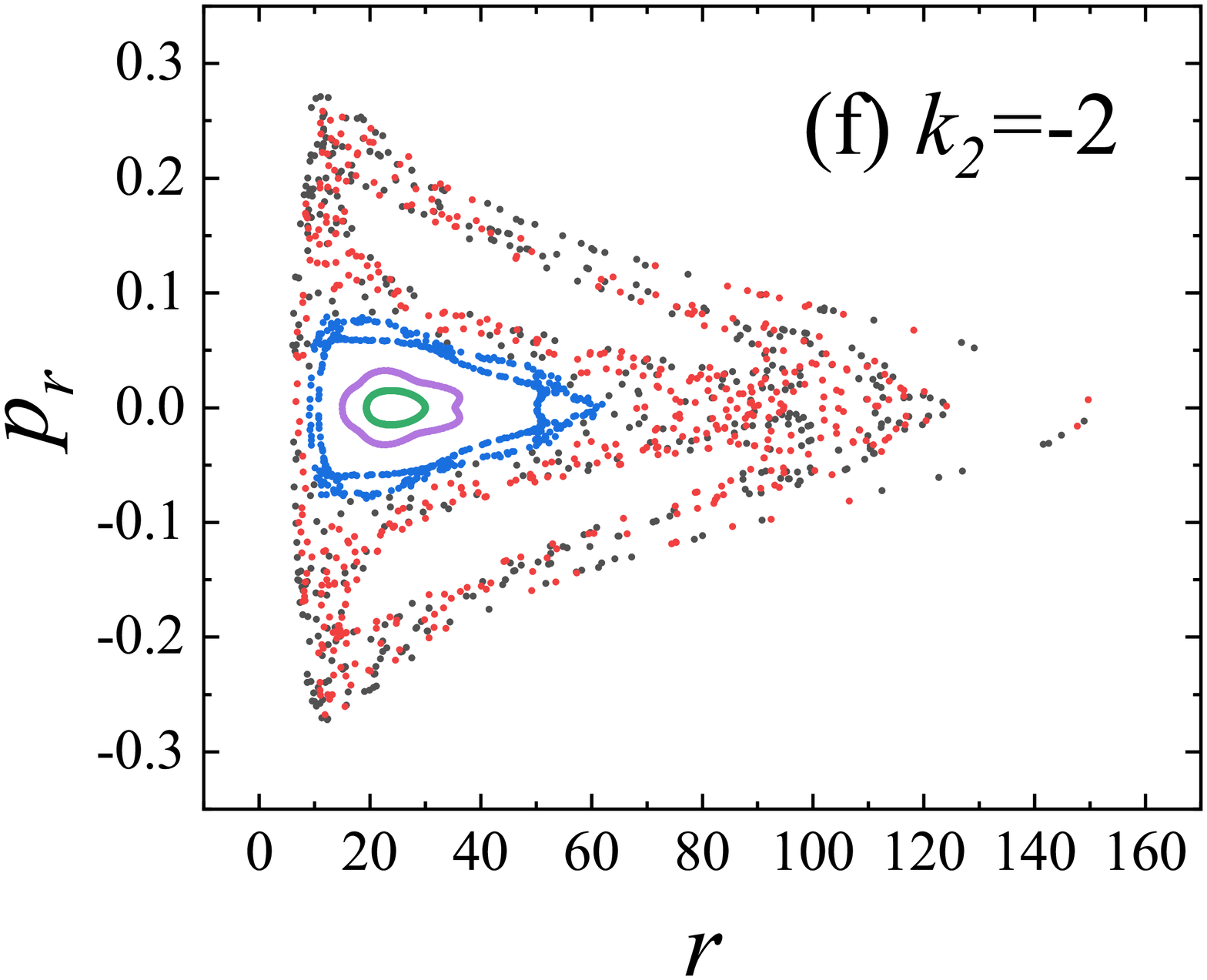}
        \caption{Poincar\'{e} sections for different values of the deformation parameter
        $k_2$. The parameters are $L=4.6$, $Q=8\times10^{-4}$, $k_0=5\times10^{-4}$, $k_1=5\times10^{-3}$ and
        $k_3=1$. (a)-(c):  $E=0.995$ and $k_2>0$. The strength of
        chaos is weakened with an increase of $k_2$. (d)-(f): $E=0.994$ and $k_2<0$.
        Chaos is enhanced as $|k_2|$ increases.
        }  }
\end{figure*}

\begin{figure*}
    \centering{
        \includegraphics[width=18pc]{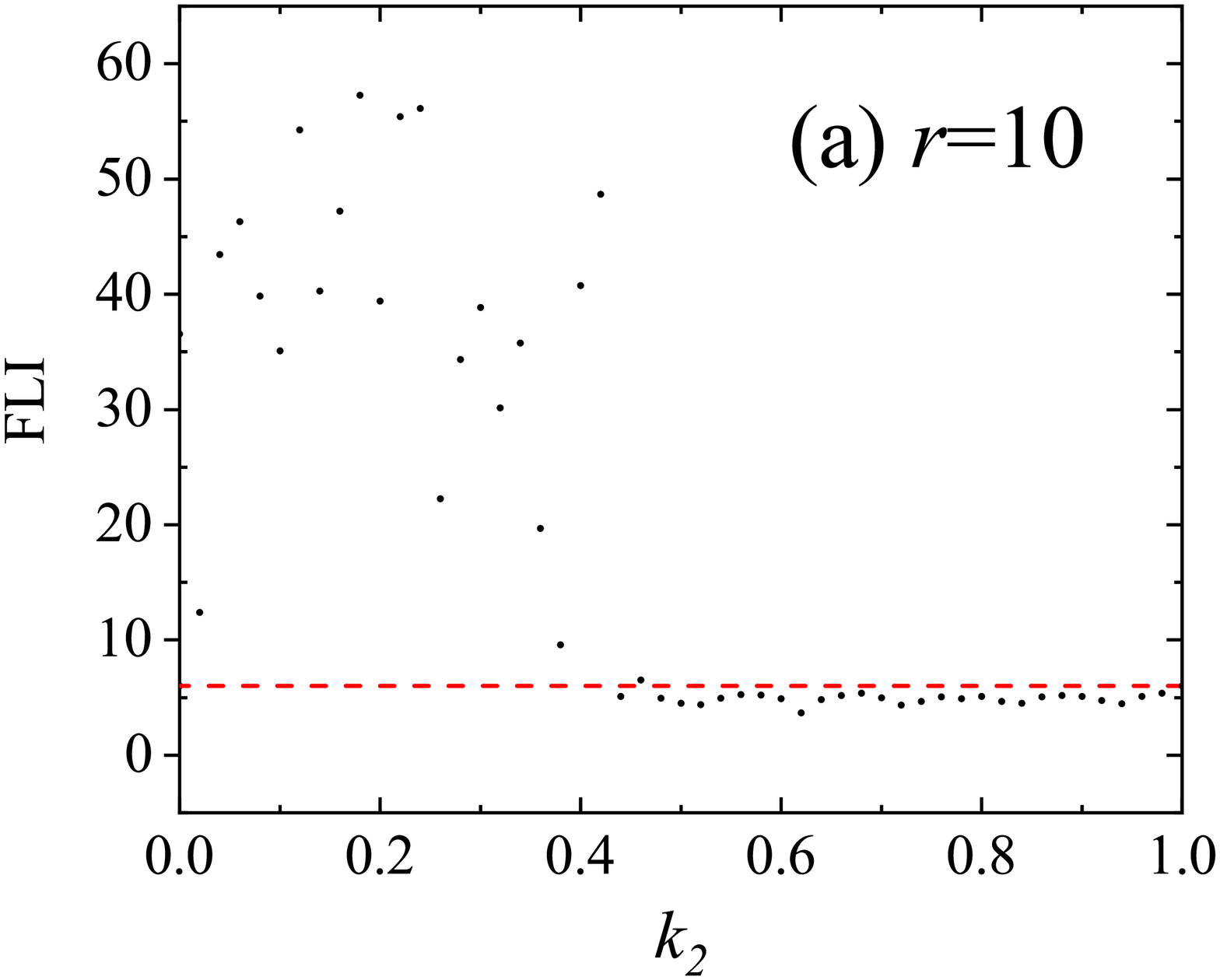}
        \includegraphics[width=18pc]{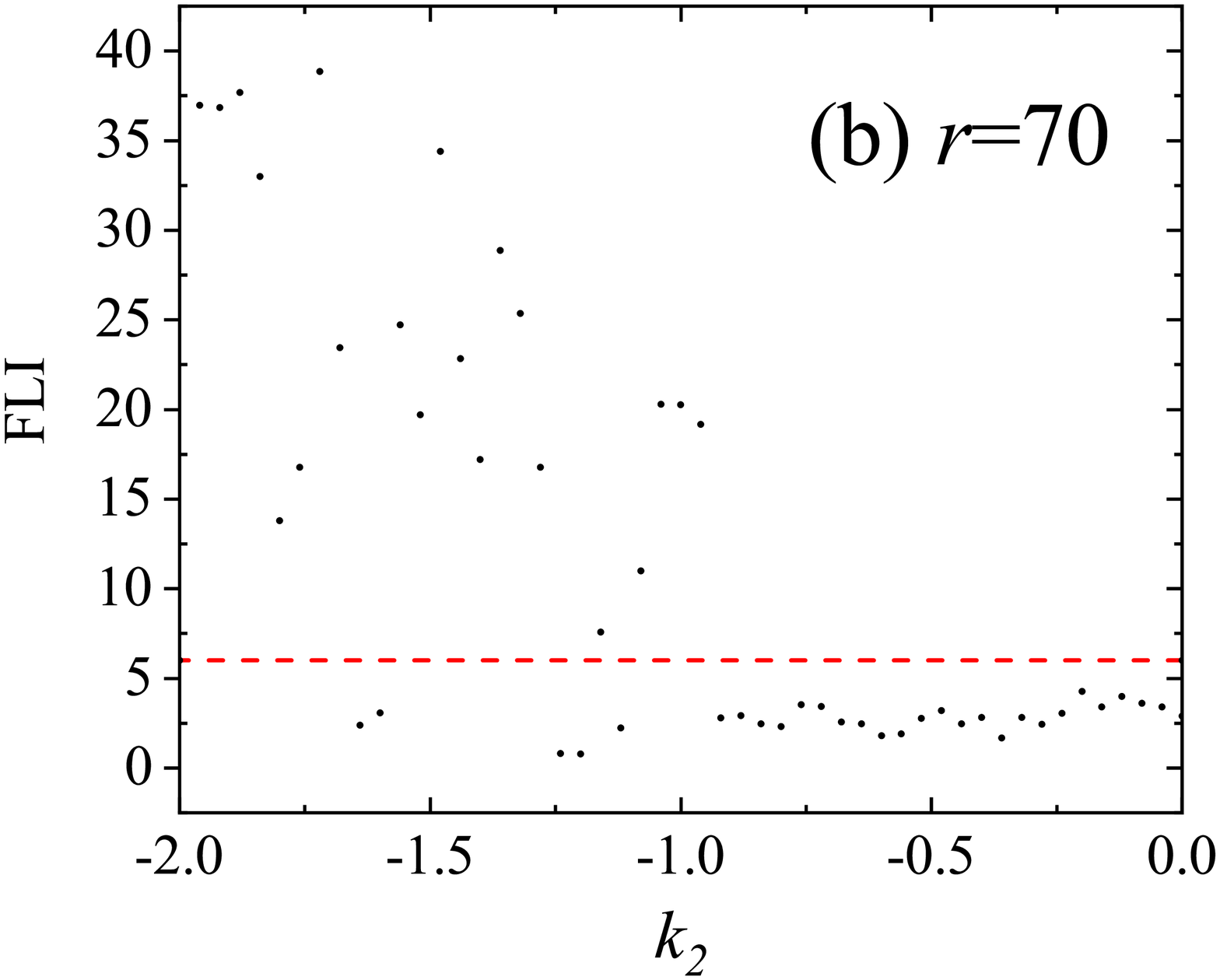}
        \caption{(a): Dependence of FLI on the positive deformation
parameter $k_2$ in Figures 11 (a)-(c). The initial separation is
$r=10$. When $k_2>0.46$, chaos is absent. (b): Dependence of FLI
on the negative deformation parameter $k_2$ in Figures 11 (d)-(f).
The initial radius is $r=70$. When $k_2<-0.96$, the chaotic
properties are strengthened.} }
\end{figure*}

\begin{figure*}
    \centering{
        \includegraphics[width=12pc]{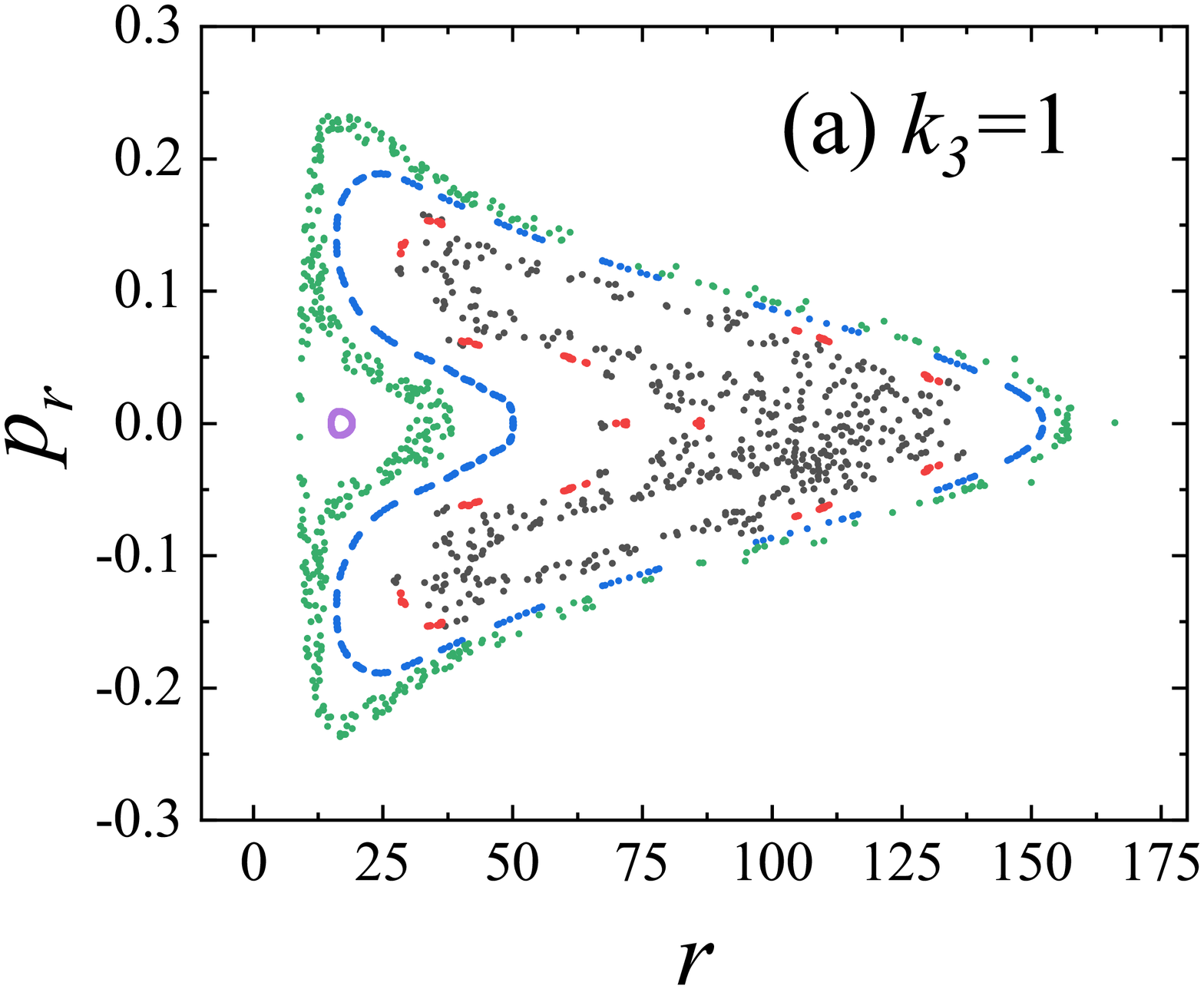}
        \includegraphics[width=12pc]{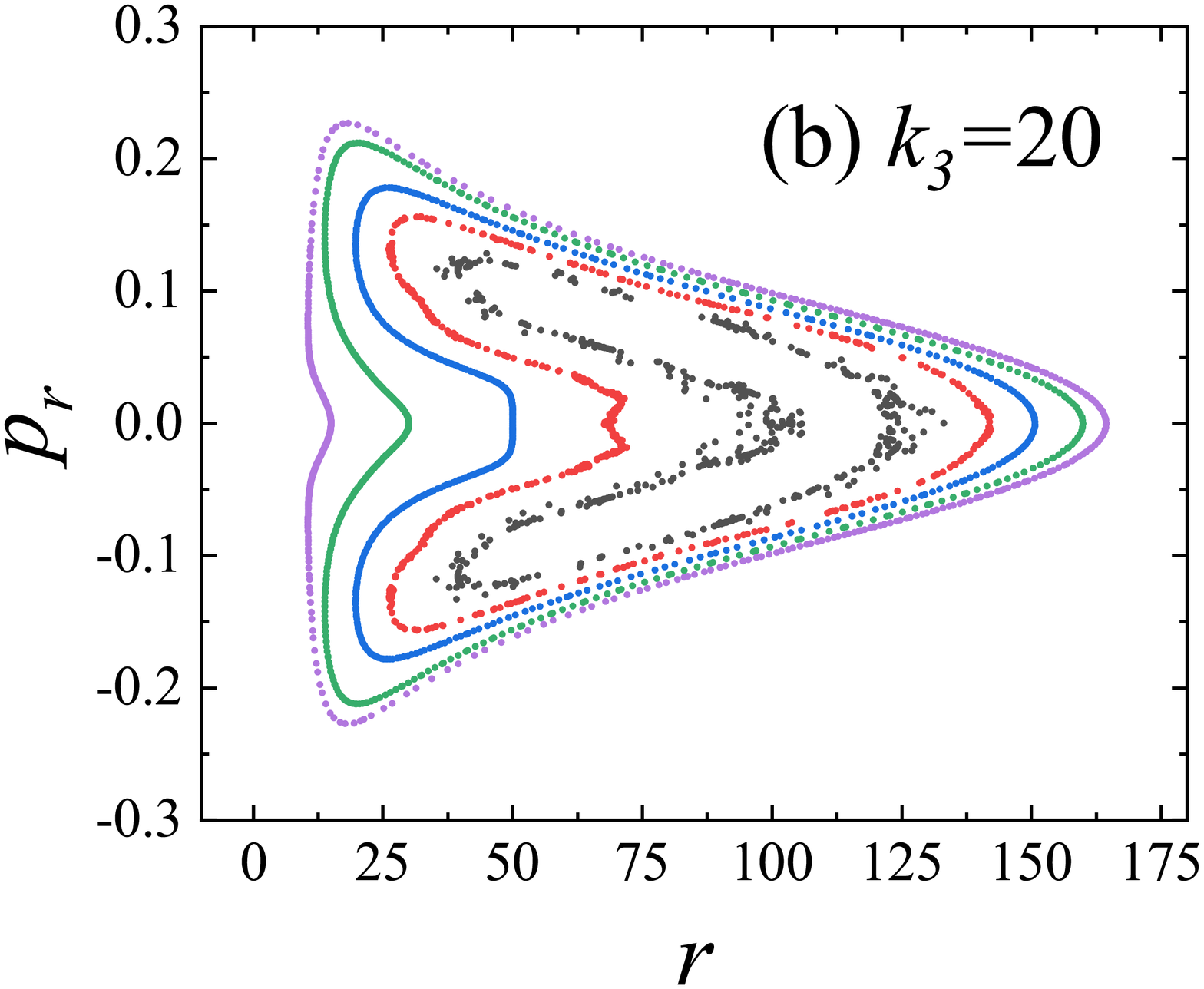}
        \includegraphics[width=12pc]{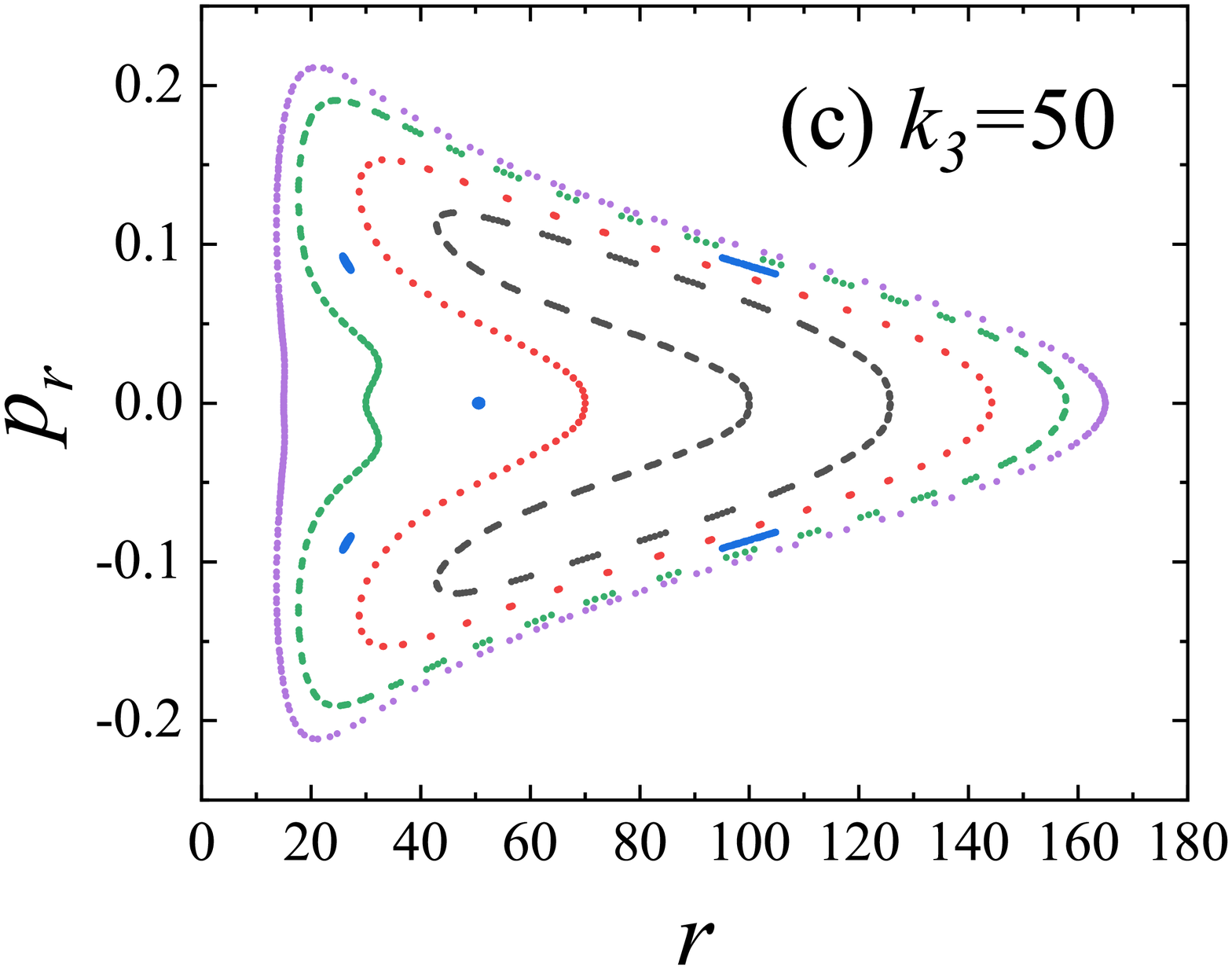}
        \includegraphics[width=12pc]{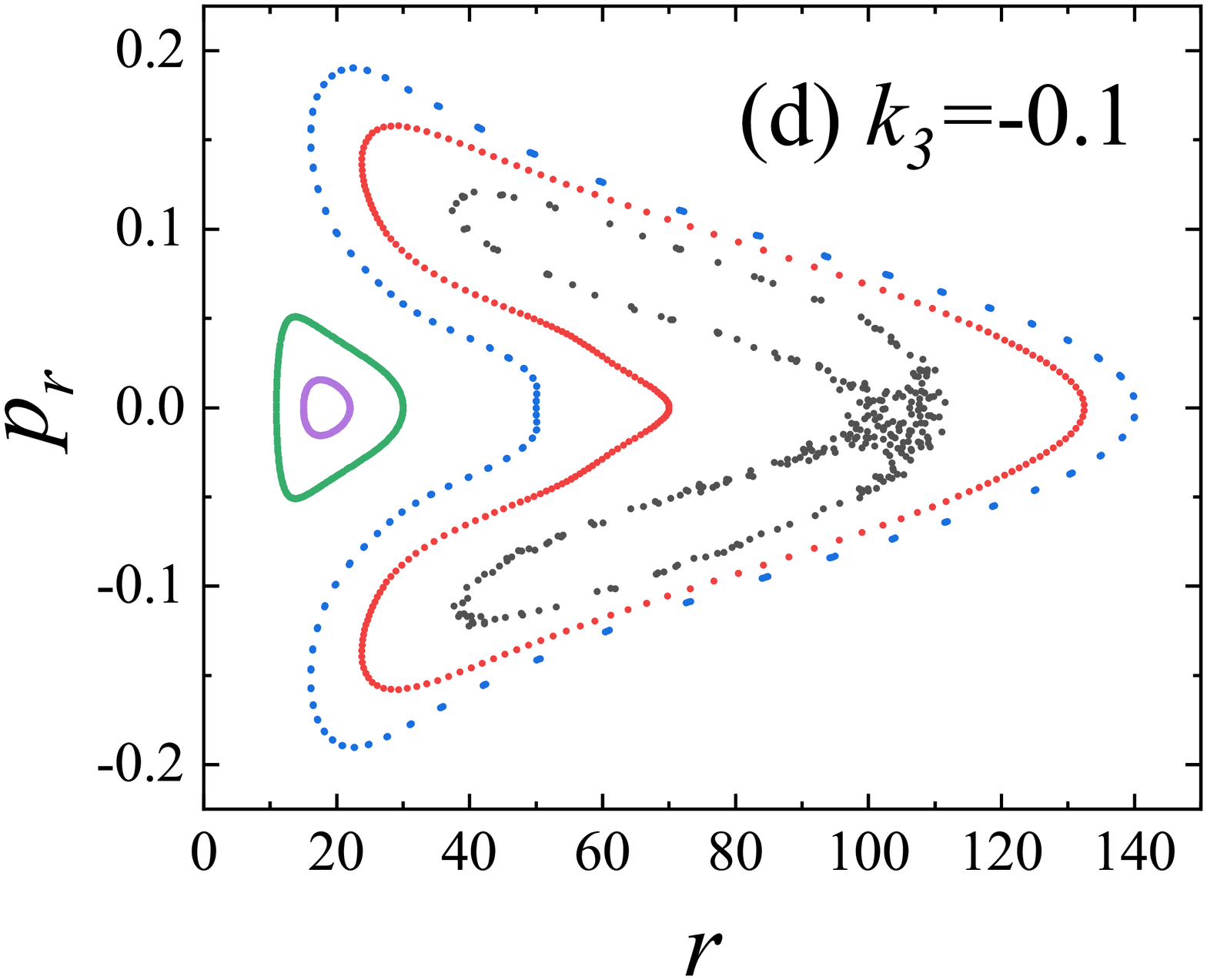}
        \includegraphics[width=12pc]{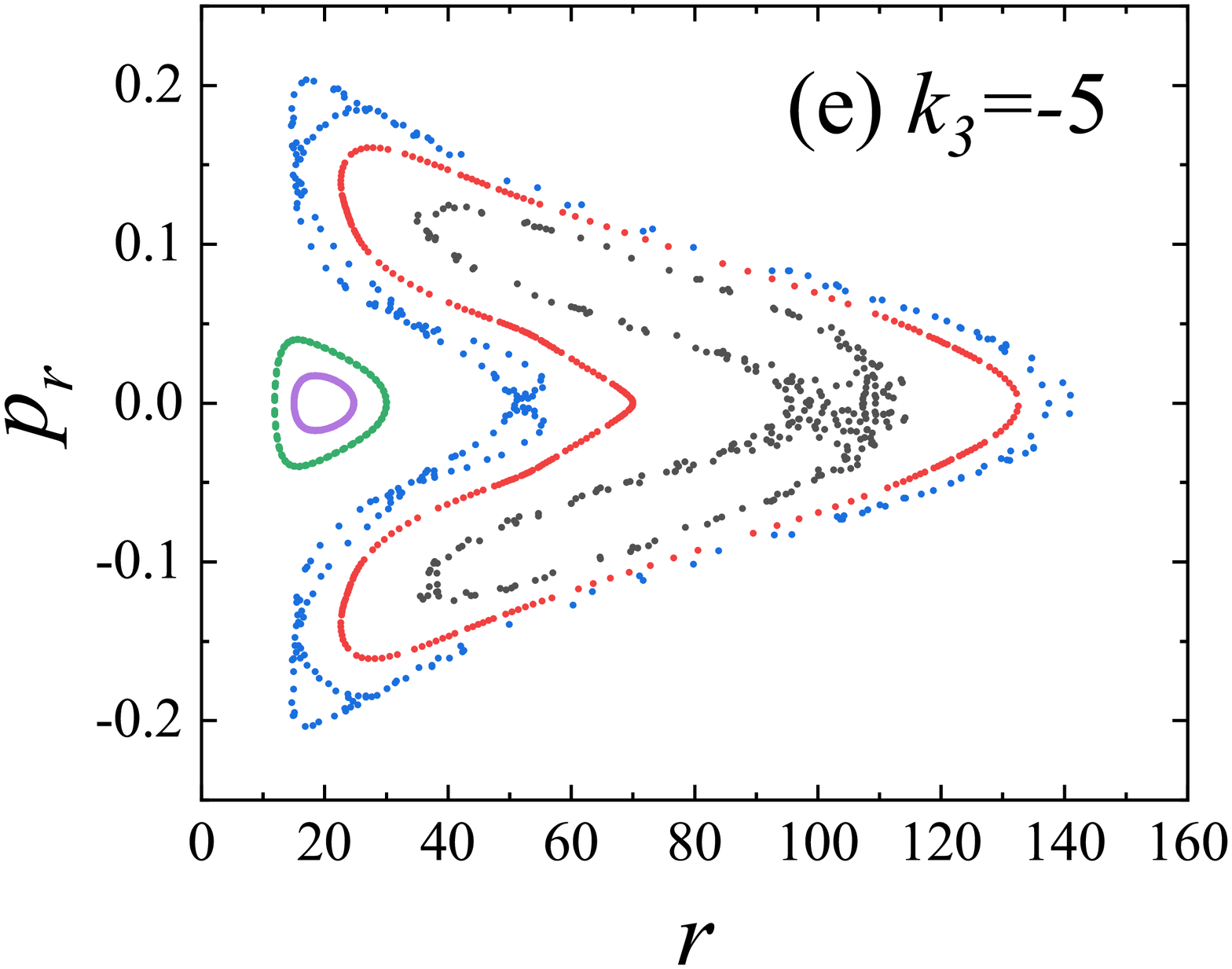}
        \includegraphics[width=12pc]{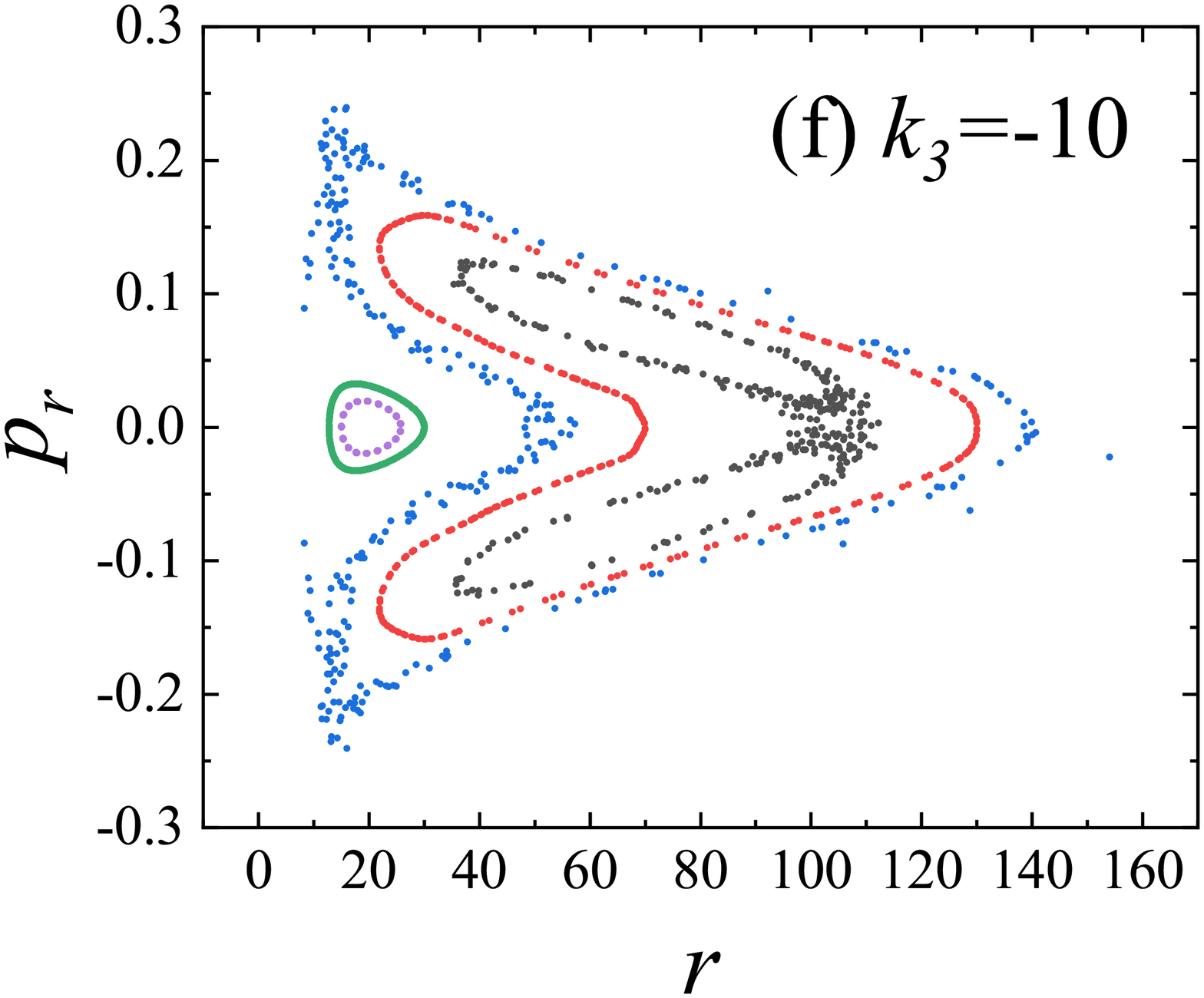}
        \caption{Similar to Figure 11, but $k_2$ in Figure 11 is replaced with $k_3$and energies $E$ are different. Here,
        $k_2=0.5$.  (a)-(c): $E=0.995$ and $k_3>0$.  (d)-(f): $E=0.9945$ and $k_3<0$.  } }
\end{figure*}

\begin{figure*}
    \centering{
        \includegraphics[width=18pc]{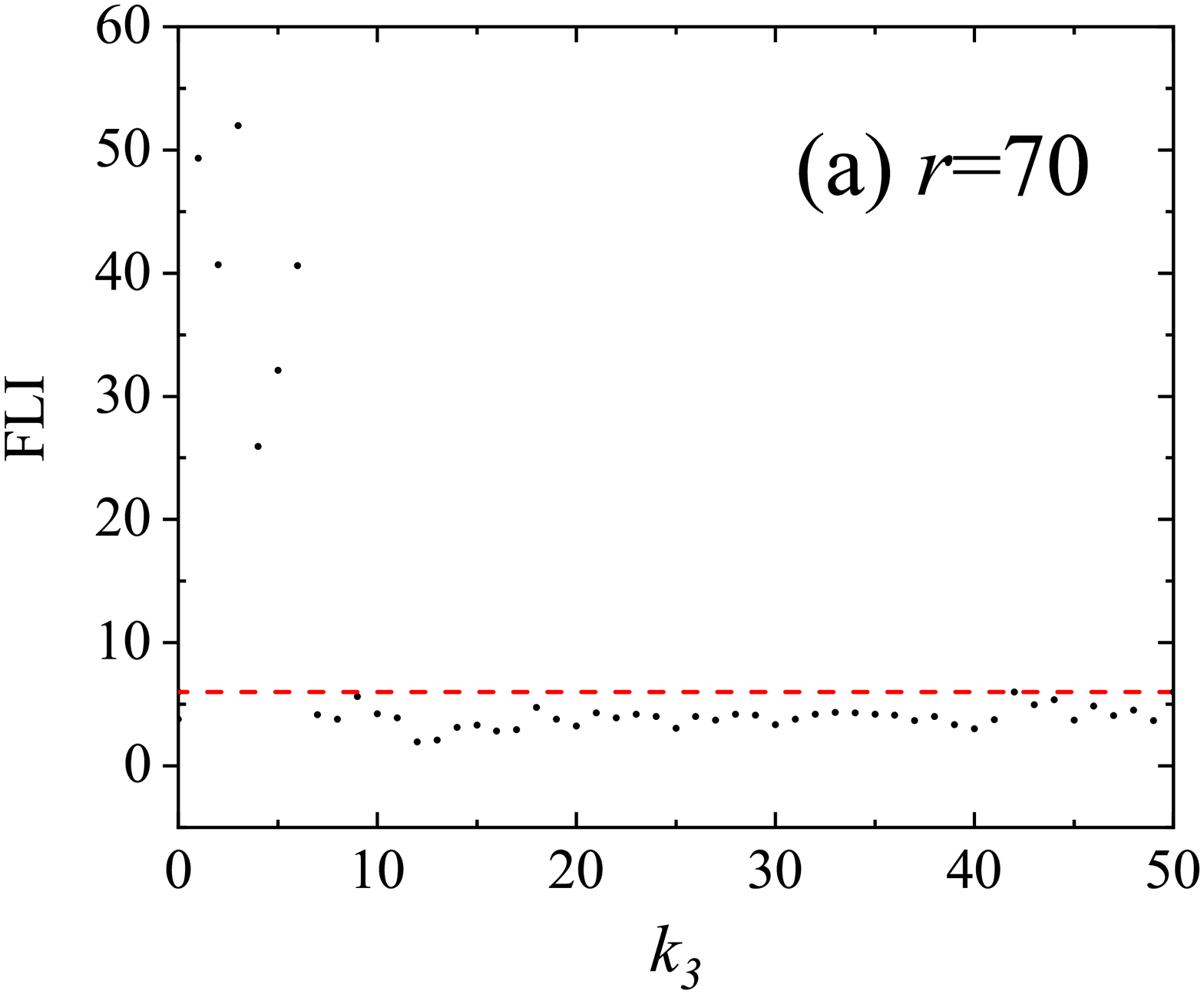}
        \includegraphics[width=18pc]{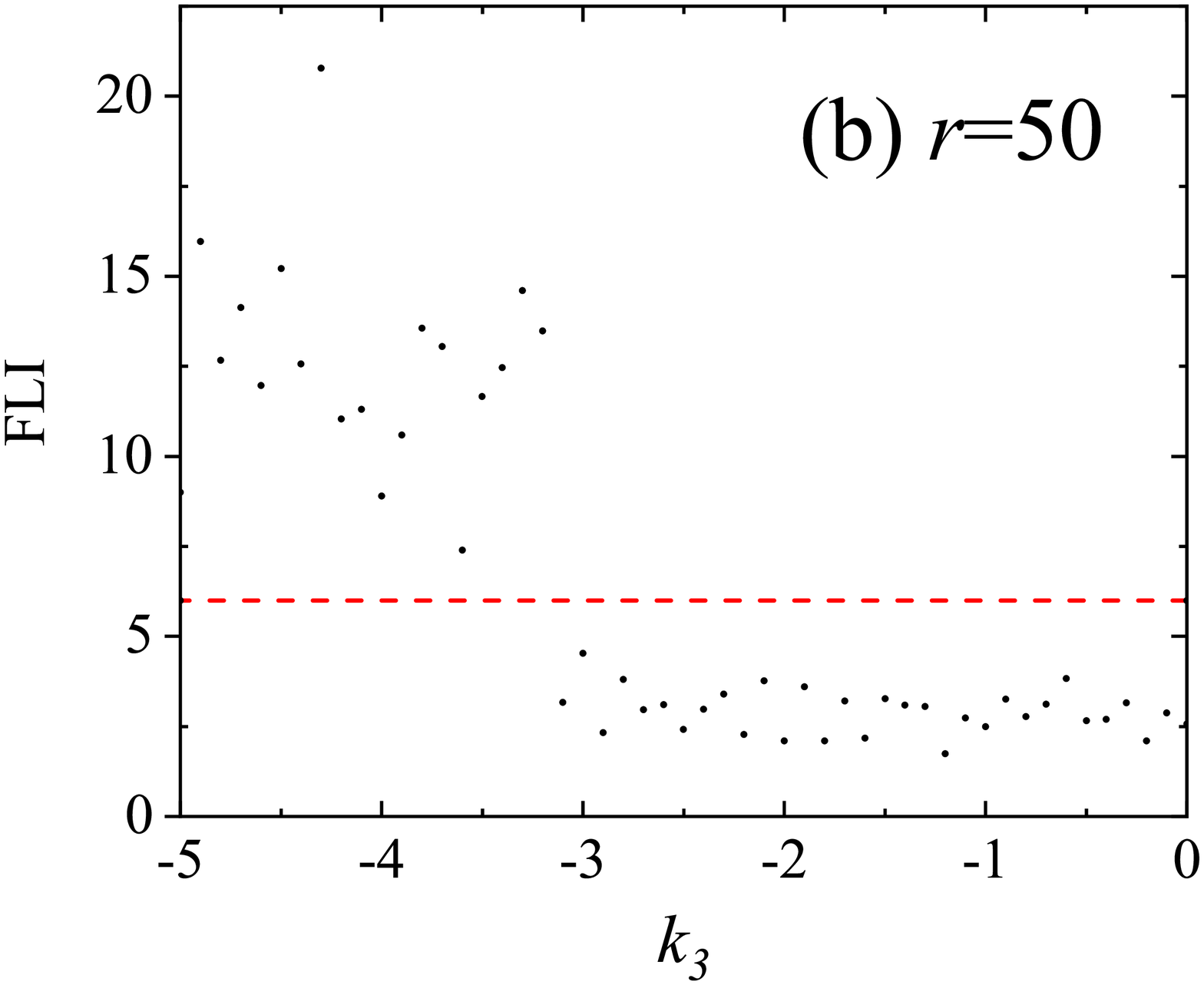}
        \caption{Dependence of FLI on the positive deformation
parameter $k_3$. The other parameters are the same as those in
Figure 13. (a): $E=0.995$, $k_3>0$ and the initial separation
$r=70$; when $k_3>6$, chaos begins to disappear. (b): $E=0.9945$,
$k_3<0$ and the initial separation $r=50$; when $k_3<-3.2$, the
chaotic properties are strengthened. } }
\end{figure*}

\end{document}